\RequirePackage{etex}
\documentclass[conf]{new-aiaa}
\usepackage[utf8]{inputenc}
\usepackage[pdftex]{graphicx}
\usepackage{stfloats}

\usepackage{url}
\usepackage{abstract}
\usepackage{bm}
\usepackage{comment}

\usepackage{float}
\usepackage[english]{babel}
\usepackage[pdftex]{graphicx}                     
\usepackage{rotating}  
\usepackage{wrapfig}
\usepackage{subcaption}
\usepackage{booktabs} 

\usepackage{indentfirst}                      
\usepackage{url}
\usepackage{lipsum}
\usepackage{gensymb} 
\usepackage{lscape} 
\usepackage[final]{pdfpages}
\usepackage{afterpage}
\usepackage{array}
\usepackage{hyperref} 
\usepackage{cleveref} 
\hypersetup{%
  colorlinks=false,
}
\usepackage{graphicx}
\usepackage[format=hang,font={small,bf}]{caption}
\usepackage{paralist}
\usepackage{multirow}
\usepackage{tabulary}
\usepackage{hhline}
\usepackage{enumerate}

\usepackage[referable]{threeparttablex}
\renewlist{tablenotes}{enumerate}{1}
\makeatletter
\setlist[tablenotes]{label=\tnote{\alph*},ref=\alph*,itemsep=\z@,topsep=\z@skip,partopsep=\z@skip,parsep=\z@,itemindent=\z@,labelindent=\tabcolsep,labelsep=.2em,leftmargin=*,align=left,before={\footnotesize}}
\makeatother

\usepackage{catchfilebetweentags} %
\usepackage{tikz}
\usepackage{xifthen}
\usepackage{xcolor}
\usepackage[export]{adjustbox}
\usepackage{pgfplots}
\pgfplotsset{compat=1.15}
\usetikzlibrary{shapes,arrows,backgrounds}

\newcommand{\bvec}[1]{\boldsymbol{#1}}

\usepackage{eurosym}
\usepackage{amstext} 
\DeclareRobustCommand{\officialeuro}{%
  \ifmmode\expandafter\text\fi
  {\fontencoding{U}\fontfamily{eurosym}\selectfont e}}
  
  \makeatletter
\renewcommand*\env@matrix[1][*\c@MaxMatrixCols c]{%
  \hskip -\arraycolsep
  \let\@ifnextchar\new@ifnextchar
  \array{#1}}
\makeatother

\usepackage[colorinlistoftodos]{todonotes}

\usepackage[nonumberlist,acronym,nogroupskip,nohypertypes={acronym, notation, symbols}]{glossaries}
\newglossary{symbols}{sym}{sbl}{Nomenclature}

\glsdisablehyper

\newglossaryentry{greekletter}{
name={\large{\textbf{Greek Symbols}}},
description={\nopostdesc},
sort=parenta,
}

\newglossaryentry{alphanum}{
name={\large{\textbf{Alphanumeric Symbols}}},
description={\nopostdesc},
sort=parentb,
}

\newglossaryentry{pi}{
type=symbols,
name=\ensuremath{\pi},
text={\ensuremath{\pi}},
sort=pi,
description={Transcendental number},
parent=greekletter,
symbol=[-]}

\newglossaryentry{rcirc}{
type=symbols,
name=\ensuremath{R},
text={\ensuremath{R}},
sort=r,
description={Radius of Circular Motion},
parent=alphanum, 
}

\newglossaryentry{fz}{
type=symbols,
name=\ensuremath{F_{z}},
text={\ensuremath{F_{z}}},
sort=fz,description={=  Linear force along aircraft z-direction acting at aircaft c.g},
parent=alphanum, 
}

\newglossaryentry{fx}{
type=symbols,
name=\ensuremath{F_{x}},
text={\ensuremath{F_{x}}},
sort=fx,description={=  Linear force along aircraft x-direction acting at aircaft c.g},
parent=alphanum, 
}

\newglossaryentry{fy}{
type=symbols,
name=\ensuremath{F_{y}},
text={\ensuremath{F_{y}}},
sort=fy,description={=  Linear force along aircraft y-direction acting at aircaft c.g},
parent=alphanum, 
symbol=[N]}

\newglossaryentry{gz}{
	type=symbols,
	name=\ensuremath{G_{z}},
	text={\ensuremath{G_{z}}},
	sort=gz,description={=  Inertial body vector, positive downward},
	parent=alphanum, 
	}

\newglossaryentry{gy}{
	type=symbols,
	name=\ensuremath{G_{y}},
	text={\ensuremath{G_{y}}},
	sort=gy,description={=  Inertial body vector, positive sideways right},
	parent=alphanum, 
	}

	\newglossaryentry{gx}{
	type=symbols,
	name=\ensuremath{G_{x}},
	text={\ensuremath{G_{x}}},
	sort=gx,description={=  Inertial body vector, positive backward},
	parent=alphanum, 
	}

\newglossaryentry{Glead}{
	type=symbols,
	name=\ensuremath{G_{lead}},
	text={\ensuremath{G_{lead}}},
	sort=fz,description={=  Instanteneous G-vector predicted by the COHAM},
	parent=alphanum, 
	}

\newglossaryentry{xdat}{
	type=symbols,
	name=\ensuremath{{}\bar{X}_{dat}},
	text={\ensuremath{{}\bar{X}_{dat}}},
	sort=fz,description={=  Dynamic Break point data vector; the first element is the constant baseline of 1.4G, the second element is the
		predicted maximum G-level varying between 1.8-3G},
	parent=alphanum, 
	}

\newglossaryentry{ydat}{
	type=symbols,
	name=\ensuremath{{}\bar{Y}_{dat}},
	text={\ensuremath{{}\bar{Y}_{dat}}},
	sort=fz,description={=  Table data vector containing the mapping from G-level to degrees of mismatch. Appropriate G-max level as a function of time required to generate either lead or lag},
	parent=alphanum, 
	}

\newglossaryentry{uroll}{
	type=symbols,
	name=\ensuremath{u_{roll}},
	text={\ensuremath{u_{roll}}},
	sort=fz,description={=  Stick signal in roll},
	parent=alphanum, 
	}

\newglossaryentry{upitch}{
	type=symbols,
	name=\ensuremath{u_{pitch}},
	text={\ensuremath{u_{pitch}}},
	sort=fz,description={=  Stick signal in pitch},
	parent=alphanum, 
	}

\newglossaryentry{Nz}{
	type=symbols,
	name=\ensuremath{N_z},
	text={\ensuremath{N_z}},
	sort=fz,description={=  Normal acceleration},
	parent=alphanum, 
	}

\newglossaryentry{Gbaseline}{
	type=symbols,
	name=\ensuremath{G_{baseline}},
	text={\ensuremath{G_{baseline}}},
	sort=fz,description={=  Baseline G-level equal to 1.4 G},
	parent=alphanum, 
	}

\newglossaryentry{Gmax}{
	type=symbols,
	name=\ensuremath{G_{max}},
	text={\ensuremath{G_{max}}},
	sort=fz,description={=  Maximum unfiltered G-level},
	parent=alphanum, 
	}

\newglossaryentry{Gmaxtilde}{
	type=symbols,
	name=\ensuremath{\tilde{G}_{max}},
	text={\ensuremath{\tilde{G}_{max}}},
	sort=fz,description={=  Maximum filtered G-level},
	parent=alphanum, 
	}

\newglossaryentry{Gact}{
	type=symbols,
	name=\ensuremath{{G}_{act}(x)},
	text={\ensuremath{{G}_{act}(x)}},
	sort=fz,description={=  Instantaneous simulator G-level as a function of time},
	parent=alphanum, 
	}

\newglossaryentry{Utrue}{
	type=symbols,
	name=\ensuremath{U_{true}(t)},
	text={\ensuremath{U_{true}(t)}},
	sort=fz,description={=  True Pitch coordination angle},
	parent=alphanum, 
	}

\newglossaryentry{Ucoham}{
	type=symbols,
	name=\ensuremath{U_{COHAM}(t)},
	text={\ensuremath{U_{COHAM}(t)}},
	sort=fz,description={=  Mismatch to the coordination angle},
	parent=alphanum, 
	}

\newglossaryentry{t}{
	type=symbols,
	name=\ensuremath{t},
	text={\ensuremath{t}},
	sort=t,description={=  time},
	parent=alphanum, 
	}

\newglossaryentry{thetacoham}{
	type=symbols,
	name=\ensuremath{\theta_{COHAM}},
	text={\ensuremath{\theta_{COHAM}}},
	sort=theta_c,description={=  Mismatch to the coordination angle},
	parent=alphanum, 
	}

\newglossaryentry{thetatrue}{
	type=symbols,
	name=\ensuremath{\theta_{true}\left( t \right)},
	text={\ensuremath{\theta_{true}\left( t \right)}},
	sort=theta_t,description={=  True cabin alignment angle},
	parent=greekletter, 
	}

\newglossaryentry{Gmaxunf}{
	type=symbols,
	name=\ensuremath{G_{maxunf}},
	text={\ensuremath{G_{maxunf}}},
	sort=Gmaxf,description={=  Maximum unfiltered G-level},
	parent=greekletter, 
	}

	\newglossaryentry{R}{
	type=symbols,
	name=\ensuremath{R},
	text={\ensuremath{R}},
	sort=R,description={=  Centrifuge session in Desdemona consisting of two rounds (evaluation phase) and one round (training phase)},
	parent=alphanum, 
	}

	\newglossaryentry{c1}{
	type=symbols,
	name=\ensuremath{C1},
	text={\ensuremath{C1}},
	sort=c,description={=  Motion condition in Desdemona featuring Rocket Man filter},
	parent=alphanum, 
	}

		\newglossaryentry{c2}{
	type=symbols,
	name=\ensuremath{C2},
	text={\ensuremath{C2}},
	sort=c1,description={=  Motion condition in Desdemona featuring COHAM filter},
	parent=alphanum, 
	}

		\newglossaryentry{c_desd}{
	type=symbols,
	name=\ensuremath{C12},
	text={\ensuremath{C12}},
	sort=c12,description={=  Alternating motion sequence during evaluation phase in Desdemona (Rocket Man-COHAM)},
	parent=alphanum, 
	}

		\newglossaryentry{c_dome}{
	type=symbols,
	name=\ensuremath{C12121},
	text={\ensuremath{C12121}},
	sort=c12,description={Alternating motion sequence during training phase in the Dome of maximum 5 rounds},
	parent=alphanum, 
	}

		\newglossaryentry{b}{
	type=symbols,
	name=\ensuremath{B},
	text={\ensuremath{B}},
	sort=b,description={=  Climb-out point of trajectory (MISC evaluation)},
	parent=alphanum, 
	}

		\newglossaryentry{a}{
	type=symbols,
	name=\ensuremath{A},
	text={\ensuremath{A}},
	sort=A,description={=  Entry point (dive-in) of trajectory (MISC evaluation)},
	parent=alphanum, 
	}

		\newglossaryentry{N}{
	type=symbols,
	name=\ensuremath{N},
	text={\ensuremath{N}},
	sort=N,description={=  Experimental days labels (N1-2 training, N4 evaluation)},
	parent=alphanum, 
	}

		\newglossaryentry{s}{
	type=symbols,
	name=\ensuremath{S},
	text={\ensuremath{S}},
	sort=S,description={=  Subject labels (1-3)},
	parent=alphanum, 
	}
\newglossaryentry{vpath}{
type=symbols,
name=\ensuremath{\nu_{\theta}},
text={\ensuremath{\nu_{\theta}}},
sort=nu,
description={Circular path Velocity},
parent=greekletter, 
}

\newglossaryentry{omegapsi}{
type=symbols,
name=\ensuremath{\Omega_{\psi}},
text={\ensuremath{\Omega_{\psi}}},
sort=omegpsi,
description={Central Yaw Rotational Velocity},
parent=greekletter, 
}

\newglossaryentry{accrad}{
type=symbols,
name=\ensuremath{a_r},
text={\ensuremath{a_R}},
sort=ar,
description={Radial or Centripetal Acceleration},
parent=alphanum, 
symbol=[$deg/s^2$]}

\newglossaryentry{acctan}{
type=symbols,
name=\ensuremath{a_{theta}},
text={\ensuremath{a_{\theta}}},
sort=at,
description={Tangential Acceleration},
parent=alphanum, 
symbol=[$deg/s^2$]}

\newglossaryentry{phi}{
type=symbols,
name=\ensuremath{\phi},
text={\ensuremath{\phi}},
sort=phi,
description={Roll angle},
parent=greekletter, 
symbol=[$deg$]}

\newglossaryentry{beta}{
type=symbols,
name=\ensuremath{\beta},
text={\ensuremath{\beta}},
sort=beta,
description={Sideslip angle},
parent=greekletter, 
symbol=[$deg$]}

\newglossaryentry{alfa}{
type=symbols,
name=\ensuremath{\alpha},
text={\ensuremath{\alpha}},
sort=alpha0,
description={Angle of attack},
parent=greekletter, 
symbol=[$deg$]}

\newglossaryentry{alfainduced}{
type=symbols,
name=\ensuremath{\alpha_i},
text={\ensuremath{\alpha_i}},
sort=alphai,
description={Induced angle of attack},
parent=greekletter, 
symbol=[$deg$]}

\newglossaryentry{deltae}{
type=symbols,
name=\ensuremath{\delta_{el}},
text={\ensuremath{\delta_{el}}},
sort=deltae,
description={Elevator angle deflection},
parent=greekletter, 
symbol=[$deg$]}

\newglossaryentry{deltaengine}{
type=symbols,
name=\ensuremath{\delta_{th}},
text={\ensuremath{\delta_{th}}},
sort=deltat,
description={Throttle setting},
parent=greekletter, 
symbol=[$-$]}

\newglossaryentry{deltarudder}{
type=symbols,
name=\ensuremath{\delta_r},
text={\ensuremath{\delta_r}},
sort=deltar,
description={Rudder angle deflection},
parent=greekletter, 
symbol=[$deg$]}

\newglossaryentry{deltaa}{
type=symbols,
name=\ensuremath{\delta_a},
text={\ensuremath{\delta_a}},
sort=deltar,
description={Aileron angle deflection},
parent=greekletter, 
symbol=[$deg$]}

\newglossaryentry{deltaflap}{
type=symbols,
name=\ensuremath{\delta_{slat_{LE}}},
text={\ensuremath{\delta_{slat_{LE}}}},
sort=deltaslat,
description={Leading edge slat deflection},
parent=greekletter, 
symbol=[$deg$]}

\newglossaryentry{uaileron}{
type=symbols,
name=\ensuremath{u_a},
text={\ensuremath{u_a}},
sort=ua,
description={Command signal for the aileron actuator},
parent=alphanum, 
symbol=[$-$]}

\newglossaryentry{urudder}{
type=symbols,
name=\ensuremath{u_r},
text={\ensuremath{u_r}},
sort=ur,
description={Command signal for the rudder actuator},
parent=alphanum, 
symbol=[$-$]}

\newglossaryentry{uelevator}{
type=symbols,
name=\ensuremath{u_{el}},
text={\ensuremath{u_{el}}},
sort=uel,
description={Command signal for the elevator actuator},
parent=alphanum, 
symbol=[$-$]}

\newglossaryentry{uengine}{
type=symbols,
name=\ensuremath{u_e},
text={\ensuremath{u_e}},
sort=ue,
description={Command signal for the engine actuator},
parent=alphanum, 
symbol=[$-$]}

\newglossaryentry{theta}{
type=symbols,
name=\ensuremath{\theta},
text={\ensuremath{\theta}},
sort=theta,
description={Pitch angle},
parent=greekletter, 
symbol=[$deg$]}

\newglossaryentry{thetaref}{
type=symbols,
name=\ensuremath{\theta_{ref}},
text={\ensuremath{\theta_{ref}}},
sort=thetar,
description={Reference Pitch angle},
parent=greekletter, 
symbol=[$deg$]}

\newglossaryentry{vt}{
type=symbols,
name=\ensuremath{V_T},
text={\ensuremath{V_T}},
sort=vt,
description={Velocity},
parent=alphanum,
symbol=[$m/s$]}

\newglossaryentry{p}{
type=symbols,
name=\ensuremath{p},
text={\ensuremath{p}},
sort=p,
description={Roll rate},
parent=alphanum,
symbol=[$rad/s$]}

\newglossaryentry{q}{
type=symbols,
name=\ensuremath{q},
text={\ensuremath{q}},
sort=q,
description={Pitch Rate},
parent=alphanum,
symbol=[$rad/s$]}

\newglossaryentry{r}{
type=symbols,
name=\ensuremath{r},
text={\ensuremath{r}},
sort=r,
description={Yaw rate},
parent=alphanum,
symbol=[$rad/s$]}

\newglossaryentry{wnsp}{
type=symbols,
name=\ensuremath{\omega_{n_{sp}}},
text={\ensuremath{\omega_{n_{sp}}}},
sort=wnsp,
description={Undamped Natural frequency of the Short Period Mode},
parent=greekletter, 
symbol=[$rad/s$]}

\newglossaryentry{wn}{
type=symbols,
name=\ensuremath{\omega_n},
text={\ensuremath{\omega_n}},
sort=wn,
description={Natural frequency},
parent=greekletter, 
symbol=[$rad/s$]}

\newglossaryentry{zeta}{
type=symbols,
name=\ensuremath{\zeta},
text={\ensuremath{\zeta}},
sort=zeta,
description={Damping ratio},
parent=greekletter, 
symbol=[$-$]}

\newglossaryentry{zetasp}{
type=symbols,
name=\ensuremath{\zeta_{sp}},
text={\ensuremath{\zeta_{sp}}},
sort=zetasp,
description={Damping ratio of the Short Period motion},
parent=greekletter, 
symbol=[$-$]}

\newglossaryentry{h}{
type=symbols,
name=\ensuremath{h},
text={\ensuremath{h}},
sort=h,
description={Altitude},
parent=alphanum, 
symbol=[$ft$]}

\newglossaryentry{period}{
type=symbols,
name=\ensuremath{P},
text={\ensuremath{P}},
sort=zeta,
description={Period},
parent=alphanum, 
symbol=[$s$]}

\newglossaryentry{t12}{
type=symbols,
name=\ensuremath{T_{1/2}},
text={\ensuremath{T_{1/2}}},
sort=ta,
description={Time to damp to half amplitude },
parent=alphanum, 
symbol=[$s$]}

\newglossaryentry{ttheta2}{
type=symbols,
name=\ensuremath{T_{\theta_2}},
text={\ensuremath{T_{\theta_2}}},
sort=tb,
description={Time constant},
parent=alphanum, 
symbol=[$s$]}

\newglossaryentry{tau}{
type=symbols,
name=\ensuremath{\tau},
text={\ensuremath{\tau}},
sort=tau,
description={Time constant},
parent=greekletter, 
symbol=[$-$]}

\newglossaryentry{an}{
type=symbols,
name=\ensuremath{a_n},
text={\ensuremath{a_n}},
sort=an,
description={Normal acceleration},
parent=alphanum, 
symbol=[$m/s^2$]}

\newglossaryentry{az}{
type=symbols,
name=\ensuremath{a_z},
text={\ensuremath{a_z}},
sort=az,
description={Vertical acceleration in the center of gravity},
parent=alphanum, 
symbol=[$m/s^2$]}

\newglossaryentry{nz}{
type=symbols,
name=\ensuremath{n_z},
text={\ensuremath{n_z}},
sort=nz,
description={ 'Incremental' normal acceleration},
parent=alphanum, 
symbol=[$g\,units$]}

\newglossaryentry{xa}{
type=symbols,
name=\ensuremath{x_a},
text={\ensuremath{x_a}},
sort=xa,
description={Accelerometer position distance relative to c.g. on $X_b$ axis},
parent=alphanum, 
symbol=[$m$]}

\newglossaryentry{qdot}{
type=symbols,
name=\ensuremath{\dot q},
text={\ensuremath{\dot q}},
sort=qd,
description={Time derivative of pitch rate},
parent=alphanum, 
symbol=[$rad/s^2$]}

\newglossaryentry{gD}{
type=symbols,
name=\ensuremath{g_D},
text={\ensuremath{g_D}},
sort=gd,
description={Standard gravity (9.80665)},
parent=alphanum, 
symbol=[$m/s^2$]}

\newglossaryentry{cg}{
type=symbols,
name=\ensuremath{c.g.},
text={\ensuremath{c.g.}},
sort=cg,
description={Center of Gravity},
parent=alphanum, 
symbol=[$m$]}

\newglossaryentry{qm}{
type=symbols,
name=\ensuremath{q_m},
text={\ensuremath{q_m}},
sort=qm,
description={Maximum pitch rate},
parent=alphanum, 
symbol=[$rad/s^2$]}

\newglossaryentry{qss}{
type=symbols,
name=\ensuremath{q_{ss}},
text={\ensuremath{q_{ss}}},
sort=qs,
description={Steady state value of pitch rate},
parent=alphanum, 
symbol=[$rad/s^2$]}

\newglossaryentry{overshootq}{
type=symbols,
name=\ensuremath{\frac{q_m}{q_{ss}}},
text={\ensuremath{\frac{q_m}{q_{ss}}}},
sort=qs,
description={Pitch rate overshoot ratio},
parent=alphanum, 
symbol=[$-$]}

\newglossaryentry{npos}{
type=symbols,
name=\ensuremath{n_{pos}},
text={\ensuremath{n_{pos}}},
sort=np,
description={North Position},
parent=alphanum, 
symbol=[$ft$]}

\newglossaryentry{gamma}{
type=symbols,
name=\ensuremath{\Gamma},
text={\ensuremath{\Gamma}},
sort=g,
description={Error Angle},
parent=greekletter, 
symbol=[$^\circ$]}

\newglossaryentry{epos}{
type=symbols,
name=\ensuremath{e_{pos}},
text={\ensuremath{e_{pos}}},
sort=ep,
description={East Position},
parent=alphanum, 
symbol=[$ft$]}

\newglossaryentry{psi}{
type=symbols,
name=\ensuremath{\psi},
text={\ensuremath{\psi}},
sort=psi,
description={Heading angle},
parent=greekletter, 
symbol=[$rad$]}

\newglossaryentry{qbar}{
type=symbols,
name=\ensuremath{\bar q},
text={\ensuremath{\bar q}},
sort=qbar,
description={Dynamic Pressure},
parent=alphanum, 
symbol=[$lb/ft^2$]}

\newglossaryentry{ps}{
type=symbols,
name=\ensuremath{p_s},
text={\ensuremath{p_s}},
sort=ps,
description={Static Pressure},
parent=alphanum, 
symbol=[$lb/ft^2$]}

\newglossaryentry{cl}{
type=symbols,
name=\ensuremath{C_L},
text={\ensuremath{C_L}},
sort=cl,
description={Lift coefficient},
parent=alphanum,
symbol=[-]}

\newglossaryentry{cl0}{
type=symbols,
name=\ensuremath{C_{L_0}},
text={\ensuremath{C_{L_0}}},
sort=cl,
description={Zero lift coefficient},
parent=alphanum,
symbol=[-]}

\newglossaryentry{Gpmax}{
type=symbols,
name=\ensuremath{G_{pmax}},
text={\ensuremath{G_{pmax}}},
sort=gpmax,
description={Maximum possible flow rate of powder through the
nozzle},
parent=alphanum,
symbol=[m/s]}

\newglossaryentry{Scr}{
type=symbols,
name=\ensuremath{S_{cr}},
text={\ensuremath{S_{cr}}},
sort=Scr,description={Area
of the supersonic nozzle throat},
parent=alphanum, 
symbol=[km]}

\newglossaryentry{fzj}{
type=symbols,
name=\ensuremath{F_{z_j}},
text={\ensuremath{F_{z_j}}},
sort=fzj,description={Force in z-direction with numerator j},
parent=alphanum, 
symbol=[N]}

\newglossaryentry{Range}{
type=symbols,
name=\ensuremath{R},
text={\ensuremath{R}},
sort=R,description={Range of the UAV},
parent=alphanum, 
symbol=[km]
}

\newglossaryentry{cruisespeed}{
type=symbols,
name=\ensuremath{V_0},
text={\ensuremath{V_0}},
sort=V0,description={Assumed ground speed for cruise},
parent=alphanum, 
symbol=[m/s]}

\newglossaryentry{tas}{
type=symbols,
name=\ensuremath{V_{tas}},
text={\ensuremath{V_{tas}}},
sort=Vtas,description={Assumed ground speed for cruise},
parent=alphanum, 
symbol=[m/s]}

\newglossaryentry{cruisepower}{
type=symbols,
name=\ensuremath{P_{cruise}},
text={\ensuremath{P_{cruise}}},
sort=Pcruise,description={The power of the engines in cruise condition},
parent=alphanum, 
symbol=[J/s]}

\newglossaryentry{maxpower}{
type=symbols,
name=\ensuremath{P_{max}},
text={\ensuremath{P_{max}}},
sort=Pmax,description={The maximum power of the engines },
parent=alphanum, 
symbol=[J/s]}

\newglossaryentry{proppower}{
type=symbols,
name=\ensuremath{P_{prop}},
text={\ensuremath{P_{prop}}},
sort=Pprop,description={Propulsive power},
parent=alphanum, 
symbol=[J/s]}

\newglossaryentry{availpower}{
type=symbols,
name=\ensuremath{P_{a}},
text={\ensuremath{P_{a}}},
sort=Pavailable,description={Total power of the aircraft available},
parent=alphanum, 
symbol=[J/s]}

\newglossaryentry{reqpower}{
type=symbols,
name=\ensuremath{P_{r}},
text={\ensuremath{P_{r}}},
sort=Pc,description={Total power of the aircraft required},
parent=alphanum, 
symbol=[J/s]}

\newglossaryentry{enginepower}{
type=symbols,
name=\ensuremath{P_{engine}},
text={\ensuremath{P_{engine}}},
sort=Pe,description={Total engine power in cruise conditions},
parent=alphanum, 
symbol=[J/s]}

\newglossaryentry{payloadpower}{
type=symbols,
name=\ensuremath{P_{payload}},
text={\ensuremath{P_{payload}}},
sort=Ppayload,description={Total power required of the payload},
parent=alphanum, 
symbol=[J/s]}

\newglossaryentry{preq}{
type=symbols,
name=\ensuremath{P_{req}},
text={\ensuremath{P_{re}}},
sort=Preq,description={Total power required },
parent=alphanum, 
symbol=[W]}

\newglossaryentry{end}{
type=symbols,
name=\ensuremath{E},
text={\ensuremath{E}},
sort=E,description={Endurance in  hours},
parent=alphanum, 
symbol=[h]}

\newglossaryentry{wingsurface}{
type=symbols,
name=\ensuremath{S},
text={\ensuremath{S}},
sort=s,description={Surface of the wing},
parent=alphanum, 
symbol=\ensuremath{[m^2]}}

\newglossaryentry{drag}{
type=symbols,
name=\ensuremath{D},
text={\ensuremath{D}},
sort=Drag,description={Drag},
parent=alphanum, 
symbol=[N]}

\newglossaryentry{cd}{
type=symbols,
name=\ensuremath{C_D},
text={\ensuremath{C_D}},
sort=Cd,description={Drag coefficient},
parent=alphanum, 
symbol=[-]}

\newglossaryentry{clmax}{
type=symbols,
name=\ensuremath{C_{L_{max}}},
text={\ensuremath{C_{L_{max}}}},
sort=clmax,description={Maximum lift coefficient},
parent=alphanum, 
symbol=[-]}

\newglossaryentry{g0}{
type=symbols,
name=\ensuremath{g_0},
text={\ensuremath{g_0}},
sort=cd,description={Gravitational acceleration },
parent=alphanum, 
symbol=[\ensuremath{m/s^2}]}

\newglossaryentry{cd0}{
type=symbols,
name=\ensuremath{C_{D_0}},
text={\ensuremath{C_{D_0}}},
sort=cd0,description={Profile drag coefficient},
parent=alphanum, 
symbol=[-]}

\newglossaryentry{cdi}{
type=symbols,
name=\ensuremath{C_{D_i}},
text={\ensuremath{C_{D_i}}},
sort=cdi,description={Induced drag coefficient},
parent=alphanum, 
symbol=[-]}

\newglossaryentry{rho}{
type=symbols,
name=\ensuremath{\rho_{air}},
text={\ensuremath{\rho_{air}}},
sort=rhoair,description={Density of air},
parent=greekletter, 
symbol=[-]}

\newglossaryentry{energytot}{
type=symbols,
name=\ensuremath{{E_{total}}},
text={\ensuremath{E_{total}}},
sort=Etot,description={Total energy of the batteries + solar power},
parent=alphanum, 
symbol=[Wh]}

\newglossaryentry{energybat}{
type=symbols,
name=\ensuremath{{E_{bat}}},
text={\ensuremath{E_{bat}}},
sort=Ebat,description={Total energy of the batteries},
parent=alphanum, 
symbol=[Wh]}

\newglossaryentry{energysolar}{
type=symbols,
name=\ensuremath{{E_{solar}}},
text={\ensuremath{E_{solar}}},
sort=Esolar,description={Energy produced by solar power},
parent=alphanum, 
symbol=[Wh]}

\newglossaryentry{thrust}{
type=symbols,
name=\ensuremath{T},
text={\ensuremath{T}},
sort=Thrust,
description={Thrust produced by engines},
parent=alphanum, 
symbol=[N]}

\newglossaryentry{eta_deg}{
type=symbols,
name=\ensuremath{\eta_{deg}},
text={\ensuremath{\eta_{deg}}},
sort=etadeg,
description={Degradation efficiency},
parent=greekletter, 
symbol=[-]}

\newglossaryentry{weight}{
type=symbols,
name=\ensuremath{W},
text={\ensuremath{W}},
sort=W,description={Weight of the aircraft},
parent=alphanum, 
symbol=[N]}

\newglossaryentry{alt}{
type=symbols,
name=\ensuremath{h},
text={\ensuremath{h}},
sort=h,description={Altitude},
parent=alphanum, 
symbol=[m]}

\newglossaryentry{ar}{
type=symbols,
name=\ensuremath{A},
text={\ensuremath{A}},
sort=AR,description={Aspect ratio},
parent=alphanum, 
symbol=[-]}

\newglossaryentry{oswald}{
type=symbols,
name=\ensuremath{e},
text={\ensuremath{e}},
sort=e,description={Oswald efficiency},
parent=alphanum, 
symbol=[-]}

\newglossaryentry{mstart}{
type=symbols,
name=\ensuremath{M_S},
text={\ensuremath{M_S}},
sort=mstart,description={Start iteration mass},
parent=alphanum, 
symbol=[kg]}

\newglossaryentry{mend}{
type=symbols,
name=\ensuremath{M_E},
text={\ensuremath{M_E}},
sort=mend,description={End iteration mass},
parent=alphanum, 
symbol=[kg]}
\newglossaryentry{mbatstart}{
type=symbols,
name=\ensuremath{M_{S_B}},
text={\ensuremath{M_{S_B}}},
sort=mbatstart,description={Start iteration battery  mass},
parent=alphanum, 
symbol={[\ensuremath{kg}]}}

\newglossaryentry{ef}{
type=symbols,
name=\ensuremath{\eta_{prop}},
text={\ensuremath{\eta_{prop}}},
sort=etaprop,description={Propulsive efficiency},
parent=greekletter, 
symbol=[-]}

\newglossaryentry{massflow}{
type=symbols,
name=\ensuremath{\dot{m}},
text={\ensuremath{\dot{m}}},
sort=mairflow,description={Mass flow of air},
parent=alphanum, 
symbol=[kg/s]}

\newglossaryentry{ld}{
type=symbols,
name=\ensuremath{\frac{L}{D}},
text={\ensuremath{\frac{L}{D}}},
sort=ld,
description={Lift over drag ratio},
parent=alphanum, 
symbol=[-]}
\newacronym{gia}{GIA}{Gravito Inertial Acceleration}
\newacronym{gif}{GIF}{Gravito Inertial Force}
\newacronym{iqr}{IQR}{Inter Quartile Range}
\newacronym{ranova}{Repeated Measures ANOVA}{Repeated Measures Analysis of Variance}
\newacronym{scc}{SCC}{Semi Circular Canals}
\newacronym{std}{SD}{Standard Deviation}
\newacronym{sv}{SV}{Subjective Vertical} 
\newacronym{iv}{IV}{Independent Variable}
\newacronym{dv}{DV}{Dependent Variable} 
\newacronym{caz}{CAZ}{Coherent Alignment Zone} 
\newacronym{coham}{COHAM}{Coherent Alignment Method} 
\newacronym{svv}{SVV}{Subjective Visual Vertical} 
\newacronym{svh}{SVH}{Subjective Visual Horizontal}
\newacronym{dof}{DOF}{Degrees-of-Freedom}
\newacronym{tcpe}{TCPE}{Toetsings-Commissie Proefpersoon-Experimenten}
\newacronym{cdf}{CDF}{Cumulative Distribution Function}
\newacronym{ecdf}{ECDF}{Empirical Cumulative Distribution Function}
\newacronym{ocr}{OCR}{Ocular Counter-Rolling}
\newacronym{vor}{VOR}{Vestibular Ocular Reflex}
\newacronym{cavok}{CAVOK}{Ceiling And Visibility OKay}
\newacronym{fov}{FOV}{Field of View}
\newacronym{misc}{MISC}{MIsery SCale}

\newacronym{mil}{MIL}{Military Power}
\newacronym{ab}{AB}{After Burner}
\newacronym{cas}{CAS}{Calibrated Airspeed }
\newacronym{msr}{MSR}{Misery Scale Recovery}
\newacronym{adi}{ADI}{Attitude Director Indicator}
\newacronym{hsi}{HSI}{Horizontal Situation Indicator}
\newacronym{ded}{DED}{Data Entry Display}
\newacronym{icp}{ICP}{Integrated Control Panel}
\newacronym{acrm}{ACRM}{Augmented Coriolis Response Model}
\newacronym{ttpf}{TTPF}{Two-Tailed Peak Flattening}
\newacronym{gloc}{G-LOCK}{G-force induced loss of consciousness}
\newacronym{empt}{EMPT}{Elementary Military Pilot Training}
\newacronym{uprt}{UPRT}{Upset Prevention and Recovery Training}
\newacronym{loci}{LOC-I}{loss of control in-flight}
\newacronym{dfs}{DFS}{Dynamic Flight Simulators}
\newacronym{jnd}{JND}{ just noticeable difference}
\newacronym{aws}{AWS}{Accumulated Well-being Score}
\newacronym{pbws}{PBWS}{Preference-Based Well-being Score}
\newacronym{ams}{AMS}{Alignment Mismatch Score}
\newacronym{alac}{ALAC}{Accumulated Level of Alignment Coherency of the G-cueing}

\newacronym{goe}{GOE}{G-onset Endurance}
\newacronym{got}{GOT}{G-onset Toughness}
\newacronym{awst}{AWST}{Accumulated Well-being Score Template}

\newcommand{\figref}[1]{Fig.~\ref{#1}}
\renewcommand{\eqref}[1]{Eq.~(\ref{#1})}
\newcommand{\tabref}[1]{Table~\ref{#1}}
\newcommand{\secref}[1]{Section~\ref{#1}}

\begin{document}

\title{Mitigating Coriolis Effects in Centrifuge Simulators\\
Through Allowing Small, Unperceived G-Vector Misalignments \\ 
}

\author{Tigran Mkhoyan,\footnote{Ph.D. student, t.mkhoyan@tudelft.nl}
        Mark Wentink,\footnote{Director technology, Desdemona B. V., mark.wentink@desdemona.eu}
        Bernd de Graaf\footnote{Director, Desdemona B. V., bernd.degraaf@desdemona.eu}
        }
\affil{Desdemona B. V., Kampweg 5, 3769 DE Soesterberg, The Netherlands.}

\author{M. M. (Ren\'e) van Paassen\footnote{Associate Professor, m.m.vanpaassen@tudelft.nl}
        and Max Mulder\footnote{Professor, m.mulder@tudelft.nl}
        }
\affil{TU Delft, Faculty of Aerospace Engineering, Control and Simulation, The Netherlands}

\maketitle

\begin{abstract}
When coupled with additional degrees of freedom, centrifuge-based motion platforms can combine the agility of hexapod-based platforms with the ability to sustain higher G-levels and an extended motion space, required for simulating extreme maneuvers. However, the false and often nauseating sensations of rotation, by Coriolis effects induced by the centrifuge rotation in combination with rotations of the centrifuge cabin or the pilot's head, are a major disadvantage. 
This paper discusses the development of a motion filter, the Coherent Alignment Method (COHAM), which aims at reducing Coriolis effects by allowing small mismatches in the G-vector alignment, reducing cabin rotations. Simulations show that as long as these mismatches remain within a region where humans perceive the G-vector as ‘coherent’, the Coherent Alignment Zone (CAZ), the cabin angular accelerations can indeed be reduced.
COHAM was tested in a high G-maneuver task with a fixed CAZ threshold obtained in a previous study. It was experimentally compared to an existing motion filter, using metrics such as sickness, comfort and false cues. Results show that sickness, dizziness and discomfort are reduced, making the centrifuge sessions more bearable. It is recommended to further improve the filter design and tuning, and test it with more fighter pilots.
\end{abstract}

\begin{center}
{\textbf{\large{Acronyms}}}\vspace{.0cm}
\end{center}
\begin{longtable}[l]{@{\hspace*{\leftmargin}} ll}
ALAC & Accumulated Level of Alignment Coherency of the G-cueing\\
AMS & Alignment Mismatch Score\\
AWS & Accumulated Well-being Score\\
CAZ & Coherent Alignment Zone\\
COHAM & Coherent Alignment Method\\
DFS & Dynamic Flight Simulators\\
DOF & Degrees-of-Freedom\\
FOV & Field of View\\
GIA & Gravito Inertial Acceleration\\
JND & just noticeable difference\\
LOC-I & loss of control in-flight\\
MISC & MIsery SCale\\
PBWS & Preference-Based Well-being Score\\
SCC & Semi Circular Canals\\
TTPF & Two-Tailed Peak Flattening\\
UPRT & Upset Prevention and Recovery Training\\
\end{longtable}

\begin{center}
{\textbf{\large{Nomenclature}}}\vspace{.0cm}
\end{center}
\begin{longtable}[l]{@{\hspace*{\leftmargin}} ll}
A, B & = entry point and climb-out point of trajectory (MISC evaluation)\\
${}^h\alpha_x$,${}^h\alpha_y$,${}^h\alpha_z$    & = angular accelerations in head-centric axes $x$,$y$,$z$, $^{\circ}$/s$^2$ \\
$a_{t}$, $a_{r}$ & = linear tangential and radial accelerations during centrifugation, m/s$^2$\\
C1, C2 &= motion condition in Desdemona, Rocket Man, COHAM filter\\
${G}_{act}$& = instantaneous simulator G-level, g-units\\
$G_{baseline}$&  = baseline G-level equal to 1.4 G, g-units\\
$G_{lead}$ & = instantaneous G-vector predicted by the COHAM, g-units\\
$G_{max}$ &  = maximum unfiltered G-level, g-units\\
$G_{maxunf}$&= maximum unfiltered G-level, g-units\\
$G_{x},G_{y},G_{z}$ & = gravitational accelerations in $x$, $y$ and $z$,  m/s$^2$\\
$\tilde{G}_{max}$ &=  maximum filtered G-level, g-units\\
$g$ & = gravitational acceleration,  m/s$^2$\\
$N_{x},N_{y},N_{z}$ & = normal accelerations in $x$, $y$ and $z$,  m/s$^2$\\
N1-N4  &= experiment evaluation days \\
$R_{c}$ & = centrifuge arm,  m\\
S1-S3  &=  experiment subjects\\
$t_{A}$, $t_{B}$ & = time point before and after the onset peak in TTPF, s\\
$U_{COHAM}$& = COHAM cabin angle deviation, $^{\circ}$\\
$u_{pitch}$, $u_{roll}$& = stick signal in pitch and roll, lbf\\
${}\bar{X}_{dat} $&= table input mapping vector G-level, g-units\\
${}\bar{Y}_{dat}$	&= table output mapping vector alignment angle mapping, $^{\circ}$	\\	
$\theta$ & = rotation around the cabin or the head,$^{\circ}$ \\
$\Omega_{c}$&  = centrifuge main axis yaw rate,$^{\circ}$/s\\
$\omega_{c}$&  = cabin rotation rate,$^{\circ}$/s\\
\end{longtable}%
\begin{flushleft} {\textit{{Subscripts}}}\vspace{.2cm} \end{flushleft}
\begin{longtable}[l]{@{\hspace*{\leftmargin}} ll}
true &= true cabin alignment deviation or angle \\
COHAM &= COHAM cabin alignment deviation or angle \\
min, max &= minimum, maximum\\
lower, upper &= lower and upper threshold of cabin alignment angle \\
$ss$ &= steady state value of velocity \\
\end{longtable}%


\section{Introduction}
\label{sec:intro}

\begin{figure}[!ht]
    \centering
    \begin{subfigure}[b]{.45\linewidth}
        \vskip 0pt
        \centering
        \includegraphics[width=\linewidth]{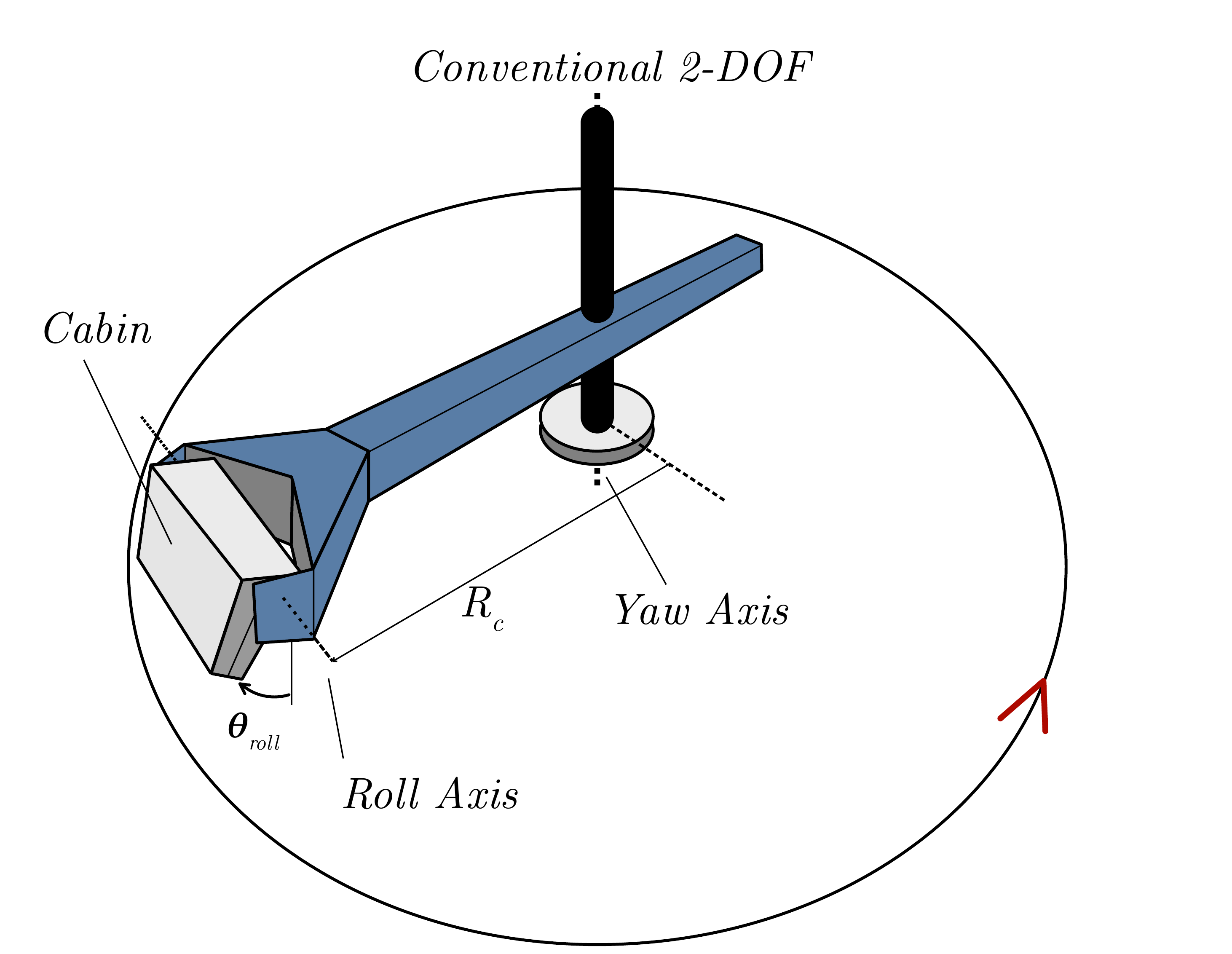}
        \caption{$2$-DOF centrifuge platform}
        \label{fig:conventional_sim}
    \end{subfigure}%
    \begin{subfigure}[b]{.5\linewidth}
        \vskip 0pt
        \centering
        \includegraphics[width=\linewidth]{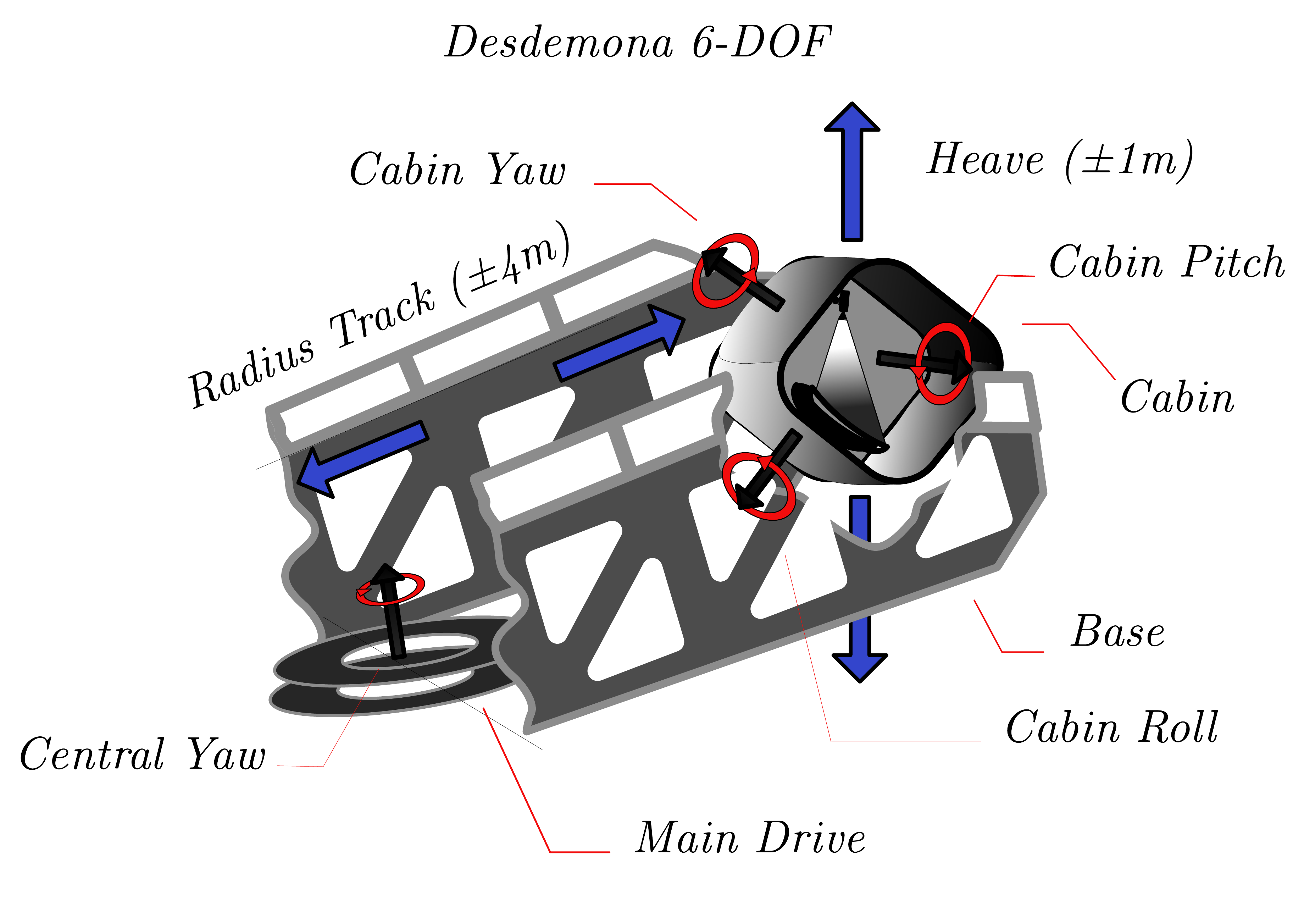}
        \caption{$6$-DOF Desdemona platform}
        \label{fig:desdaxis}
    \end{subfigure}
    \caption{Conventional and Desdemona centrifuge platforms.}
    \label{fig:sims_compare}
\end{figure}

\lettrine{P}{assive} G-training has been the main application of centrifuge-based simulators. In this type of training, pilots are required to sustain a prescribed G-plateau (up to 9G for fighter pilots) for a fixed duration of time ($10$ seconds), without losing consciousness (known as G-lock \cite{burton1988g}). \figref{fig:conventional_sim} shows an example of a typical human centrifuge used for passive training. Active G-training, in contrast, would allow to simulate more realistic dynamic flight scenarios where pilots have \emph{active} control over the \gls{dof}. Dynamic Flight Simulators (DFS) have been in development for the past four decades \cite{cp:Bischoff1985,cp:Eyth1992,cp:Kiefer1992,cp:Vasiletz1993}. An example of such a dynamic flight simulator is Desdemona, \figref{fig:desdaxis}, which has $6$-\gls{dof} and allows pilots to control three or more \gls{dof}s in a simulation \cite{cp:Bles2000}.

Fighter pilots are must be capable of carrying out mentally demanding tasks, while simultaneously enduring significant impact from G-forces on their bodies during maneuvering. The strain resulting from the high G-levels has shown to deteriorate pilot performance and decision-making \cite{gtrack,mt:Masica2009}. Significant training is required to build resistance and the ability to perform in these conditions \cite{jp:Bell1997}. Active G-training has the potential to provide a realistic platform where pilots can safely practice dangerous scenarios at a much lower cost \cite{cp:Spenny2000,mt:Masica2009}; G-suits or G-seats are an alternative \cite{cp:Sabourin2000,cp:Ng2001}. Ref.~\cite{Vidakovic2021} provides a comprehensive review of research carried out in \gls{dfs} and its usefulness in various simulation scenarios.

\gls{uprt} is one example of such scenario \cite{cp:Comtois2010}, where adequate simulation training can help reduce \gls{loci}, a significant contributor to fatal accidents \cite{IATA2019,Belcastro2017}. 
Various efforts have been undertaken to reduce \gls{loci}, such as predictive cueing \cite{Stepanyan2017}, haptic feedback \cite{VanBaelen2020} and adaptive flight control systems \cite{Klyde2017,Smaili2017}.  In the European research project `Simulation of Upset Recovery in Aviation' (SUPRA), centrifuge-based simulation has shown to significantly improve pilot's ability to recover from adverse flight attitudes \cite{Ledegang2012,Nooij2017,Abramov2019}. Apart from enabling psycho-physical studies on human perception \cite{jp:Tribukait2006,jp:Tribukait2013}, the artificial gravity generated by short-radius centrifuges is a promising countermeasure to halt the deterioration of astronauts' health in space \cite{Sides2005,Pavy-LeTraon2007,jp:Mahmoodi2020}.
 
The main limitation to the application of centrifuges is the Coriolis cross-coupling -- also referred to as {\em Coriolis effects} -- induced at the pilot’s head, during simultaneous rotations of the cabin (or gondola) and the centrifuge's central yaw rotation. This cross-coupling can trigger various sensory reactions such as eye movements, illusions, disorientation and motion sickness \cite{jp:Stewart65,reason75,cp:Glaser2013,Bertolini2016}. While adaptation in training has been demonstrated to the vestibulo-ocular reflex \cite{Young2003} and the Corolis-inducing head movements \cite{Newman2013}, (severe) motion sickness symptoms remain a limitation to the tolerably of centrifuges \cite{mt:Masica2009,Frett2020}.

Coriolis rotations can be induced passively {\em and} actively. In `passive Coriolis', the pilot's head is fixed relative to the cabin and the cross-coupling is caused by the cabin motion required to align the G-vector. In `active Coriolis', the cross-coupling is caused by the pilot rotating the head. Whereas active Coriolis cannot be controlled without stringent pilot instruction or adaptation training \cite{Young2003}, passive Coriolis {\em can}, this is the subject of this paper.

\begin{figure}[!ht]
\begin{center}
\includegraphics[width=.8\linewidth]{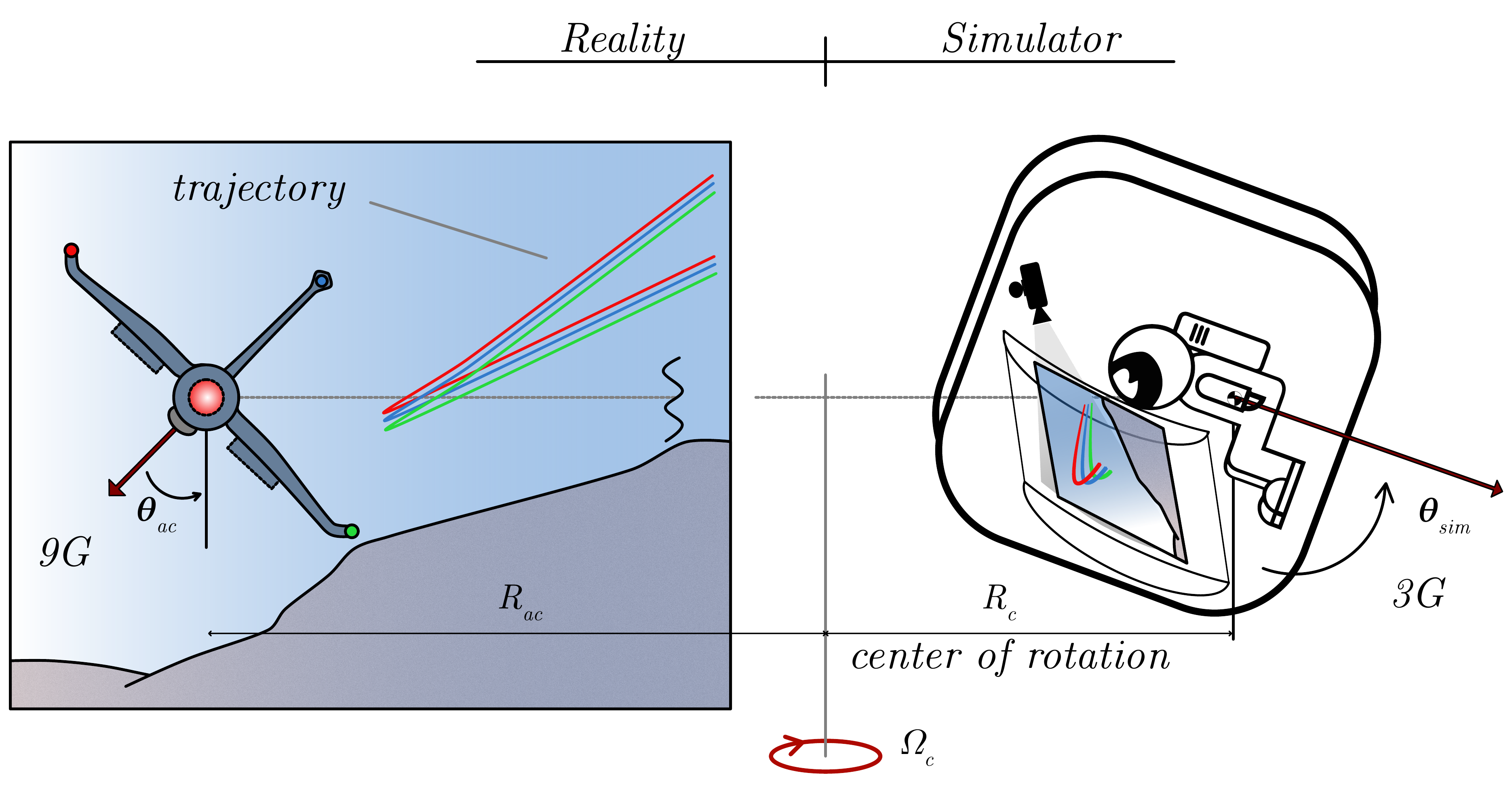}
\caption{G-cueing during F-16 high-G maneuver in Desdemona.}
\label{fig:gcueing}
\end{center}
\end{figure}

In most practical flight scenarios, for example pulling sharp turns in an F-16, one aims to replicate the G-forces experienced in reality as well as possible in the centrifuge (within the allowable limits, 3G for Desdemona), \figref{fig:gcueing}. For this purpose, the subject must be {\em aligned} with the resultant G-vector along the vertical axis, through rotating the cabin towards the centrifuge center. The sharp pull-up maneuver can peak as high as 9G (maximum for F-16), which corresponds to 3G and a cabin roll or pitch angle (here $\theta_{sim}$) in the order of $70$ degrees.

The magnitude of the G-level is dictated by the rotation rate $\Omega_c$ of the centrifuge central yaw-axis. For each change in G-level, a transient referred to as ``G-onset'', the cabin must rotate accordingly to maintain a correct alignment of the resulting G-vector. This combination of cabin alignment rotation and the centrifuge's yaw rotation causes the passive Coriolis effect introduced above. The experienced Coriolis rotation is proportional to both the central yaw and the cabin alignment rotations: the higher these rotations, the stronger the Coriolis effect. Hence, to mitigate the passive Coriolis effect, either one or both rotations would need to be made smaller. 

Since the central yaw rotation rate is controlled to properly simulate the required G-level, the {\em only} degree of freedom that remains in mitigating the Coriolis effect is the cabin alignment rotation. In this paper, it is hypothesized that small deviations from the proper cabin alignment angle are acceptable, as long as these deviations are not perceived by pilots. The concept of manipulating the cabin rotation angle and rotation rate will lead to a novel motion cueing filter, the Coherent Alignment Method, or COHAM. 

The principle of COHAM is to align the centrifuge cabin with a minimum amount of rotation, while not allowing the pilot to notice the mismatch in the G-alignment during each G-onset. The positive (upward) and negative (downward) limits of the mismatch are determined by the hypothetical Coherent Alignment Zone (CAZ), defined as a region where the deviation from the correct cabin angle is still perceived as coherent and consistent by pilots. Details on how this CAZ zone is determined are discussed in Mkhoyan et al. \cite{mkhoyan2019mitigating}, only the main results will be discussed here.

COHAM advances the state-of-the-art in centrifuge simulator motion cueing \cite{cp:Crosbie1985,cp:Wentink2005,cp:Glaser2011}, a problem even more challenging than motion cueing for common 6-DOF Stewart platforms which are typically based on classical washout \cite{jp:Sturgeon1981,jp:Nahon1990,jp:Grant1997}, robust control \cite{jp:Beccera2012}, tuned using insights from pilot models or behavior \cite{jp:Schroeder2000,jp:Hosman2005,jp:Hess2009}, and modal analysis \cite{jp:Miletovic2021}. 

This paper describes the development and experimental evaluation of the COHAM filter and is structured as follows. 
In \secref{sec:back} some background information is provided on centrifuge cueing and Coriolis cross-couplings. 
The concept of the CAZ coherence zone and results from a first experiment to determine the mismatch limits are summarized in \secref{sec:caz}. 
The COHAM motion cueing algorithm rationale is discussed in \secref{sec:coham}. 
A second experiment has been performed to test COHAM, \secref{sec:exp}, the results of which are discussed in \secref{sec:results}. The paper ends with a discussion, \secref{sec:disc} and conclusions in \secref{sec:conc}.

\section{Background}
\label{sec:back}

\subsection{Centrifuge Platforms}

\begin{figure}[!ht]
\centering
\begin{subfigure}[b]{.5\textwidth}
\vskip 0pt
\centering
\includegraphics[width=\linewidth]{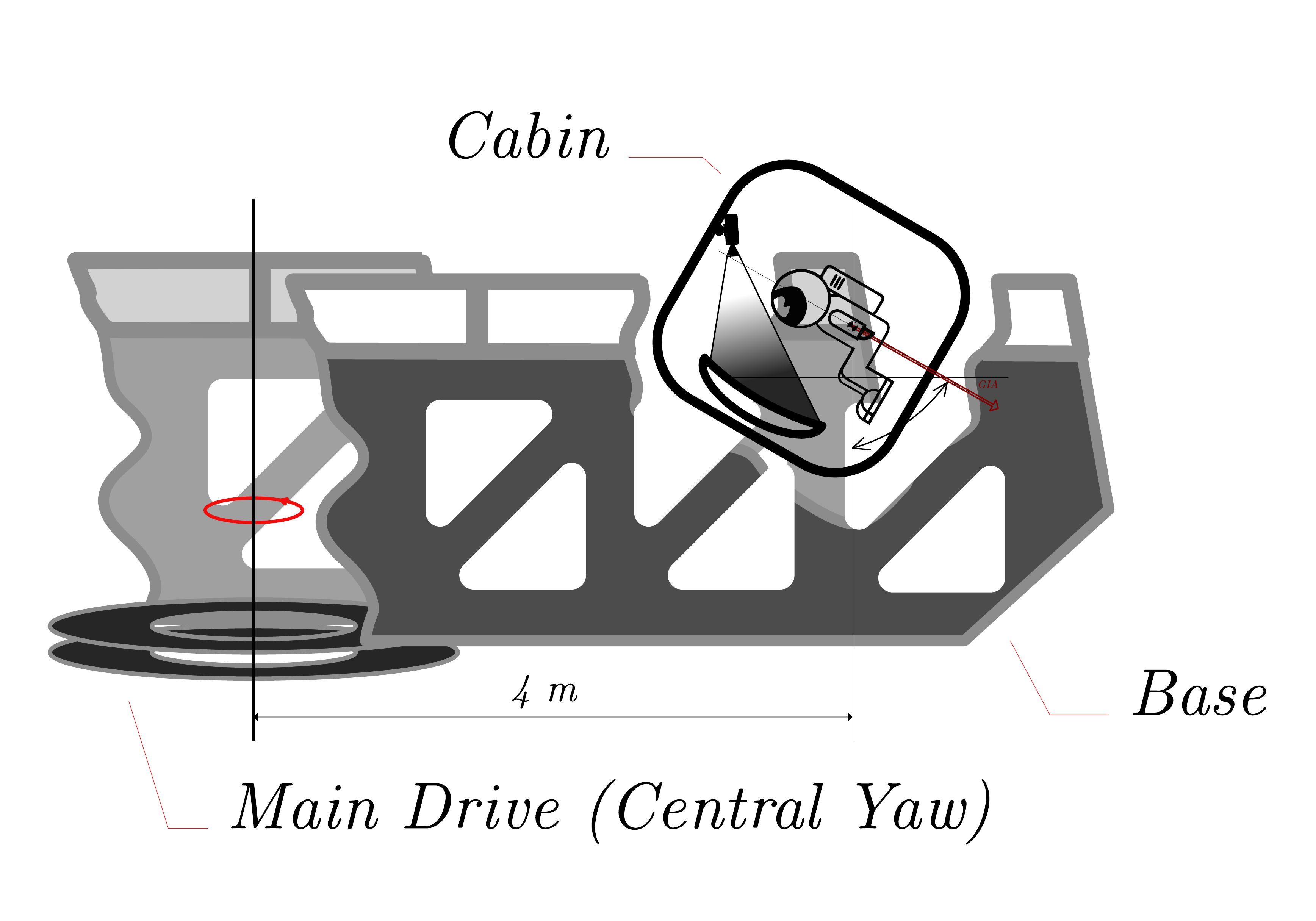}
\caption{Schematic overview of the subject's orientation}
\label{fig:fb2g}
\end{subfigure}%
\qquad
\begin{subfigure}[b]{.3\textwidth}
\vskip 0pt
\centering
\includegraphics[width=\linewidth]{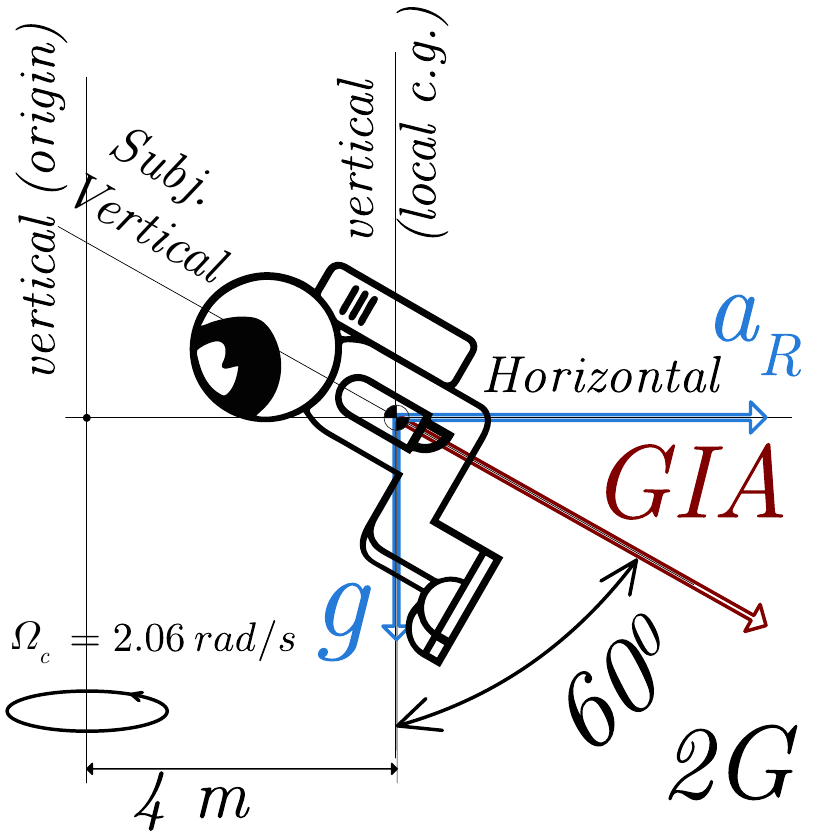}
\caption{Free Body Diagram of linear accelerations at constant centrifuge yaw rotation of $2.06$ rad/s}
\label{fig:fb2g_feneral}
\end{subfigure}
\caption{Schematics of subject's orientation in centrifuge cueing in the Desdemona simulator (here configured for the `Rocket Man' cueing, i.e., alignment of the cabin in pitch).}
\label{fig:2g}
\end{figure}

To attain sustained G-levels, the centrifuge base needs to spin along the central yaw axis, \figref{fig:fb2g}. This yaw velocity, denoted as $\Omega_c$ can be as high as $150$ degrees/s \cite{roza2007performance} in Desdemona. The desired constant G-level, usually referred to as the magnitude of the resultant \gls{gia} corresponds to a steady-state value $\Omega_{c_{ss}}$. From the perspective of the pilot situated in the cabin, the \gls{gia} coincides with the gravitational acceleration z-direction \gls{gz} for most practical simulation scenarios. A free body diagram of accelerations is shown in \figref{fig:fb2g_feneral} for the case of constant centrifugation at a 2G level.

The resultant G-vector, also referred to as \gls{gia}, can be expressed as follows:
\begin{equation}
GIA = \sqrt{{a_t}^2 + {a_R}^2 + {g}^2},
\label{eq:GIA}
\end{equation}
where ${a_t}$ and ${a_R}$ are the tangential and radial linear accelerations resulting from centrifugation, respectively, and $g$ the gravitational constant. Here, ${a_R}$ can be expressed as: 
\begin{equation}
a_R = \Omega_c^2\cdot R_c,
\label{eq:centripetal}
\end{equation}
with $R_c$ the distance from the cabin to the centrifuge central axis ($4$ m in Desdemona).


During the centrifugation phase, the cabin needs to `swing out' to keep the true orientation of the human-vertical axis $G_z$ concerning the resultant G-vector, see \figref{fig:fb2g_feneral}. For a constant centrifuge rotation the correct steady-state cabin angle required for alignment, $\theta_{true}$, can be computed as:
\begin{equation}
\gls{thetatrue} = \arctan\left( \frac{\gls{accrad}}{g}\right) = \arctan \left( \frac{\Omega_c^2 R_c}{g}\right) 
\label{eq:thtrue}
\end{equation}
Higher G-levels require increasingly higher yaw-rates $\Omega_c$, which in turn require larger cabin rotation angles $\theta_{true}$. Both relations are quadratic in nature.


\subsection{Cabin Rotation Options}

\begin{figure}[!ht]
\centering
\begin{subfigure}[b]{.35\textwidth}
\centering
\vskip 0pt
\includegraphics[width=\linewidth]{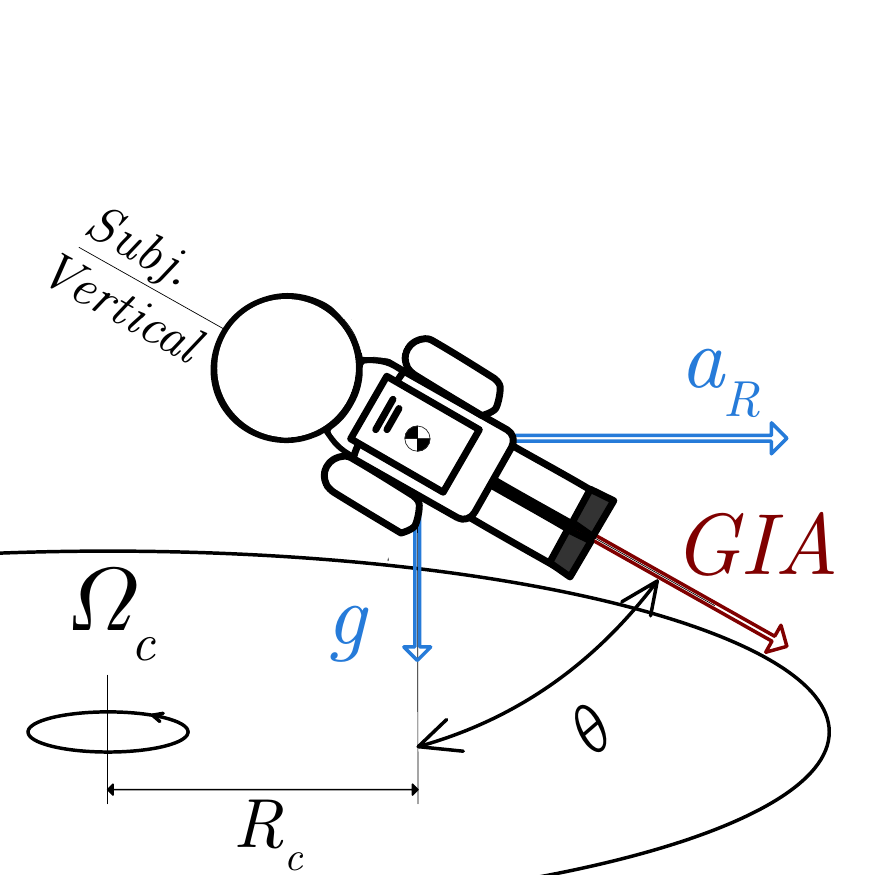}
\caption{Conventional cueing (alignment in roll)}
\label{fig:compare_conventional}
\end{subfigure}%
\qquad
\qquad
\begin{subfigure}[b]{.35\textwidth}
\centering
\vskip 0pt
\includegraphics[width=\linewidth]{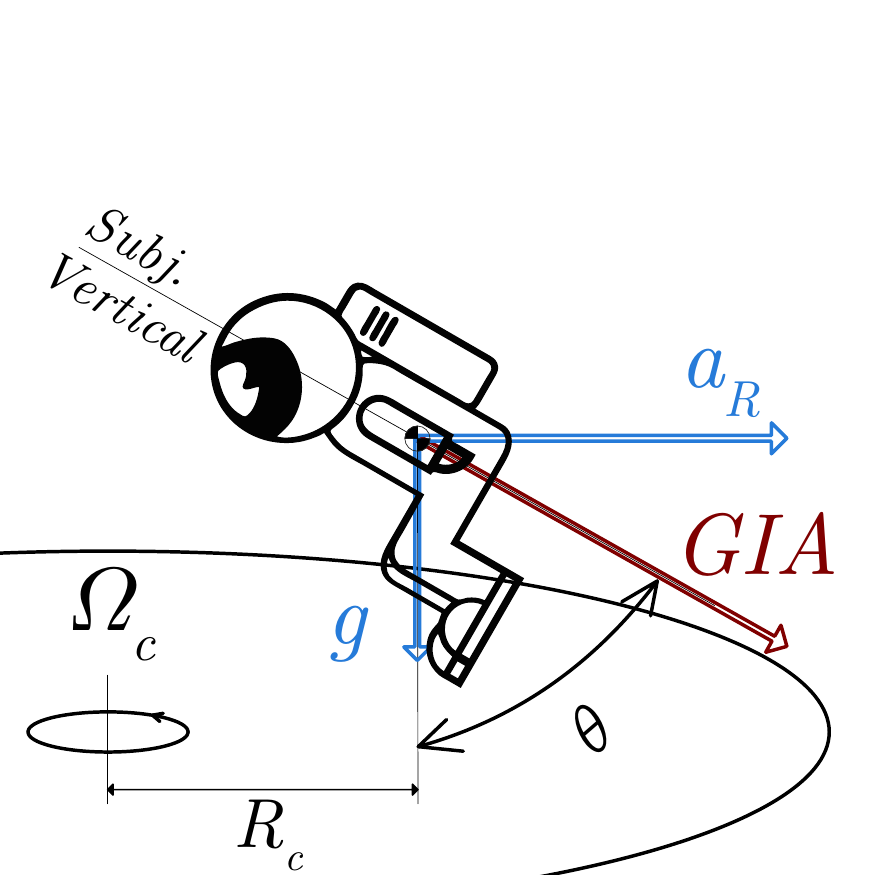}
\caption{`Rocket Man' cueing (alignment in pitch)}
\label{fig:compare_rocketman}
\end{subfigure}
\caption{Illustration of the two cueing orientations used in Desdemona.}
\label{fig:cabin}
\end{figure}

Over the years, two G-force alignment cueing solutions were developed for the Desdemona motion simulator \cite{desdemona}, see \figref{fig:cabin}. The first is a conventional solution where the pilot faces tangential to the rotation and where $G_z$ is aligned through rotating the cabin in {\em roll}, \figref{fig:compare_conventional}. The second is a solution where the pilot faces inward towards the center of rotation and where $G_z$ is aligned through rotating the cabin forward in {\em pitch}, \figref{fig:compare_rocketman}. This latter solution, nicknamed the ``Rocket Man'', or RM for short, is reported by pilots to be more comfortable and appears to cause less motion sickness for centrifuge-based \gls{uprt} \cite{Nooij2017}.

\subsection{Coriolis Cross-coupling}

When the cabin is allowed to have a \gls{dof} of angular rotation (roll or pitch), in an axis {\em other} than that of the main yaw-axis, simultaneous rotations will induce a Coriolis effect \cite{schubert1932physiologischen}. From the discussion above it is clear that, since cabin rotations are required to align the subjective vertical, or G-vector, this cross-coupling is inevitable. For pilots situated inside the cabin, this cross-coupling can be experienced as a sensation of tumbling, nausea and or dizziness \cite{guedry1974psychophysics}.

\begin{figure}
 \centering
 \includegraphics[width=0.2\linewidth]{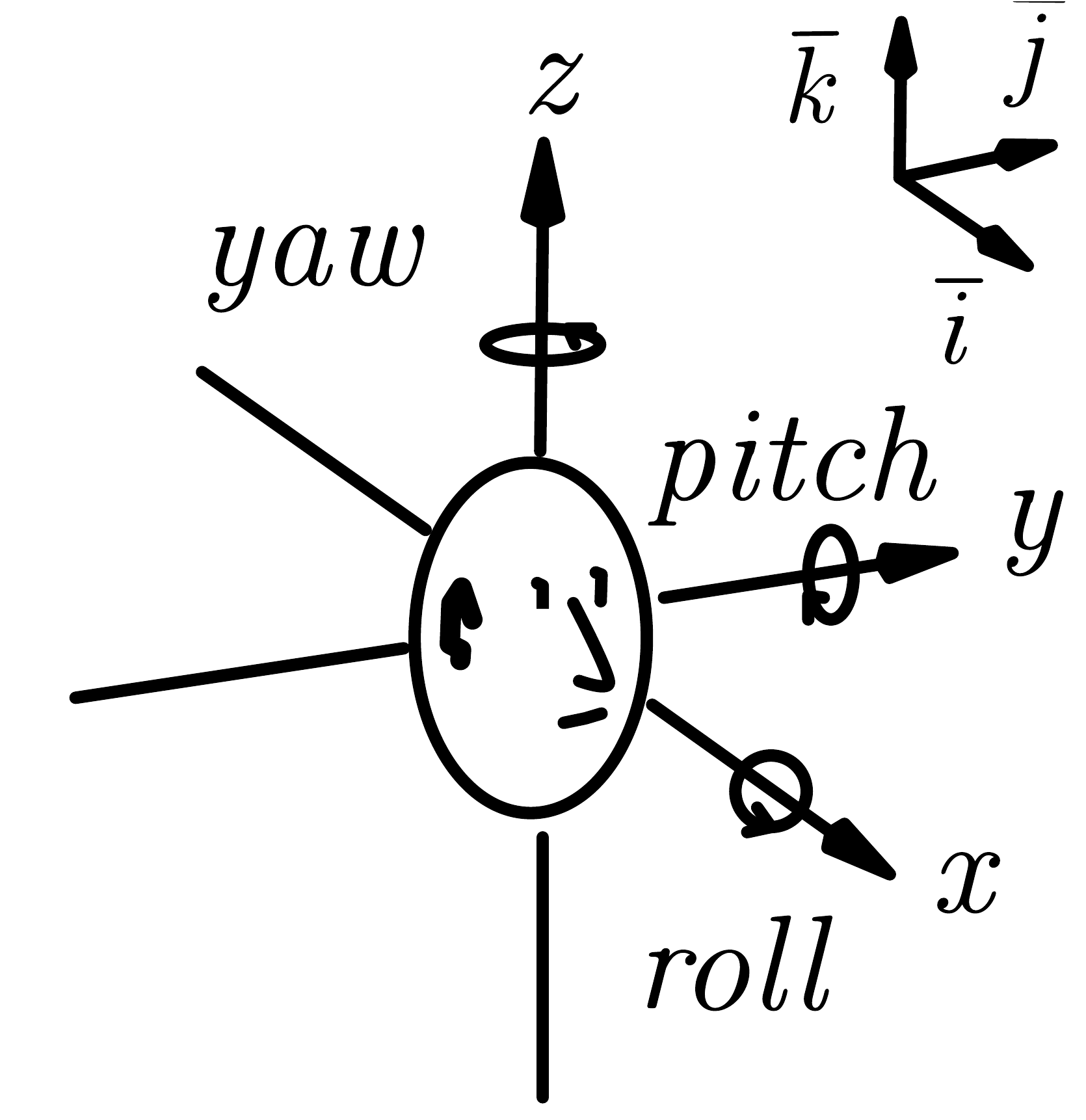}
 \caption{Standard coordinates and unit vectors for specification of head angular motion (from \cite{holly1996subject}).}
   \label{fig:headaxis}
\end{figure}

Although the perceived Coriolis effect is related to the (inaccurate) vestibular responses of the \gls{scc} -- and therefore varies between subjects -- its kinematic cause can be computed. Considering that the pitch, roll and yaw \gls{scc} canals have a specific orientation\footnote{The exact orientation differs from person to person. The yaw-axis is, in fact, slightly tilted downward to accommodate for our default orientation of the head (slightly looking downward).} with respect to the horizontal plane, the angular acceleration arriving at the pilot's head can be decomposed into three axes as shown in \figref{fig:headaxis} \cite{holly1996subject}. The resulting angular accelerations are the {\em head}-centric, ${}^h\alpha_X$, ${}^h\alpha_Y$ and ${}^h\alpha_Z$, accelerations expressed by Holly \cite{holly2008whole,holly1996subject,holly2004vestibular}:
\begin{equation}
{}^h\bvec{\alpha} = \begin{bmatrix} {}^h\alpha_x \\ {}^h\alpha_y \\ {}^h\alpha_z \end{bmatrix} = 
\begin{bmatrix}
\ddot \theta \\
\Omega \dot \theta \cos(\theta) + \dot \Omega \sin(\theta) \\
-\Omega \dot \theta \sin(\theta) + \dot \Omega \cos(\theta) 
\end{bmatrix}\\
\label{eq:holly}
\end{equation}

This set of equations describe a typical case of head tilt, rotation around the  head-centric-axis, in the presence of simultaneous whole-body rotation. The latter represents the default, `rotating chair', configuration for fundamental Coriolis-related experiments \cite{holly2004vestibular,Guedry1978}. Here, $\theta$ represents the rotation around the head-$x$-axis, $\dot{\theta}$ the angular head rotation rate, $\ddot{\theta}$ the head rotational acceleration; $\Omega$ represents the angular rate of the rotating reference frame (the chair, or, in our case, the centrifuge) and, $\dot \Omega$, the angular acceleration of the rotating frame. 

\begin{figure}[!ht]
\centering
\begin{subfigure}[b]{.55\textwidth}
\vskip 0pt
\centering
\includegraphics[width=\linewidth]{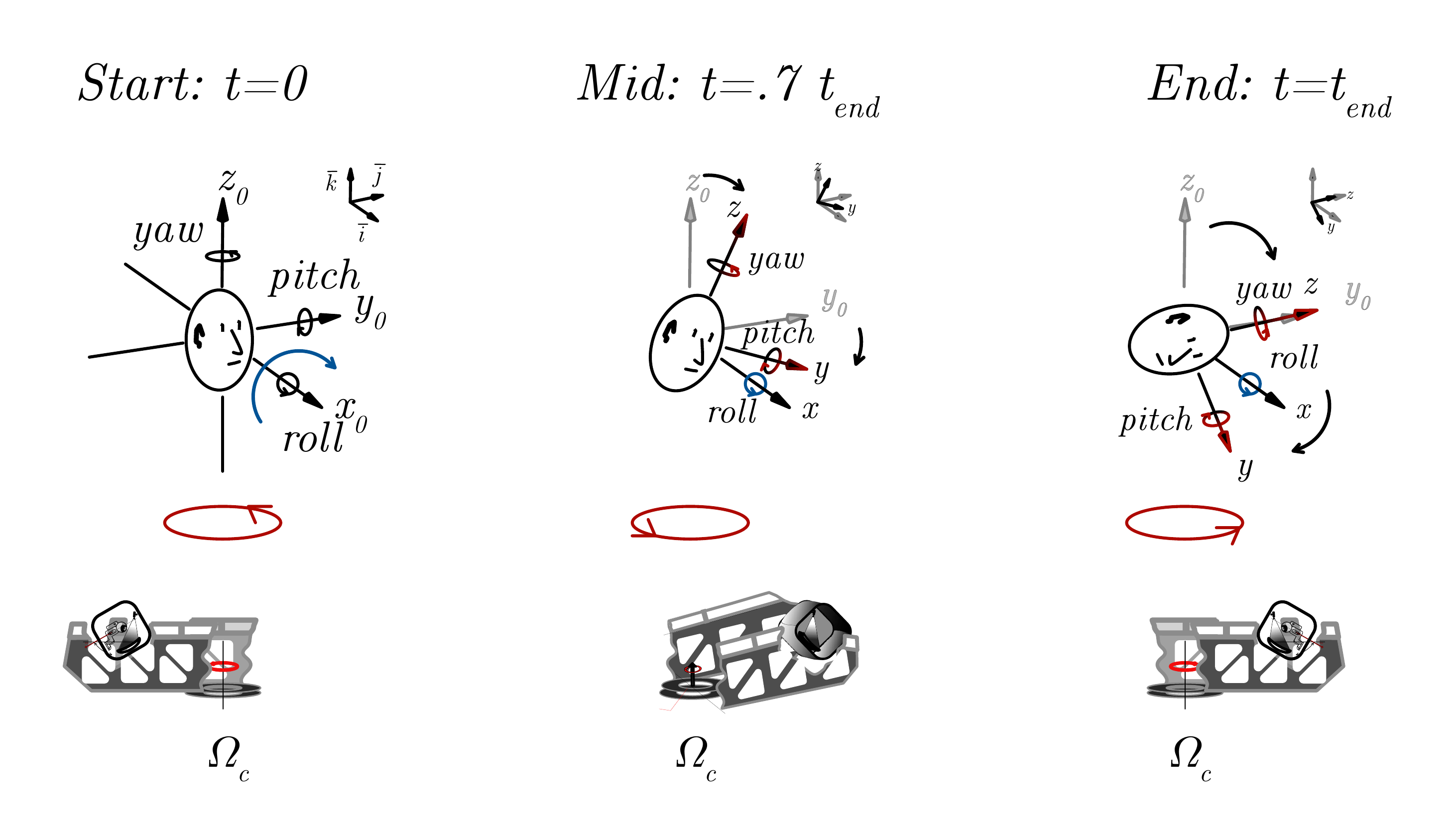}
\caption{Orientation of the subject' head under simultaneous head and centrifuge rotation}
\label{fig:coriolis_ilustration}
\end{subfigure}%
\quad
\begin{subfigure}[b]{.425\textwidth}
\vskip 0pt
\centering
\includegraphics[width=\linewidth]{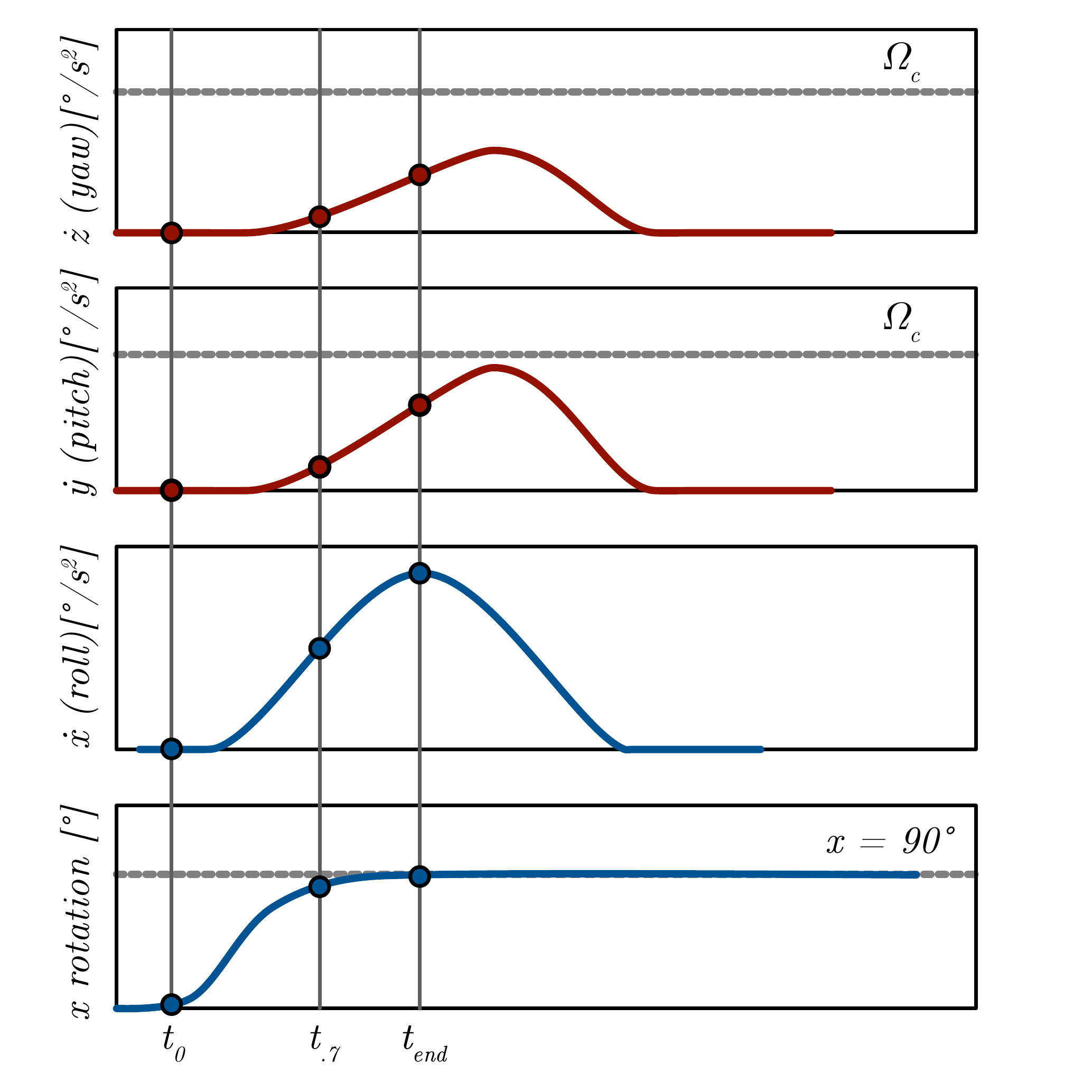}
\caption{Responses of the subject $x$, $y$ and $z$ SCC channels resulting from head rotation in a centrifuge}
\label{fig:coriolis_responses}
\end{subfigure}
\caption{Illustration of the Coriolis cross-coupling effect in a human centrifuge.}
\label{fig:coriolis}
\end{figure}

\figref{fig:coriolis_ilustration} exemplifies Coriolis cross-coupling, where the subject initiates $90$ degrees of head tilt in roll ($x$-axis) while the centrifuge is rotating with a constant yaw rate $\Omega_c$. Due to the rotation around the $x$-axis, the subject's $y$ and $z$ axes are tilted into the centrifugal rotation plane and components of $\Omega_c$ appear in both axes. The resulting cross-coupled stimulus is sensed by the \gls{scc} in these two axes, causing a sensation of tumbling, nausea and dizziness \cite{guedry1974psychophysics}. This form of {\em active} Coriolis can be mitigated by instructing subjects to keep their head straight-up relative to the cabin.

But even in case the subject's head is aligned with the cabin, the G-force alignment requires the cabin to rotate, causing {\em passive} Coriolis. A similar situation as \figref{fig:coriolis_ilustration} occurs in the conventional roll rotation of the cabin. This cabin alignment rotation causes a coupling in the two other axes, the magnitude of which depends on the {\em product} of the cabin rotation (in the form of angular position and angular rate) and the centrifuge yaw rotation (in the form of angular velocity and angular acceleration), Eq.~\ref{eq:holly}.

The passive Coriolis effect can thus be minimized by reducing the centrifuge rotation, the cabin rotation, or both. Given that the centrifuge yaw rotation rate $\Omega_c$ is required to create the G-level required for the simulation, Eqs.~\ref{eq:GIA}-\ref{eq:centripetal}, the cabin rotation, its rate and acceleration are the primary means to mitigate passive Coriolis. Then, what would be the effects of rotating the cabin with a (slightly) different angle than $\theta_{true}$ on the perceived subjective vertical? When subjects would not be able to perceive the G-force alignment perfectly, which seems a reasonable assumption given the limitations in human self-motion perception, this opens up a possibility to exploit this perceptual inaccuracy, and manipulate the cabin rotation such to reduce passive Coriolis effects. The next section describes an experiment which investigated perceptual thresholds in the human perception of G-force alignment.
\section{Coherent Alignment Zone: Concept and Evaluation}
\label{sec:caz}

\subsection{Coherence Zone Concept}

The coherence zone indicates the range where visual and vestibular motion, even if not identical, are perceived as coherent, as studied in \cite{van1998self,valente2010perception}. 
Beyond the limits of the coherence zone, humans perceive the simulated motion to be incorrect. Results of these studies have been applied in flight and driving simulator motion cueing tuning \cite{valente2010perception,jp:CorreiaGracio2013}.

\begin{figure}[!ht]
\centering
\includegraphics[width=0.6\textwidth]{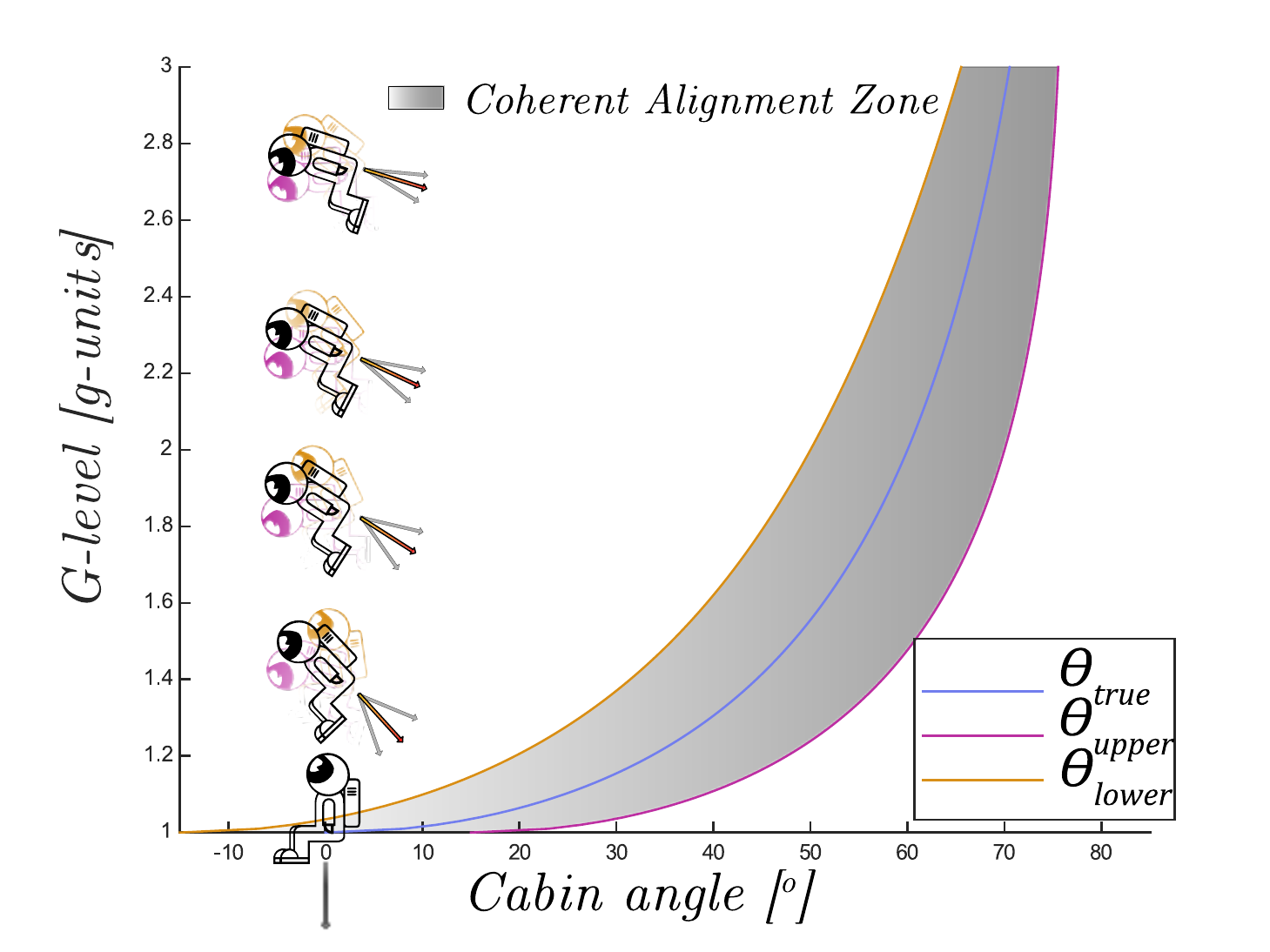}
\caption{Coherent Alignment Zone.}
\label{fig:caz}
\end{figure}

Regarding the centrifuge's G-vector, its magnitude and direction (alignment) can be manipulated in the centrifuge, and ideally the motion cueing algorithm presents the pilots with a G-vector that is equal to the real one. But if the pilot would not notice small deviations from this G-vector, in its magnitude and alignment, this allows harmful Coriolis effects to be mitigated. Gracio et al. \cite{cp:Gracio2009} studied the \gls{jnd} in the perception of G-load {\em magnitude} and found that this \gls{jnd} is related to the G-load intensity.
In the alignment of the centrifuge's G-vector, it is likely that small deviations from the true alignment angle, $\theta_{true}$, will also {\em not} be perceived by pilots. A region of small alignment angles around $\theta_{true}$ may be defined within which the pilot would still perceive the alignment of the G-vector to be `correct', or coherent. Hence, it is hypothesized that a `Coherent Alignment Zone' (CAZ) exists, illustrated in \figref{fig:caz} for the RM cueing condition. 

\figref{fig:caz} shows the typical quadratic relationship between the cabin rotation angle $\theta_{true}$ (horizontal axis) and the desired G-level (vertical axis), which in turn requires a centrifuge rotation rate $\Omega_c$. A constant 1.4G G-level means that the cabin needs to be rotated forward with $\theta_{true}\approx 44$ degrees. The CAZ indicates the set of possible cabin rotation angles $\theta$ which yield a coherent perception of the magnitude and angle of the G-vector alignment. In this hypothetical example, rotating the cabin between $\theta_{lower} \approx 30$ deg and $\theta_{upper} \approx 57$ deg would yield the same perceived subjective vertical by the pilot. An experiment has been conducted to measure this zone, discussed next.

\subsection{Experiment}
\label{sec:exp1}

The objective of the experiment, discussed in detail in Ref.~\cite{mkhoyan2019mitigating}, was to establish the body tilt perception thresholds under elevated G-levels ($1$-$1.4$ G) in both pitch (RM cueing) and roll (conventional cueing) axes. Twelve naive subjects (without piloting experience) were introduced to sub-threshold pitch and roll cabin tilt motion ($0.4$ deg/s and a maximum of $20$ degrees) in random order and asked to indicate the perceived direction of tilt with respect to the perfectly upright sitting position. The motion conditions that were tested, were: $1$G (no elevated gravity), $1$G with central yaw rotation and no elevated gravity, and $1.4$ G in an elevated G-environment.

The hypothesis was that the pitch axis cabin rotation (RM cueing) would result in higher tilt thresholds, because of differences in the body  somatosensory pressure interfaces stimulated with pitch or roll tilt. That is, it was hypothesized that subjects `measured' roll misalignment using a differential pressure interface (left and right-hand side of the body along the symmetry axis), providing a more `immediate' comparison between pressure differences. Pitch misalignment was expected to be measured using absolute (time-dependent) somatosensory cues, where the pressure cues were sampled from a dissipating pressure region (backside of the head and back), and compared against an absolute reference point (headrest at start of the simulation) for the duration of the tilting motion. The latter was believed to lead to higher ambiguity in the pitch axis, making it more difficult to `perceive' the tilt. These are explained in detail in Ref.~\cite{mkhoyan2019mitigating}.

\begin{figure}[!htb]
 \centering
 \includegraphics[width=0.5\textwidth]{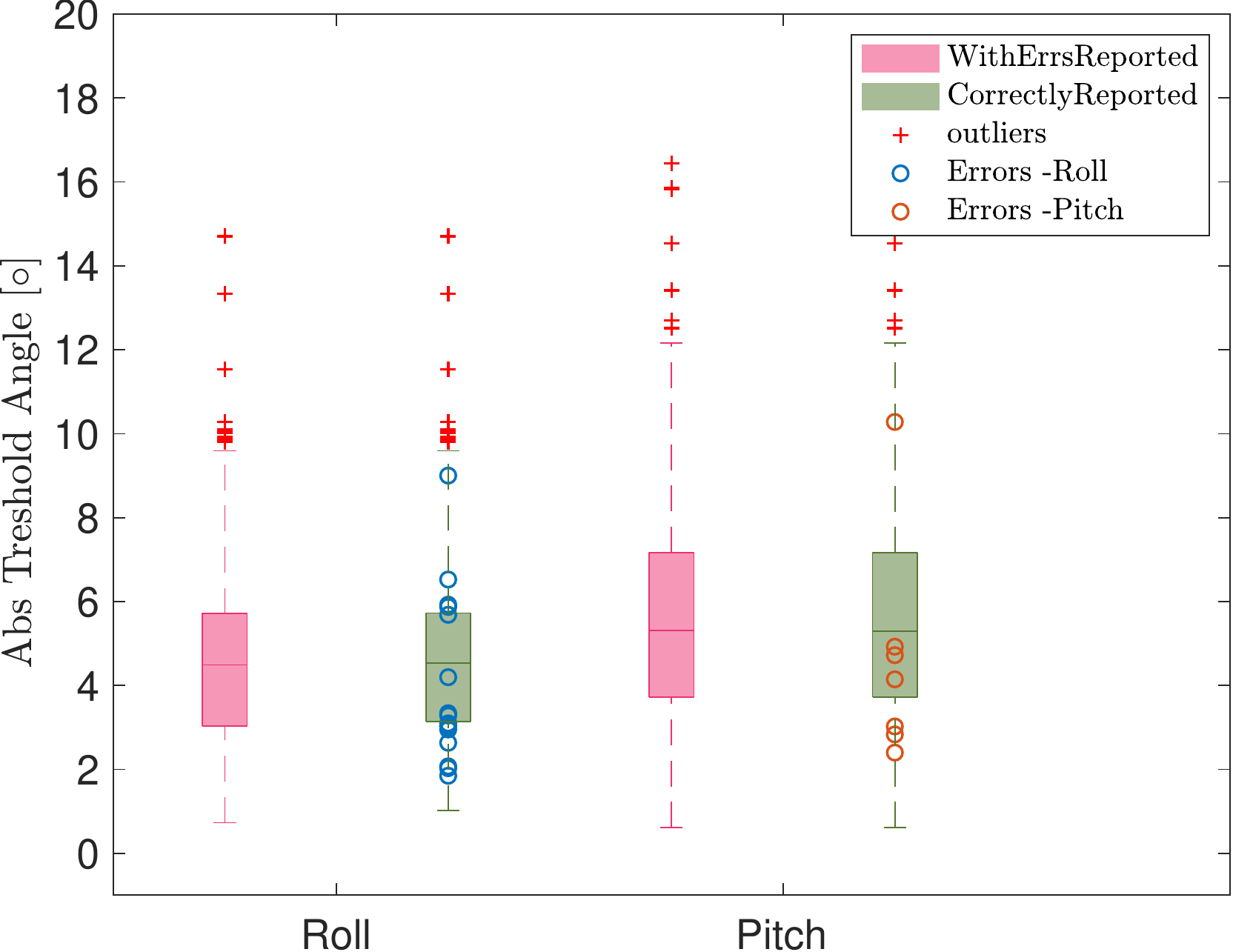}
 \caption{Comparison of group median thresholds for axis, including and excluding `errors', Ref.~\cite{mkhoyan2019mitigating}.}
 \label{fig:cazresults1}
\end{figure}

\begin{figure}[!htb]
 \centering
 \includegraphics[width=0.5\textwidth]{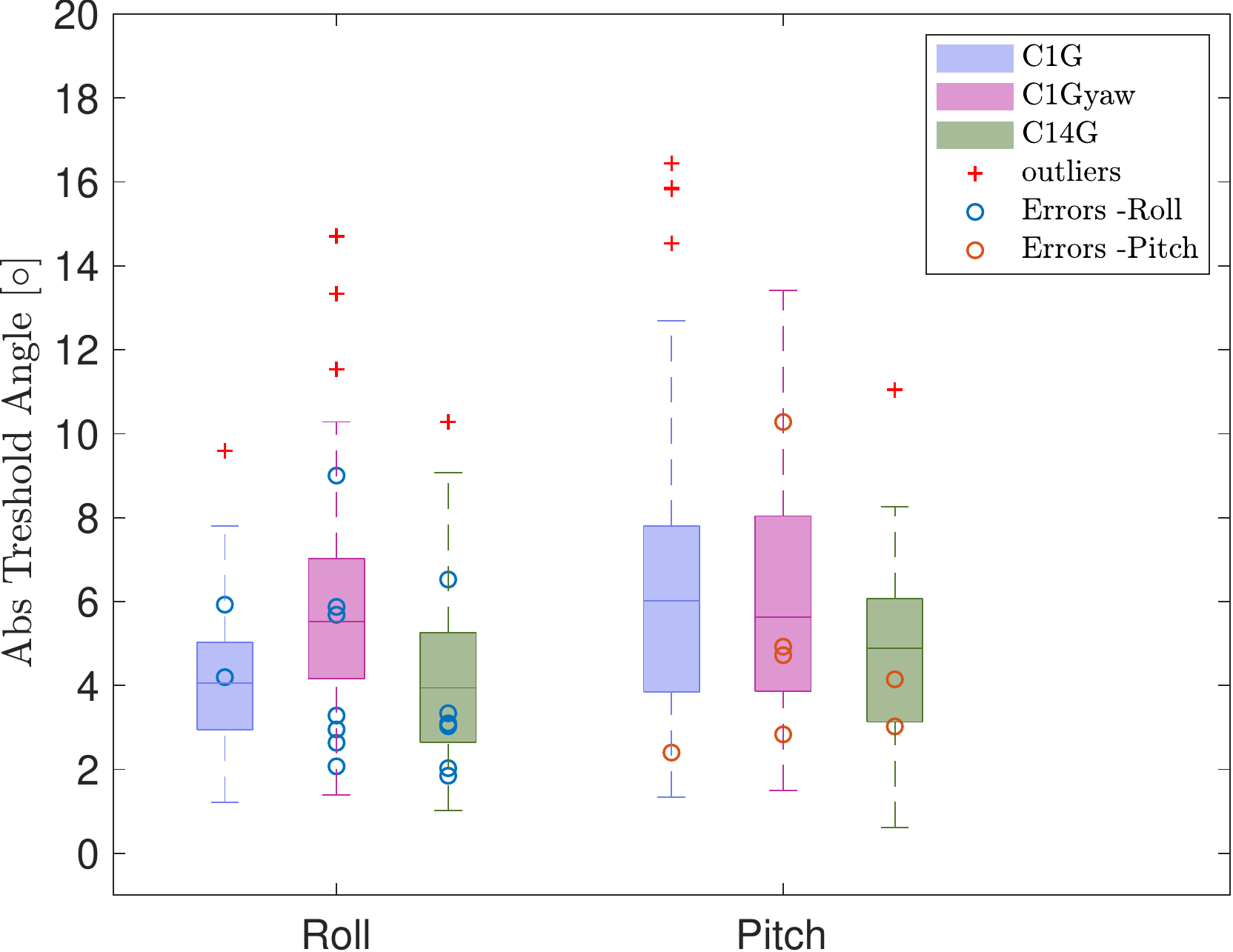}
 \caption{Average thresholds in two axes (roll and pitch) for Conditions 1G, 1G yaw and 1.4 G, see Ref.~\cite{mkhoyan2019mitigating}.}
 \label{fig:cazresults2}
\end{figure}

Experimental results, shown in \figref{fig:cazresults1}, support the hypothesis that cabin pitch rotation has a larger ambiguity, leading to slightly higher thresholds and also larger variations in pitch (average $\approx 5$ degrees) than roll (average $\approx 4$ degrees). No differences were found in the direction of the pitch (down, up) or roll (left, right) rotations. The `errors' included in the figure refer to subject ratings that were erroneous in either the direction (roll or pitch) or sign of the cabin tilt, or both, see Ref.~\cite{mkhoyan2019mitigating} for details. The (slightly) higher thresholds in pitch suggest that the RM cueing condition allows for more room to manipulate the cabin alignment angle without the subject inside the cabin noticing it. This corroborates experience in the Desdemona simulator that the RM cueing is preferred by pilots. In the following, the Rocket Man cueing will therefore be selected for further study of the COHAM motion cueing development.


\figref{fig:cazresults2} shows the threshold values for the three motion conditions separately. No clear trend can be observed from this plot as regarding whether the G-level has an effect on the reported threshold. Preliminary experiments at higher G-levels had to be aborted because of a lack of subjects, and also because the subjects involved quickly experienced motion discomfort. Although this could indicate the CAZ thresholds to become {\em smaller} when G-level increases, more experiments are needed to find more evidence and to exclude  potential confounds. 


Concluding the results of the CAZ experiment of Ref.~\cite{mkhoyan2019mitigating}: although a dynamic, i.e., G-level -dependent threshold may exist, a {\em fixed} $5$-degree CAZ will be used for the development of the novel cueing method here. We concentrate on pitch, as there the thresholds were found to be (slightly) larger, and only study the `Rocket Man' condition. The rationale and operation of the resulting COHAM filter will be discussed in the next section. 



\section{Cueing Method: COHAM}
\label{sec:coham}

\subsection{Coherent Alignment Rationale}

The presence of a perceptual coherence zone and the fact that minimizing cabin alignment rotations reduces Coriolis effects led to the development of the \glsfirst{coham}. Its main purpose is to align the simulator cabin with the least amount of motion possible, without the pilot noticing a mismatch in the G-alignment. COHAM operates within the maximum allowable mismatch defined by the \gls{caz}. 

\begin{figure}[!htb]
 \centering
 \includegraphics[width=.7\linewidth]{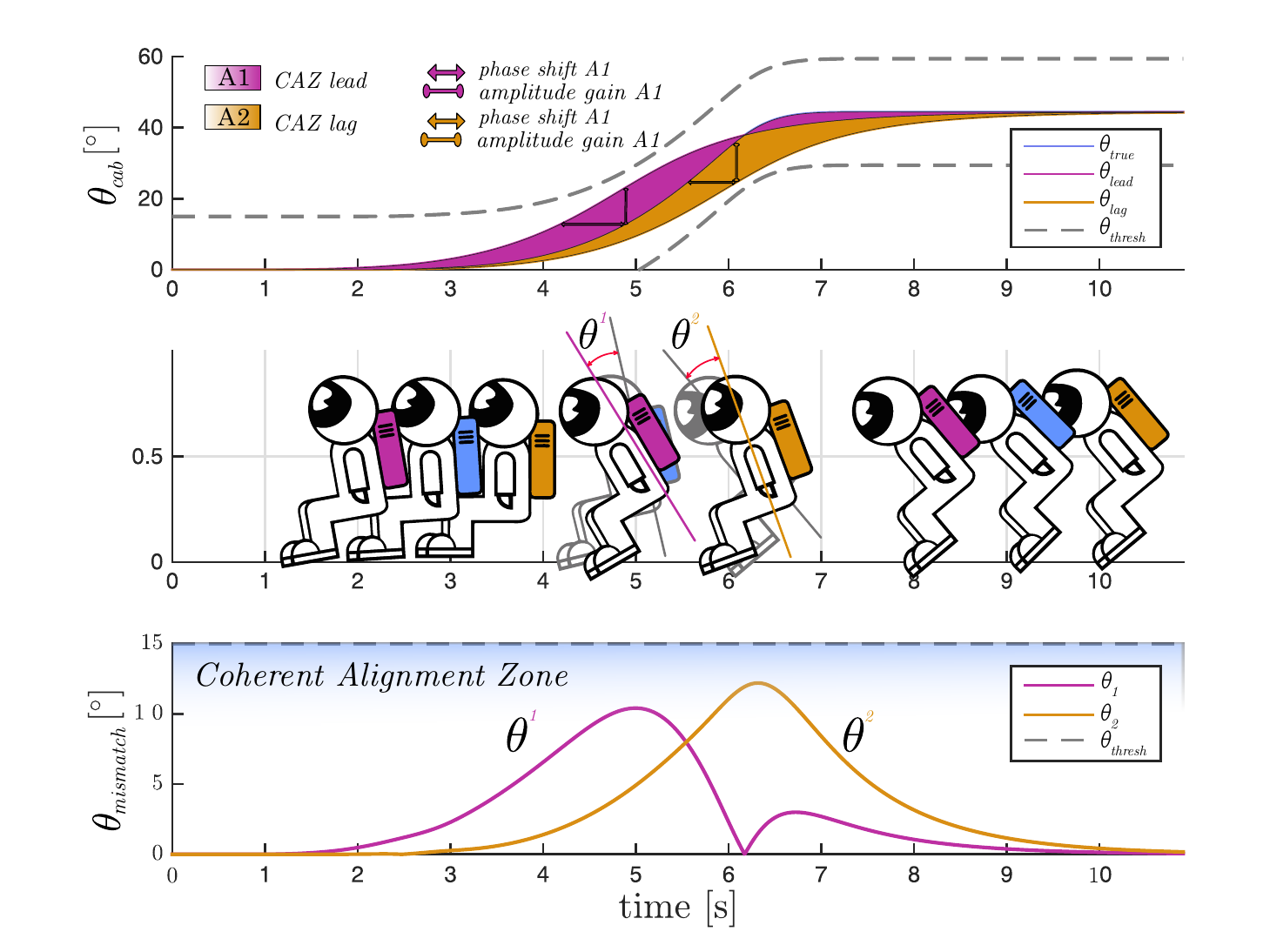}
 \caption{Illustration of the COHAM for a single cabin alignment in pitch from $1$G to a baseline level of $1.4$G \cite{prelim}}
 \label{fig:coham}
\end{figure}

The principle of COHAM can be explained using \figref{fig:coham}, which shows a typical G-onset from 1G to a steady-state level of 1.4 G (top), and a hypothetical CAZ boundary of fifteen degrees pitch (bottom). The figure illustrates three possible alignments of the cabin angle: (i) using the nominal rotation (blue) $\theta_{true}$ which follows from \eqref{eq:thtrue}, (ii) a leading mechanism (magenta) that starts rotating earlier than the nominal rotation, and (iii) a lagging mechanism (yellow) that rotates the cabin slightly later than the nominal rotation. The pilot icons (center plot) represent the subject's orientation in the cabin for the three cases of rotation. As long as the cabin alignment trajectories lie in-between the dashed lines of the top plot, showing the hypothesized $15$ degree CAZ boundary around $\theta_{true}$, the cabin misalignment (the absolute value of which is shown in the bottom plot) will be {\em unnoticed} by pilots.

\begin{figure}[!htb]
 \centering
 \includegraphics[width=0.6\linewidth]{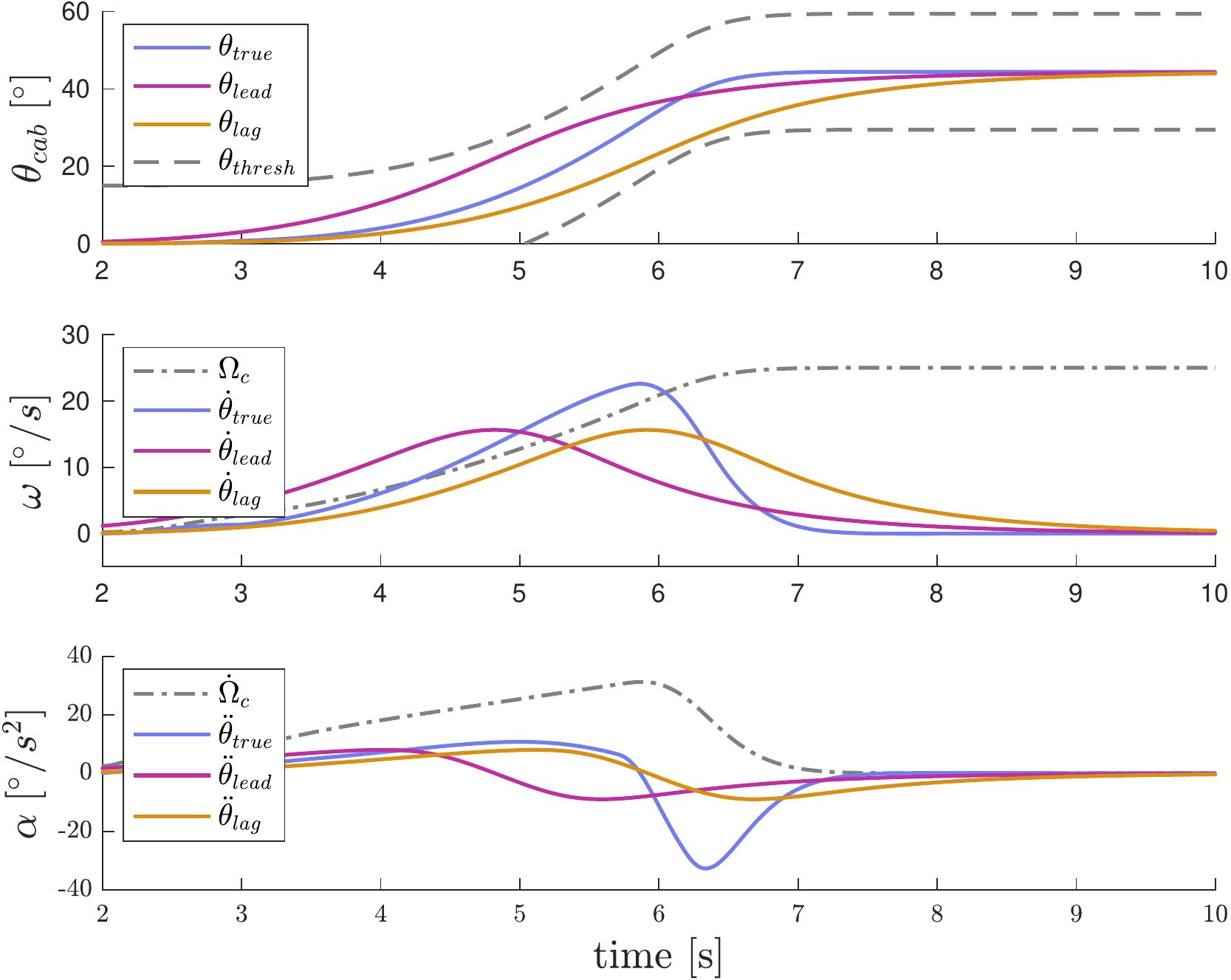}
 \caption{Response characteristics of nominal (blue), leading (magenta) and lagging (yellow) cabin rotation pitch motions.}
 \label{fig:cabresp}
\end{figure}

\figref{fig:cabresp} illustrates the response characteristics for these three cabin rotation trajectories (true, leading, lagging), in terms of the achieved cabin rotation angle (top), rotation rate (center) and rotation acceleration (bottom). The top plot shows the CAZ limits using dashed lines. The center and bottom plots also show, respectively, the centrifuge rotation rate $\Omega_c$ and acceleration $\dot{\Omega}_c$ using dash-dot lines. From these latter plots, it is clear that the cabin rotation rate and acceleration are significantly reduced in both the leading and lagging mechanisms. In addition, the {\em onset} of the cabin angular acceleration peaks is shifted to a region with lower centrifuge yaw rate magnitude, further decreasing the product of two angular rotations, \eqref{eq:holly}, reducing kinematic cross-couplings. A significant reduction is achieved in cabin rotation rate and acceleration with respect to the nominal case. Both leading and lagging mechanisms sweep through the CAZ region, introduce a mismatch, but as long as this mismatch does not exceed the CAZ threshold, the motion is perceived to be coherent.

\subsection{COHAM Building Blocks}

The analysis so far was performed while having full knowledge of the to-be-simulated G-onset, and with that the cabin rotation angle. For any practical application of these findings in an {\em on-line} motion filter the cabin must be coordinated within the CAZ from a baseline G-level (typically for Desdemona 1.4 G, corresponding with $\approx 44.5$ degrees of cabin rotation) to any G-level initiated by pilot control actions, which are unknown beforehand. The main challenge therefore is to predict the required simulator G-level and dynamically adapt the cabin rotation to always remain within the CAZ.

To address this challenge, COHAM has three components. First, COHAM applies a lead-lag filter, the \gls{ttpf}. Second, it includes an algorithm to {\em predict} the G-onset (and with that the required centrifuge rotation rate) from pilot control actions. Third, it applies a Dynamic Lookup Table method \cite{dlt} to guarantee that the mismatch always remains within the CAZ limits. We briefly discuss these components in the following.

\subsubsection{Lead/lag filter: the TTPF}
\label{sec:TTPF}

Leading or lagging mechanisms applied to the cabin rotation can reduce the angular velocity and acceleration associated with cabin alignment. One could argue that both mechanisms characterize a `one-tailed' use of the allowed mismatch. For example, using the lead mechanism the angular velocity and acceleration are increased up to the G response peak approaching from the left (positive mismatch), whereas for the lag mechanism this goes from the peak towards the steady-state value (negative mismatch), see the bottom plot of \figref{fig:coham}. Both mechanisms feature a reversal of mismatch while traversing through the CAZ, and act mainly in one half of the onset curve, hence, one-tailed. 

\begin{figure}[!htb]
\centering
\includegraphics[width=.6\linewidth]{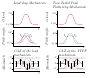}
\caption{Comparison of lead or lag mechanism (left) versus the Two-Tailed Peak Flattening Mechanism (right) for a hypothetical mismatch $\pm U$ in cabin pitch alignment.}
\label{fig:ttpf}
\end{figure}

To fully exploit the CAZ boundaries, however, a combination of both lead and lag mechanisms would be desired, as this could result in `flattening' and `spreading' out of the response over the two tails of the peak. Ideally, this would allow for a twice as wide spread over time and a further reduction of cabin alignment angular velocity and acceleration. This concept is referred to as \glsfirst{ttpf}, and can be characterized as follows (see \figref{fig:ttpf}): 
\begin{compactitem}
 \item the TTPF rotation {\em leads} the true cabin rotation prior to the onset peak $t< t_A$ in \figref{fig:ttpf};
 \item the TTPF rotation has a lower magnitude than the true rotation ($t_A < t < t_B$); and
 \item the TTPF rotation {\em lags} the true cabin rotation after the onset peak $t> t_B$.
\end{compactitem}
\noindent The second characteristic of the TTPF, that the cabin angle rotation can be {\em lower} than the true rotation at the peak of the G simulation, alludes to another opportunity of using the CAZ, 
by having the cabin not return to the nominal rotation after every onset, but actually return to a {\em higher} angle. Again, as long as the cabin alignment mismatch remains within the CAZ, this will not be perceived by pilots. The TTPF filter therefore applies a certain offset from the baseline value (here: $5$ degrees), allowing the controller to lead and transition to lag over a longer time span, and rotate the cabin even less. The result is a \emph{lazy but punctual} control mechanism, which is in phase with the predicted onset peak but always operates within the specified CAZ limits.

\begin{figure}[!htb]
\centering
\includegraphics[width=0.6\linewidth]{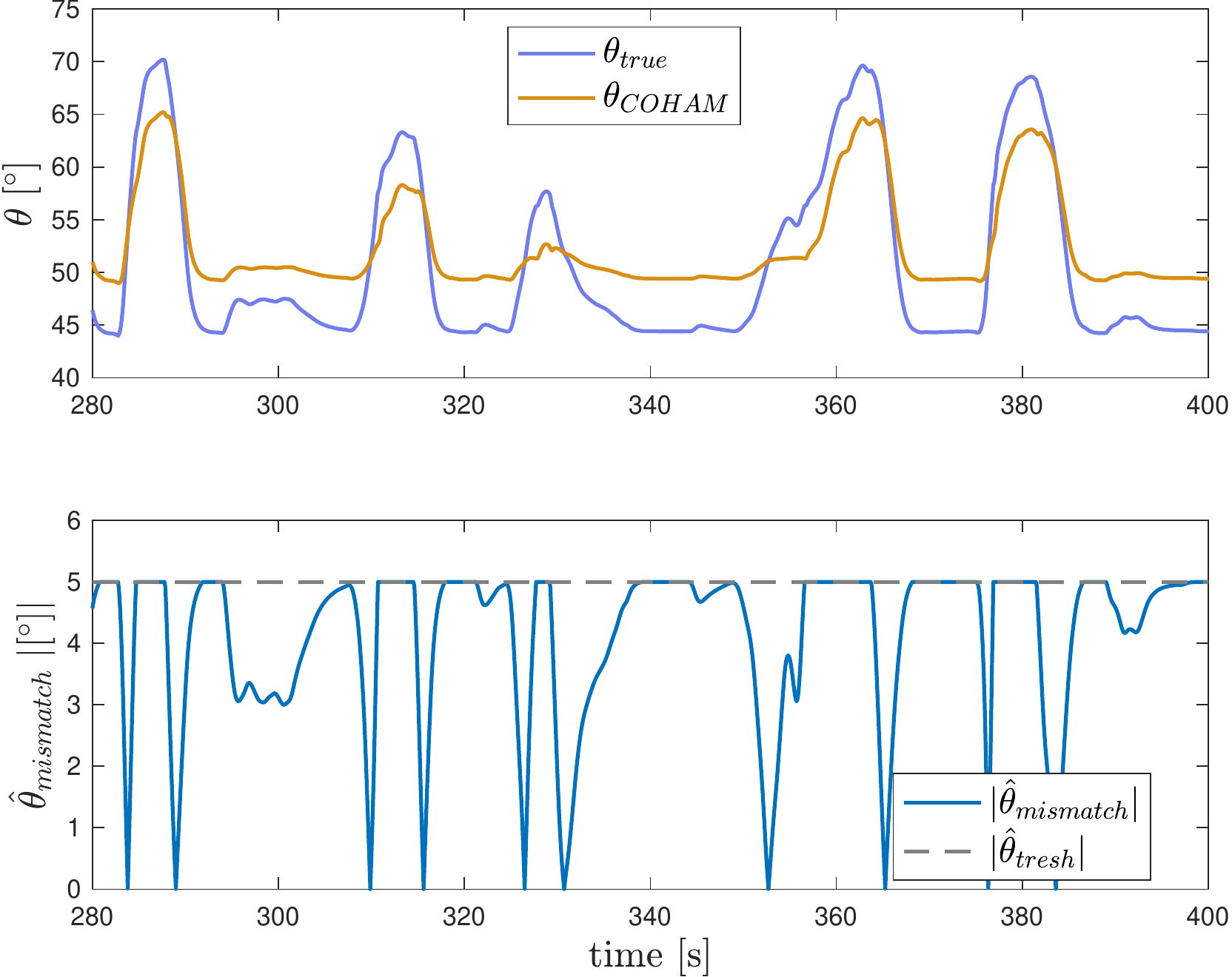}
\caption{Time traces of nominal and COHAM cabin alignment (top), and the mismatch between the two (bottom).}
\label{fig:G_lead}
\end{figure}

The result is illustrated in \figref{fig:G_lead}, which shows the cabin rotation angle (top) and mismatch (bottom) for a recorded sequence of G-pulls in a Desdemona test run. At the baseline level of 1.4G, the cabin is tilted $5$ degrees {\em more} than required, in anticipation of a pilot pulling larger G-forces than 1.4G. Once the pilot pulls a G-force, the cabin pitches forward to align the pilot onto the \gls{gz} vector, but at the maximum G-level ideally remains $5$ degrees {\em short} of the true alignment angle. After the G-pull, the simulator cabin returns to a tilt angle that is $5$ degrees more than nominal.
The mismatch is always smaller than the $5$ degrees limit of the CAZ. In this way, the smallest pitch rotation is required to align the cabin without the pilot noticing the difference.

\subsubsection{Predicting the G-onset}

The ability to predict the G-load pulled by the pilot is crucial for the COHAM filter. Whereas adding $5$ degrees of cabin pitch to the baseline $\theta_{true}$ is trivial, arriving $5$ degrees short at the maximum of the G-maneuver is not. This requires prediction of the G-load exerted on the aircraft model, and consequently a leading response to the instantaneous simulator G-level, $G_{act}$. 


COHAM uses the pilot stick input as a means to predict the G-load resulting from this input. This stems from the fact that any aircraft response can be characterized as `low-pass filtering' the pilot input, causing a lag. In the F-16 fighter aircraft, the stick input controls normal accelerations ($N_x,N_y,N_z$) in all flight conditions apart from landing. The stick input therefore always leads the $N_z$, and therefore the requested G-load.


\begin{figure}[!hb]
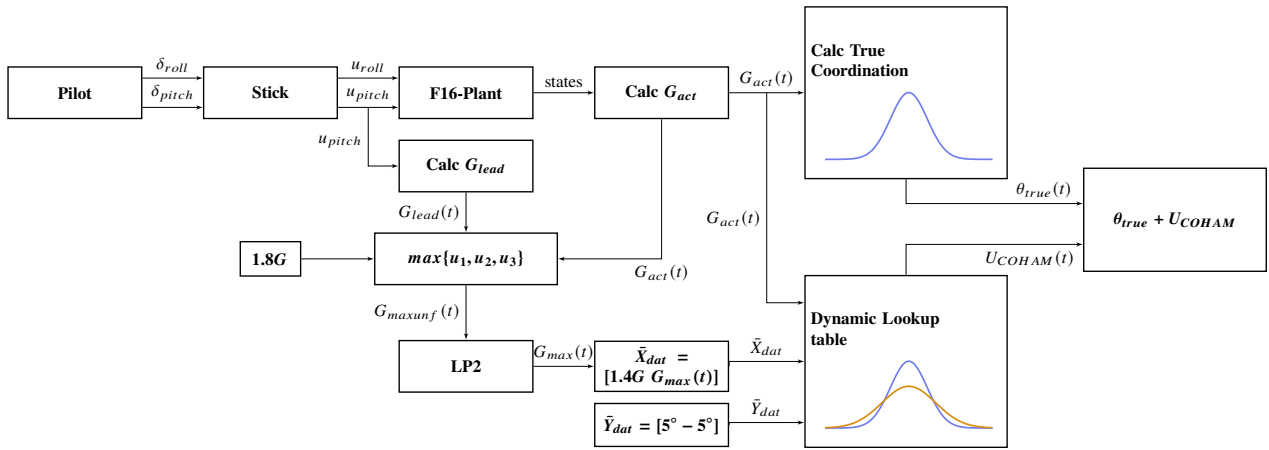

\begin{center}
\scalebox{.65}{%
\ExecuteMetaData[tikz/diagram.tex]{tikzcode}%
}%
\caption{Schematics of the pitch mismatch coordination by COHAM, illustrating the G-onset prediction and use of a Dynamic Lookup Table.}
\label{fig:COHAMscheme}
\end{center}
\end{figure}

\figref{fig:COHAMscheme} illustrates the COHAM prediction mechanism. Starting in the top-left, the pilot exerts forces on the stick, leading to stick inputs in roll and pitch which enter the F-16 dynamics (block F-16 Plant). COHAM uses the pitch input $u_{pitch}$ to directly compute $G_{lead}$ (block ``Calc G-lead''), whereas the simulator software computes the actual F-16 G force using the model (block ``Calc Gact''). This latter block also includes a G-onset timing mechanism based on the maximum G-value that ensures realistic onset even at reduced G-levels, given that the Desdemona simulator has a maximum of 3G and the F-16 model a maximum of 9G. That is, the $Nz$ from the aircraft model (1-9G), is mapped to an instantaneous simulator G-level $G_{act}$ (1-3G) within the simulator performance limits \cite{roza2007performance}. The conventional cueing, illustrated in the top-right, then uses $G_{act}$ to compute the true cabin angle rotation $\theta_{true}$ (block ``Calc True Coordination'').

COHAM, on the other hand, as illustrated in the bottom left of \figref{fig:COHAMscheme}, takes the maximum of two signals, the actual G-level of the F-16 model $G_{act}$ and the leading G-level $G_{lead}$ obtained from the pilot input, and a baseline level of 1.8G. Typical lead times observed in $G_{lead}$ vs. $G_{act}$ are in the order of 1 second. Hence, the maximum G-level will be taken at any time based on the F-16 simulation and pilot inputs, but not lower than 1.8G. This latter minimum will take care of the fact that in case of no or small maneuvers, the cabin rotates back to a higher -- but not perceived to be higher -- pitch angle than would correspond to the baseline 1.4G. The resulting output signal $G_{maxunf}$ is low-pass filtered (second order filter, damping $1$ and natural frequency $1$ rad/s), to obtain the maximum required G-level at that instance, $G_{max}$, which will enter the ``Dynamic Lookup Table'' block (bottom right), discussed next.

\begin{figure}[!htb]
\centering
\includegraphics[width=0.65\linewidth]{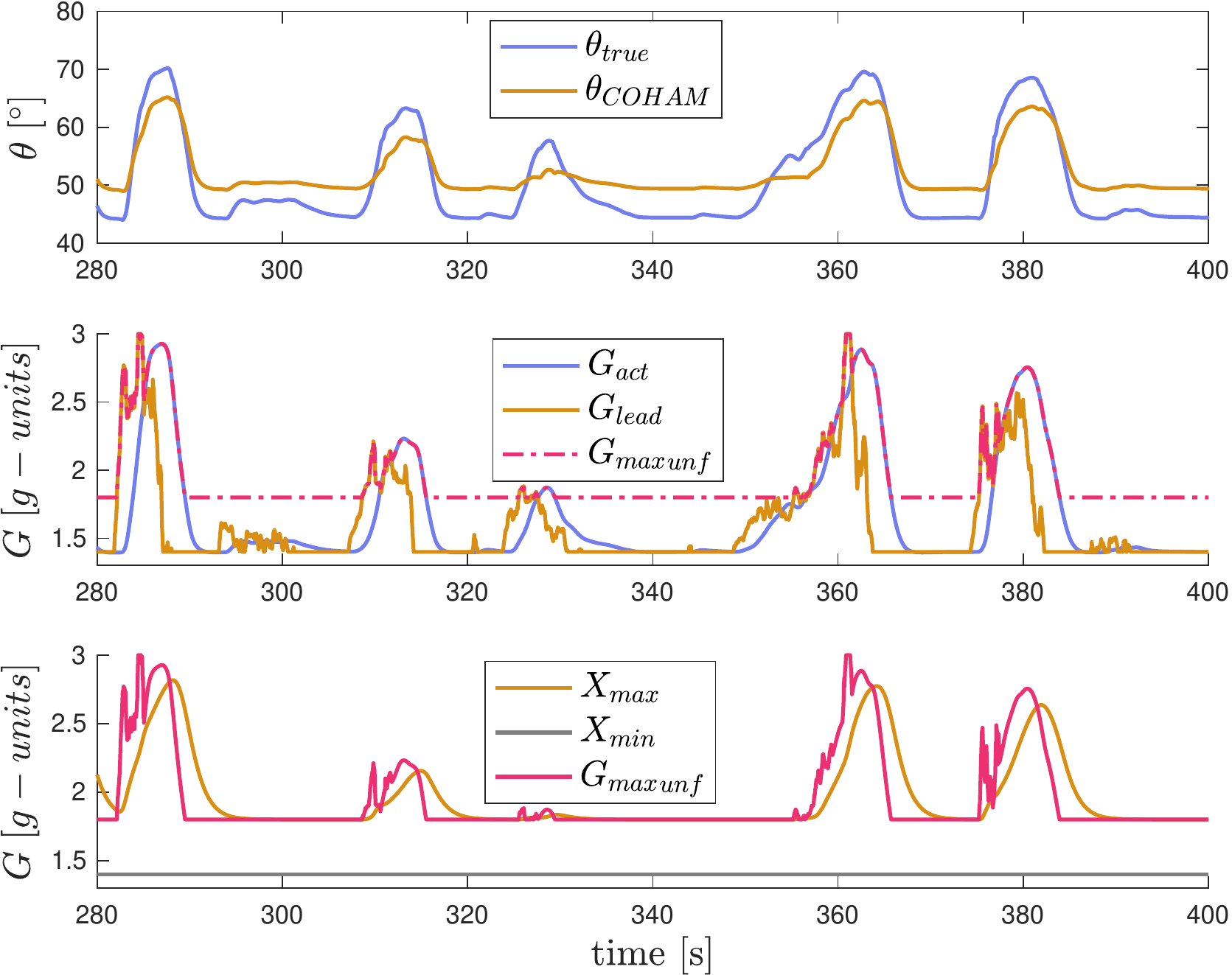}
\caption{Time traces of G-coordination for the test run with the COHAM using the predicted signal.}
\label{fig:Gcalc}
\end{figure}

\figref{fig:Gcalc} illustrates the same maneuver as \figref{fig:G_lead}. The middle figure shows the signal $G_{lead}$ (yellow) clearly leading the actual G-level $G_{act}$ (blue) used in the conventional cueing. The effect of taking the maximum of these two signals and then prohibiting the resulting signal to get below 1.8 G, as well as the relatively high frequency content in $G_{maxunf}$ (magenta) are clear from \figref{fig:Gcalc}. Filtering this signal and applying it to the dynamic look-up table (described below) eventually results in the cabin rotation $\theta_{COHAM}$ shown in the top figure.

There are two additional noteworthy comments on the filtering.
First, because of the substantial lead, $G_{lead}$ arrives at the \emph{estimated} maximum G-load (approximately $1$ second) earlier than the actual response of the aircraft $G_{act}$. It is used to predict what the actual maximum G-level will be at any time to maintain 5 degrees less cabin pitch after passing $t=t_A$ in \figref{fig:ttpf}. Second, given that the F-16 has a (velocity-dependent) approximately proportional relationship between pilot pitch inputs and $N_z$, the $G_{lead}$ signal shape roughly corresponds to the pilot pitch stick input, explaining the relatively high frequency content and the need to filter $G_{maxunf}$.

\subsubsection{Dynamic Lookup Table}
\label{sec:lookup}



To guarantee a smooth transition from ``5 degrees additional pitch at baseline'' to ``5 degrees less at max. G'', a Dynamic Lookup Table is used, illustrated in the bottom right of \figref{fig:COHAMscheme}. 
This function block first creates a dynamic, required G-level interval, from a fixed 1.4G lower bound to a maximum $\tilde{G}_{max}$, i.e., [1.4G $\tilde{G}_{max}$] ($\bar{X}_{dat}$ in \figref{fig:COHAMscheme}), and maps it to the static minimum and maximum cabin angle rotation CAZ thresholds [+5 -5] degrees ($\bar{Y}_{dat}$ in \figref{fig:COHAMscheme}). It then uses the actual aircraft G-level $G_{act}$ to compute the actual cabin pitch angle deviation $U_{COHAM}$ to be added to the true cabin pitch angle $\theta_{true}$ from the conventional simulator cueing. In other words, the dynamic look-up table adapts the required cabin rotation $\theta_{true}$ such that the discrepancy never deviates outside the CAZ, while allowing the COHAM TTPF and predictive blocks to work as designed, i.e., be lazy but punctual.

The inputs to the Dynamic Lookup Table are illustrated in the bottom figure of \figref{fig:Gcalc}. The unfiltered $G_{maxunf}$ signal (red) is filtered by the second order low-pass filter (LP2) yielding the yellow signal $\tilde{G}_{max}$ which is used as the maximum value of the $\bar{X}_{dat}$ vector, $X_{max}$. The minimum value $X_{min}$ is set at 1.4 G. Based on the actual G-level, $G_{act}$, the look-up table dynamically assigns an allowable cabin angle rotation discrepancy ranging between $+5$ degrees ($G_{act}$ lower than 1.4G, so tilting too much forward) and $-5$ degrees ($G_{act}$ higher than the maximum predicted G-level at that time, so tilting less than needed).

\subsection{Recapitulation and Main Results}

Summarizing, the essence of the COHAM dynamic mapping mechanism lies in the following: 
\begin{compactenum}
 \item{Since the saturation value of 1.8G is higher than the baseline 1.4G, for low values of $G_{act}$ the cabin mismatch will always be 5 degrees above the baseline as shown in the top subplot in \figref{fig:Gcalc}. The result is a \emph{lazy} pitch coordination that always leads prior to a G-onset: \gls{thetacoham}$>$\gls{thetatrue}  before $t_A$.}
 \item{Since the maximum G-level is estimated ahead of time, the mismatch flips to a negative mismatch as the G-level rises. The result is a lower coordination onset at the peak: \gls{thetacoham}$<$\gls{thetatrue} between $t_A$ and $t_B$.}
 \item{When the G-level drops back to the baseline, the dynamic input value \gls{Gmaxtilde} (the last element of the \gls{xdat} vector) drops back to 1.8 G, ensuring the cabin moves back to the baseline mismatch: \gls{thetacoham}$>$\gls{thetatrue} after $t_B$.}
\end{compactenum}

\begin{figure}[ht]
 \centering
 \includegraphics[width=.8\linewidth]{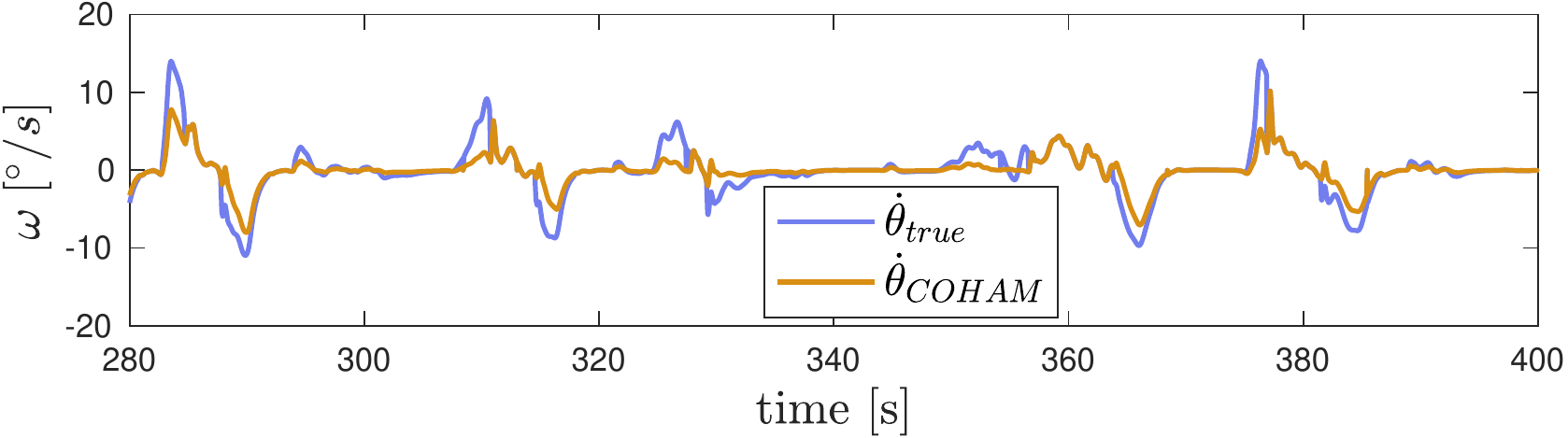}
 \caption{Comparison of true cabin pitch rate and COHAM cabin pitch rate for the same recorded test run.}
 \label{fig:pitch_rate}
\end{figure}

\figref{fig:pitch_rate}, corresponding to the time traces of \figref{fig:G_lead}, shows the cabin rotation rates with or without the COHAM filter. With COHAM, the magnitude at the peaks is reduced up to a factor of two, significantly reducing the Coriolis effect, without the pilot noticing. The COHAM cueing is evaluated experimentally and discussed next.
\section{COHAM Evaluation: Method}
\label{sec:exp}

\subsection{Goal and Description}

A small experiment was performed which compared the Desdemona ``Rocket Man'' motion cueing (RM) to COHAM, in terms of the experienced Coriolis effect, comfort and false cues. The main goal was to evaluate whether COHAM can mitigate Coriolis effects, while still providing a realistic G-cueing experience in high-G piloting tasks. Constraints on the available number of fighter pilots (three) and simulator hours (one week) were severe, so the experiment is a first qualitative evaluation, no statistical analyses were performed; results should be interpreted with caution.

The evaluation lasted four days. During the first three days, pilots were trained to fly an F-16 along a prescribed trajectory with a given airspeed, resulting in a sequence of high G-pulls. Two simulators were used for training, Desdemona and the Dome, both facilitated by the Desdemona B.V.~\cite{desdemona}. Training was important for pilots to produce comparable G-levels and allow for a fair  comparison between RM and COHAM during the fourth evaluation day.


\subsection{METHOD}
\label{sec:method}

\subsubsection{Apparatus}
\label{sec:apparatus}

\begin{figure}[!htb]
\centering
\begin{subfigure}[t]{.45\linewidth}
\vskip 0pt
\centering
\includegraphics[width=\linewidth]{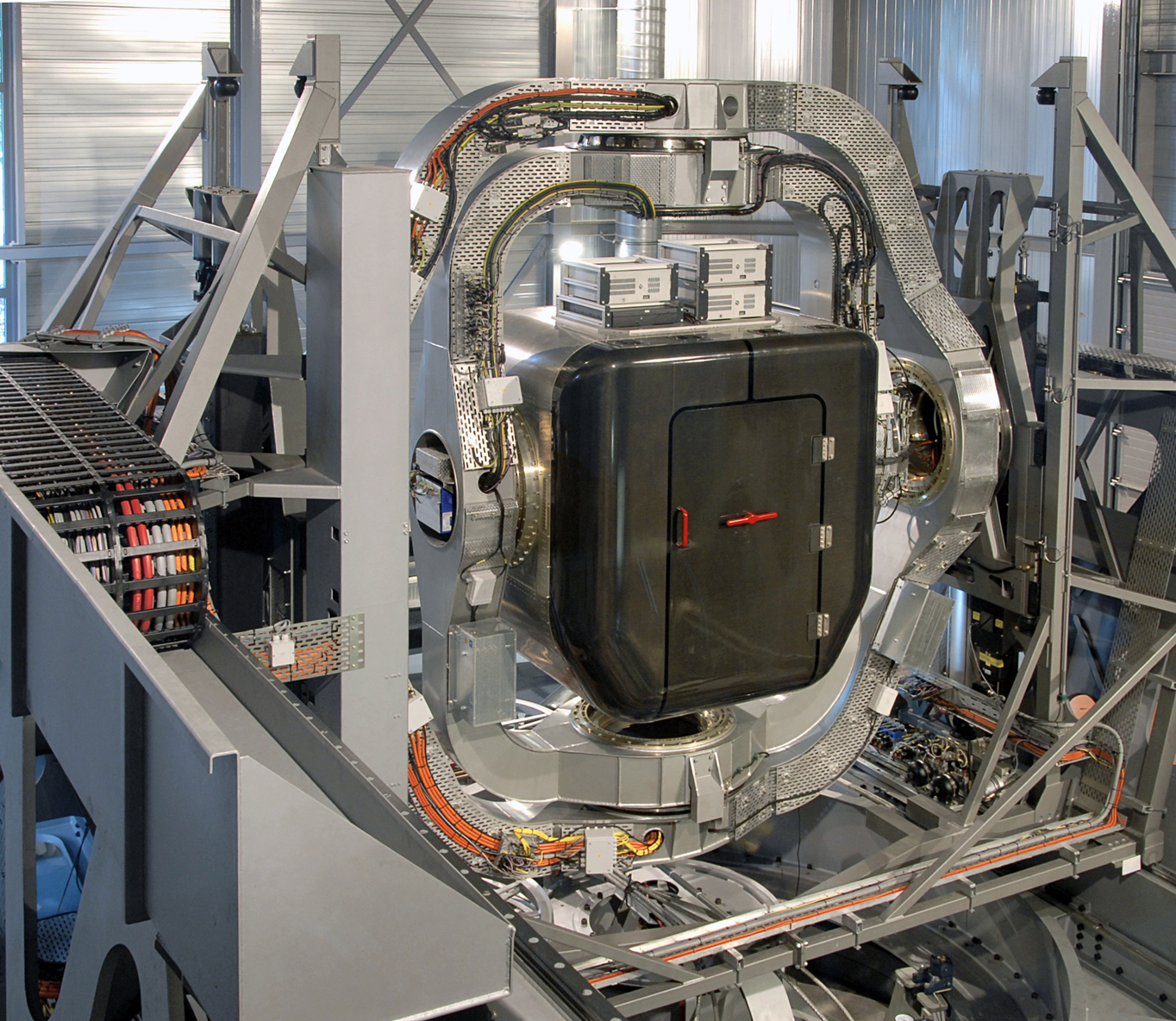}
\caption{Desdemona}
\label{fig:desdemonasim}
\end{subfigure}%
\hskip 1pt
\begin{subfigure}[t]{.445\linewidth}
\vskip 0pt
\centering
\includegraphics[width=\linewidth]{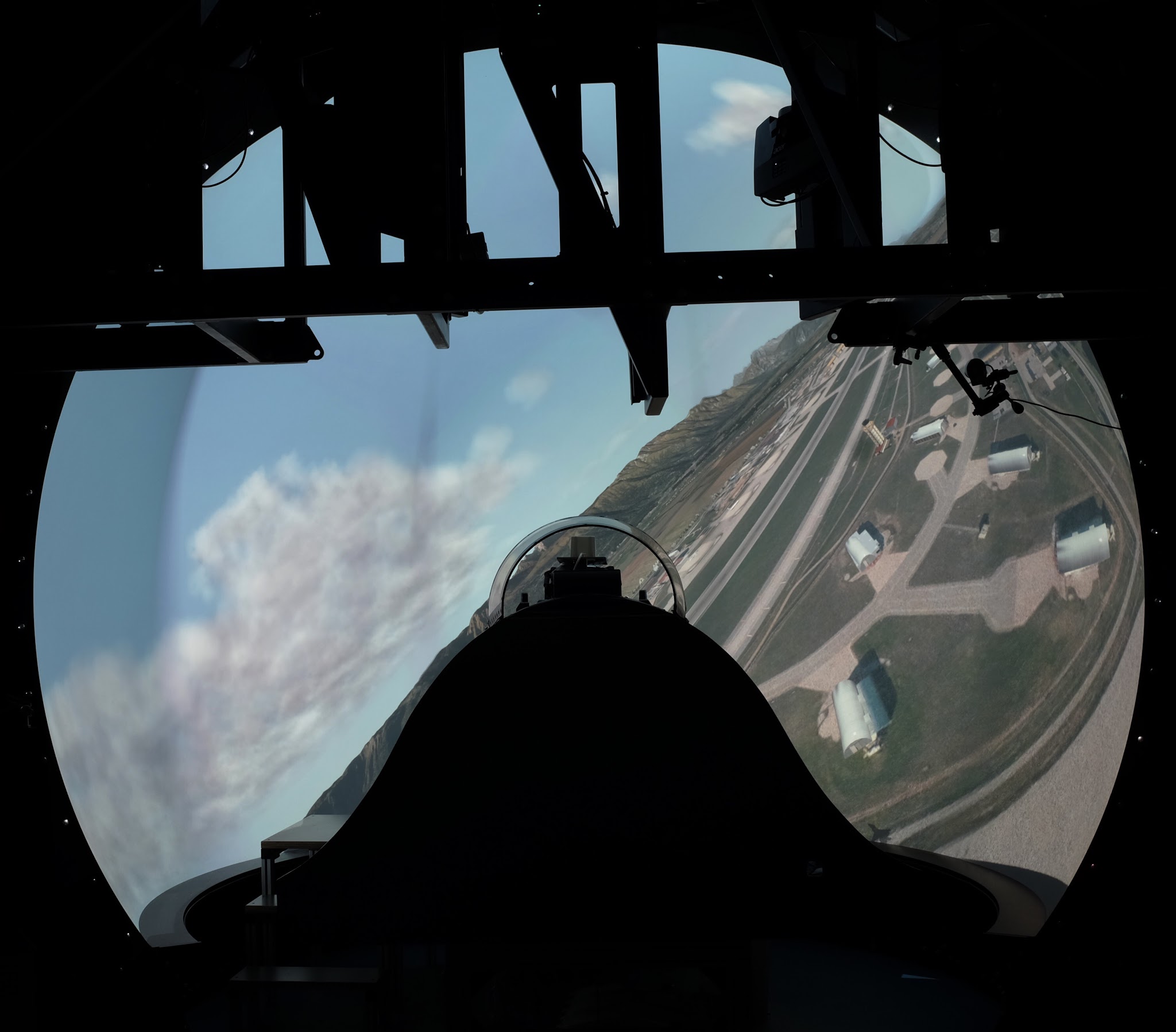}
\caption{Dome}
\label{fig:dome}
\end{subfigure}
\caption{Apparatus used in the evaluation (based in Soesterberg, The Netherlands).}
\label{fig:exp_setup}
\end{figure}

The apparatus used in the experiment were the 6-degrees of freedom, centrifuge-based simulator Desdemona \cite{desdemona}, and the fixed-based F-16 BARCO Dome simulator \cite{barco_dome}, see \figref{fig:exp_setup}. Both simulators were equipped with an F-16 cockpit configuration with nearly identical controls, but a more realistic F-16 cockpit interior for the Dome. The Dome features a 240 x 270 degrees visual \gls{fov}; Desdemona has a 120 x 30 degrees \gls{fov}.

The same simulation environment (Grand Canyon scenery) and trajectory were used in both simulators. 

\subsubsection{Piloting Task}
\label{sec:ptask}

\begin{figure}[!htb]
\centering
\includegraphics[width=0.6\textwidth]{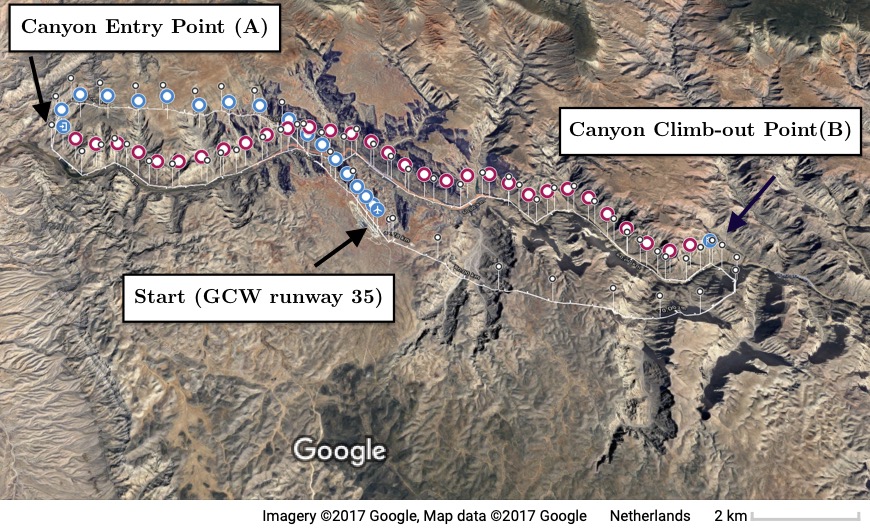}
\caption{Map of the Grand Canyon and the trajectory flown.}
\label{fig:trajectory}
\end{figure}

\begin{figure}[t]
\centering
\begin{subfigure}[t]{.3\linewidth}
\vskip 0pt
\centering
\includegraphics[width=\linewidth]{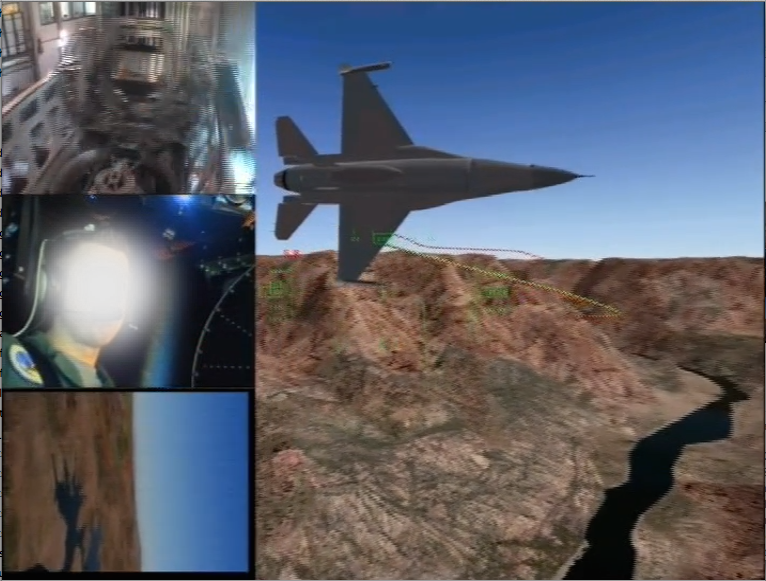}
\caption{S1 just before Dive-In (A).}
\label{fig:divein}
\end{subfigure}%
\hskip 1pt
\begin{subfigure}[t]{.3\linewidth}
\vskip 0pt
\centering
\includegraphics[width=\linewidth]{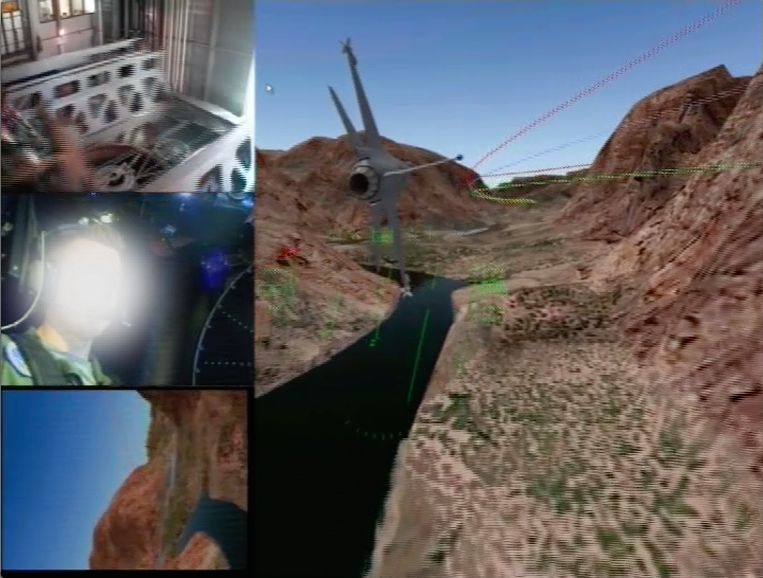}
\caption{S2 mid-trajectory.}
\label{fig:mid}
\end{subfigure}%
\hskip 1pt
\begin{subfigure}[t]{.3\linewidth}
\vskip 0pt
\centering
\includegraphics[width=\linewidth]{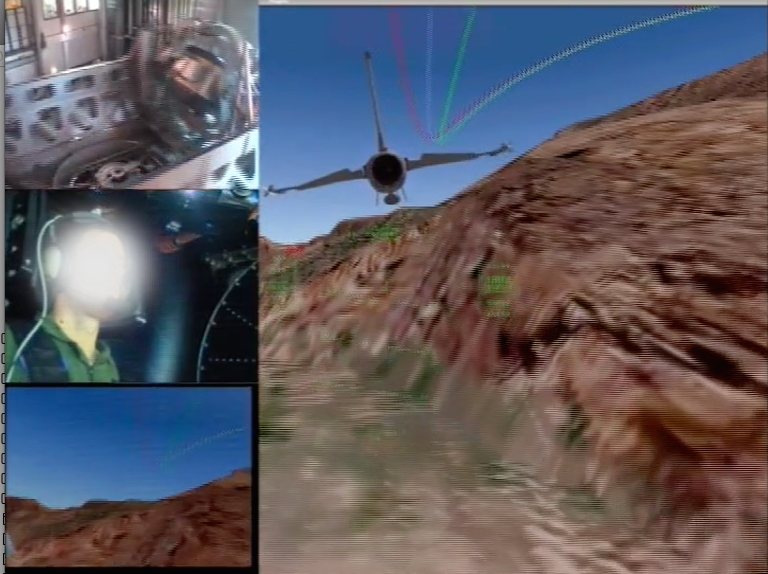}
\caption{S3 just before Climb-Out (B).}
\label{fig:}
\end{subfigure}
\caption{Outside view of F-16 and cockpit camera, showing subjects flying the F-16 through the Grand Canyon.}
\label{fig:traj_pilots}
\end{figure}

The piloting task was to follow a trajectory of approximately 52 miles ($\approx 84 km$) over and along the Grand Canyon, see \figref{fig:trajectory}. A run, or `round' in the following, started with taking-off from Grand Canyon West Airport runway 35, climbing to 3,000 ft, after which pilots entered the Grand Canyon (Dive-In) at entry point (A) and followed its contours at 100 feet AGL until they left the canyon (Climb-Out) at exit point (B). The round was considered completed when the aircraft returned safely to the airport waypoint, flying at 3,000 ft again. For every subsequent round, subjects were asked to proceed with the trajectory (no landing), fly comfortably to the entry point (A) and re-enter the valley for another round. \figref{fig:traj_pilots} illustrates the pilot outside view at three positions during a round.

The trajectory consisted of a high-G phase with a sequence of high pulls (indicated by the red circles in \figref{fig:trajectory}) and a low-G cruise phase (blue circles) allowing subjects sufficient time to recover from the high-G strain in-between rounds. The G-peaks of the high-G phase of the trajectory varied between 3 G to 5 G with two extremes at 7 G. 

The target trajectory was shown on the outside visual as a three-dimensional contour built up by red, green and blue lines, corresponding to the traces of the aircraft left wing tip, right wing tip and body axis, respectively. The red and green lines served to control the required amount of bank, thus the amount of G-force pulled at the given airspeed, see \figref{fig:traj_pilots}. Airspeed was kept constant by subjects at approximately 400 kt, and they were trained to accurately reproduce the desired pattern of G-pulls during the high G-phase. 


\subsubsection{Independent Variable}

The independent variable was the Desdemona centrifuge simulator motion cueing condition, with two levels: Rocket Man and COHAM. In the following these are referred to as C1 and C2, respectively.

\subsubsection{Control Variables}

All runs featured the same trajectory. The COHAM filter was set to accommodate a maximum of 5 degrees mismatch in cabin pitch rotation. The cockpit was kept in constant F-16 configuration and 30-degree seat inclination (from the vertical). Subjects were outfitted with the same equipment: headphones with two-way communication. Lastly, the briefing, questionnaire, experimental procedure and protocol were the same for all subjects.

\subsubsection{Participants and Instructions}
\label{sec:subj}

The three participants (all males, aged between 22 and 28 years) were Dutch air-force freshmen, in the midst of completing their elementary military pilot training. All had piloting experience with light trainer aircraft (PC-7) and limited ($<$2 hours) centrifuge experience (conventional centrifuge and Desdemona).

Prior to and during the training phase (days N1-N3), subjects were told that they participated in a study of long term G-training; they were not made aware of the existence of two motion filters. 
On the evaluation day (N4), subjects were told that two motion filters were being compared and got familiarized with the experiment questionnaire. They were given no specific details about the type of motion filter, nor were they aware which of the two was active at any given time. Each subject performed an assessment and two comparisons (referred to as R1 and R2) of the two motion filters, using a set of criteria including comfort, dizziness, sickness and G-alignment, all discussed below.

\subsubsection{Experiment Set-up and Procedure}


\noindent{\em Experiment Set-up} \hspace*{10mm} Table \ref{tab:schedule} illustrates the training and evaluation schedule for the three subjects S1-S3. In this table, labels D and C refer to simulation in the Dome and Desdemona, respectively, with the number of runs in  parentheses. For the Desdemona runs, `1' and `2' refer to the RM and COHAM motion cueing, respectively. In each row, a vertical bar indicates the transition from the Dome to the Desdemona simulator. 
The first day of the training phase (N1) consisted of five rounds in the Dome and one round in the Desdemona simulator with Rocket Man cueing C1. This was done to get subjects accustomed to the controls and the trajectory. Days 2 and 3 were used to also familiarize the pilots to the COHAM motion and train their trajectory-following skills.

\begin{table}[htbp]
  \centering
  \caption{Training and evaluation schedule.}
  \small
    \begin{tabular}{llll}
    \hline
          & Subject 1 (S1)   & Subject 2 (S2) & Subject 3 (S3)\\\hline
          \multicolumn{4}{c}{\em Training days} \\
    \hline
    Day 1 (N1) & D($\times$5) $\vert$ C1     & D($\times$5) $\vert$ C1 & D($\times$5) $\vert$ C1 \\
    Day 2 (N2) & D($\times$1) $\vert$ C12121 & D($\times$1) $\vert$ C21212 & D($\times$1) $\vert$ C12121 \\
    Day 3 (N3) & D($\times$1) $\vert$ C12121 & D($\times$1) $\vert$ C21212 & D($\times$1) $\vert$ C12121 \\ \hline
          \multicolumn{4}{c}{\em Evaluation day} \\
    \midrule
    Day 4 (N4) 
    & D($\times$1) $\vert$ $ \underbrace{\mbox{C12}}_{\mbox{R1}},\underbrace{\mbox{C21}}_{\mbox{R2}}$ 
    & D($\times$1) $\vert$ $ \underbrace{\mbox{C21}}_{\mbox{R1}},\underbrace{\mbox{C12}}_{\mbox{R2}}$ 
    & D($\times$1) $\vert$ $  \underbrace{\mbox{C12}}_{\mbox{R1}},\underbrace{\mbox{C21}}_{\mbox{R2}}$  \\
    \hline
    \end{tabular}%
  \label{tab:schedule}%
\end{table}%

\noindent{\em Procedure} \hspace*{10mm} In the final evaluation session on day 4, the R1 and R2 refer to the two evaluations each subject performed of the two motion conditions in Desdemona, RM and COHAM, applied in a different order. The evaluation was done by means of a questionnaire. To reduce simulator time and exposure to motion, the questionnaire was evaluated over the intercom, during the time subjects flew from the canyon exit (point B) back to the canyon entry (A).

During the whole experiment, subjects were asked to report their well-being scores, using the \gls{misc} score (0-10). \gls{misc} is developed by Wertheim et al. \cite{wertheim2001predicting,hale2014handbook,bos2010effect}, see \tabref{tab:misc}. Subjects provided their \gls{misc} score  twice per round: just before the initial dive into the canyon (point A) and right after the climb-out (point B). Any training or evaluation session would be aborted whenever a MISC score of 5 or higher was given.

\begin{table}[ht]
\centering
\caption{MIsery SCale, the MISC score.}
\label{tab:misc}
\small
\begin{tabular}{p{5.5cm}p{1.5cm}c}
\hline
   Symptom & Degree & Score  \\
\hline
    No problems     &   & 0  \\
     Uneasiness (no typical symptoms) & &   1\\
       & &  \\
      \multirow{4}{4cm}{Dizziness, warmth, headache,\\
    stomach awareness, sweating}& Vague & 2\\ 
      & Slight & 3 \\
      & Fairly & 4\\
      & Severe & 5\\
      & & \\
     \multirow{5}{4cm}{Nausea} &  Slight & 6 \\
     &Fairly & 7 \\
     &Severe & 8\\
     &Retching & 9\\  
     & Vomiting & 10 \\
\hline
\end{tabular}%
\end{table}%

\subsubsection{Questionnaire}

The questionnaire was completed twice (R1 and R2) after flying two rounds, with the motion cueing filters RM and COHAM, in different orders, see \tabref{tab:schedule}. It consisted of two types of assessment: the subject well-being, and the experienced G-alignment. All assessments must be regarded as a `weighted' score, as they are all based on the `accumulated experience over time' when flying with {\em both} filters. Further, note that for all scores, low is better. 

\noindent{\em Well-being Scores and Preference Assessment} \hspace*{10mm} Subjects were asked to rate both motion filters using two metrics.

First, in the \gls{aws}, \tabref{tab:aws}, subjects assessed the experienced motion sickness, dizziness and bearability, using a scale similar to the Griffin and Newman Scale \cite{griffin2004visual}.

\begin{table}[ht]
  \centering
  \caption{Accumulated Well-being Scale (AWS).}
  \small
    \begin{tabular}{lc}
    \hline
    Definition & Score \\
    \hline
    no symptoms experienced & 0 \\
    slight symptoms experienced (not always recognizable) & 1 \\
    mild effect experienced (recognizable) & 2 \\
    moderate to mildly experienced & 3 \\
    moderate but was able to continue & 4 \\
    unbearable wanted to stop & 5 \\
    \hline
    \end{tabular}%
  \label{tab:aws}%
\end{table}%

Second, subjects were asked to indicate which of the two cueing settings they preferred in the \gls{pbws}, using the same three categories as in the AWS (sickness, dizziness, bearability) extended with a fourth category, comfort. In the PBWS, subjects simply had to choose which of the two cueing settings they preferred, or neither. Because subjects did not know what cueing setting was active in what round, they indicated it using `first filter', or `second filter', in the questionnaire. These were then coupled to C1 or C2 by the experimenter.

\noindent{\em G-alignment coherency metrics} \hspace*{10mm} The effects of the motion filter on the perceived G-alignment were assessed using two metrics.
First, in the \gls{ams}, subjects were asked to indicate whether during G-onsets they experienced any mismatch in G-alignment relative to the true G-vector. AMS is expressed in mismatch direction (sideways left, sideways right, forward and backward) and (relative) magnitude (range of angular mismatch in degrees: $0^o$, $0-5^o$, $5-10^o$, $>10^o$). 


Second, the \gls{alac} was defined for the {\em entire} centrifuge session, see \tabref{tab:alac}. ALAC has a numeric scale ranging from 0 to 5, and represents the accumulated level of alignment coherency for a given motion condition; here 0 stands for motion cueing that is absolutely coherent throughout the session (G-vector is `spot-on') and 5 for cueing that is not coherent (G-vector always too far titled).

\begin{table*}[ht!]
  \centering
  \caption{Accumulated Level of Alignment Coherency of the G-cueing (ALAC).}
  \small
  \label{tab:alac}%
    \begin{tabular}{lllc}
    \hline
    Level of Coherency  & Alignment G-vector  & ability to & Score  \\
    \ & \ & detect deviation & \ \\
    \hline
    absolutely coherent &  G-vector spot on & impossible & 0 \\
    almost always coherent & G-vector almost always spot on  & very hard & 1 \\
    fairly coherent  & G-vector is true most of the time sometimes too far tilted  & hard  & 2 \\
    moderately coherent & G-vector is true most of the time sometimes too far tilted  & moderate & 3 \\
    fairly incoherent & G-vector is not true most of the time & easy  & 4 \\
    not coherent & G-vector always too far tilted & very easy & 5 \\
    \hline
    \end{tabular}%
\end{table*}%

\noindent{\em Overall Preference Verdict} \hspace*{10mm} As a final metric, at the end of each comparison subjects were asked their overall preference (first filter, second filter, no preference). These were labeled C1 or C2 by the experimenter.

\subsubsection{Dependent Measures}

A number of dependent measures were obtained throughout the whole experiment: the MISC scores introduced above, the F-16 model G-levels, the simulator G-levels, and the cabin rotations and rotation rates. The F-16 model G-levels higher than 2 G (G-peaks) were processed to obtain their probability distributions. 

All other dependent measures were obtained only in the evaluation phase (N4) in the four rounds flown in Desdemona, using the questionnaire. These were the AWS, PWBS, AMS and ALAC scores defined above, and the overall preference.

\subsection{Hypotheses}
\label{sec:hypos}

The first hypothesis is that the COHAM cueing is preferred by our subjects. The lower cabin G-alignment rotations and rotation rates with COHAM mitigate the Coriolis cross-couplings, yielding lower MISC scores, lower AWS scores, higher PBWS scores and an overall preference for COHAM. The second hypothesis is that COHAM will be perceived as coherent within 5 degrees of CAZ, as indicated by low AMS and ALAC scores.

The third hypothesis is that the considerable time for training (three days) will lead to comparable G-levels in the Dome and Desdemona simulators, despite the high G-loads in Desdemona. The latter is because earlier experience with training pilots on Desdemona has shown that pilots learn to cope with the G-environment quickly. Our fourth hypothesis is that this will also be the case in this experiment, that the prolonged exposure to elevated G-levels will increase pilot endurance. It means that we expect the well-being ratings to decrease during the experimental days, perhaps only to increase again when many runs in Desdemona are done on the same day.
\section{COHAM Evaluation: Results}
\label{sec:results}

Results of the first evaluation of COHAM are presented, using the subjective and objective measures defined above.

\subsection{Subjective Measurements}

\subsubsection{MISC Scores}

\tabref{tab:misc_scores} lists all MISC scores given during the Desdemona sessions conducted in the last three days (N2-N4) of the experiment. Recall that the MISC scores were collected twice for each round, right after the `Dive-In' maneuver at the canyon entry point (A), and right after the `Climb-Out' maneuver at the canyon exit point (B).

\begin{table}[ht!]
    \centering
    \caption{All MISC scores given (note: $\star$ means that no rating was given).}
    \label{tab:misc_scores}
    \small
    \begin{tabular}{llcccccccccccccccccccc}
    \hline
    \setlength{\tabcolsep}{0.0mm}
    \ & \ & \multicolumn{6}{c}{N2} & \ 
    & \multicolumn{6}{c}{N3} & \ 
    & \multicolumn{6}{c}{N4}\\
    \cline{3-8} \cline{10-15} \cline{17-22}
    \ & \  & C1 & C2 & C1 & C2 & C1 & C2 & \ 
           & C1 & C2 & C1 & C2 & C1 & C2 & \ 
           & C1 & C2 & C1 & C2 & C1 & C2\\
    \hline
    S1 & A & 1  & 1  & 1  & 1  & 2  &  &  \ 
           & 1  & 1  & 1  & 1  & 2  &  &  \ 
           & 1  & 1  &    & 1  & 1  &  \\ 
    \  & B & 2  & 2  & 2  & 2  & 2  &  &  \ 
           & 1  & 1  & 1  & 1  & 2  &  &  \ 
           & 1  & 1  &    & 1  & 1  &  \\ 
    S2 & A &    & 1  & 6  & $\star$ & $\star$ & $\star$ & \ 
           &    & 1  & 1  & 1  & 1  & 2 & \ 
           &    & 1  & 1  &    & 1  & 1 \\ 
    \  & B &    & 3  & 5  & $\star$ & $\star$ & $\star$ & \ 
           &    & 1  & 1  & 1  & 1  & 2 & \ 
           &    & 1  & 1  &    & 2  & 1 \\ 
    S3 & A & 2 & 3 & 4 & 5 & $\star$ & & \ 
           & 1 & 2 & 2 & 2 & 2 & & \ 
           & 1 & 2 &   & 2 & 3 & \\ 
    \  & B & 2 & 3 & 4 & 4 & $\star$ & & \ 
           & 1 & 2 & 2 & 3 & 2 & & \ 
           & 1 & 2 &   & 2 & 3 & \\ 
    \hline      
    \multicolumn{16}{c}{} &
    \multicolumn{3}{c}{R1} &
    \multicolumn{3}{c}{}\\
    \cline{17-19}
    \multicolumn{16}{c}{} &
    \multicolumn{3}{c}{} &
    \multicolumn{3}{c}{R2}\\
    \cline{20-22}
    \end{tabular}
\end{table}

Several trends can be read from this table. First, the MISC scores are largest in the second day (N2) and become much smaller at the last day (N4). On the second day, a few runs had to be aborted for subjects S2 and S3 because of increasing motion sickness. Second, the MISC scores are generally a little higher after the exit point (B), which can be expected as between (A) and (B) the canyon was followed requiring several high G maneuvers. Third, in the two comparisons (R1 and R2) Subjects 1 and 2 reported the lowest MISC score for almost all rounds, whereas Subject S3 reported slightly lower MISC scores for the RM cueing in R1, then higher MISC scores for RM cueing in R2. 

Adding all scores (all subjects, days, points (A) and (B)) yields totals of 72 and 62 for RM (38 rounds) and COHAM (36 rounds), respectively. Adding all scores for the final day (N4) yields 17 and 16 for RM (12 rounds) and COHAM (12 rounds), respectively, (R1: 6 for RM, 8 for COHAM; R2: 11 for RM, 8 for COHAM). Note that adding MISC scores is a questionable step and merely yields a qualitative, first indication.

These MISC scores are evidence that subjects became accustomed with the centrifuge-based G cueing; a small difference emerged in favor for the COHAM cueing.





\subsubsection{Accumulated Well-being Score (AWS)}

The AWS scores for metrics sickness, dizziness and bearability are summarized in \tabref{tab:aws_scores}. Recall that these scores were given by pilots once for each session (R1, R2) flown with both two motion conditions (C1 and C2).

\begin{table}[ht]
\centering
\caption{Accumulated Well-being Scores (AWS).}
\small
\label{tab:aws_scores}
\begin{tabular}{lccccccccc}
\hline
\ & \ & \multicolumn{2}{c}{R1} & & \multicolumn{2}{c}{R2} & & \multicolumn{2}{c}{\em ``best''}\\
\ & subject & C1 & C2 & & C1 & C2 & & R1 & R2 \\
\hline
sickness & S1 & 1 & 1 & & 1 & 0 & & -  & C2 \\
\        & S2 & 1 & 1 & & 1 & 2 & & -  & C1 \\
\        & S3 & 1 & 2 & & 3 & 2 & & C1 & C2 \\
\hline
dizziness & S1 & 0 & 0 & & 1 & 0 & & - & C2 \\
\         & S2 & 1 & 1 & & 1 & 2 & & - & C1 \\
\         & S3 & 1 & 1 & & 3 & 2 & & - & C2 \\
\hline
bearability & S1 & 1 & 1 & & 2 & 1 & & -  & C2 \\
\           & S2 & 2 & 3 & & 1 & 2 & & C1 & C1 \\
\           & S3 & 2 & 2 & & 3 & 2 & & -  & C2 \\
\hline
\ & \ & 10 & 12 & & 16 & 13 & & C1 & C2 \\
\cline{3-4} \cline{6-7}
\ & \ & \multicolumn{2}{c}{22} & & \multicolumn{2}{c}{29} &  & \ & \ \\ 
\end{tabular}
\end{table}

Each row in the table shows the score given for C1 and C2 in R1, C1 and C2 in R2, and based on a comparison of these scores the ``best'' cueing was distilled by the authors for each of the two comparisons R1 and R2 (last two columns). Note that subjects were also asked explicitly for their preference; these scores are discussed in the next subsection.

For all scores, it can be seen that in the first comparison R1, pilots gave quite similar ratings, with perhaps a slight preference for C1 (added score 10 relative to 12 for C2). In the second comparison, R2, the scores were slightly higher (which means worse, 29 in total for R2 versus 22 for R1), especially for dizziness, and two of the three pilots gave lower (=better) ratings for motion condition C2 (added score 13 relative to 16 for C1). Note that adding AWS scores is a questionable step and merely yields a qualitative, first indication.

The lower scores for COHAM in the last session R2 is evidence that pilots can longer sustain the G training, which is supported by verbal comments (not shown). However, the number of pilots in this exploratory experiment is too low to draw definite conclusions.

\subsubsection{Preference-Based Well-being Scores (PBWS)}

\begin{table}[ht]
\centering
\caption{Preference-Based Well-being Scores (PBWS).}
\small
\label{tab:pbws_scores}
\begin{tabular}{lccccccc}
\hline
\ & \multicolumn{3}{c}{R1} & & \multicolumn{3}{c}{R2}\\
\cline{2-4} \cline{6-8}
\ & C1 & C2 & neither & & C1 & C2 & neither \\
\hline
sickness    & 3 &   &   & & 1 & 1 & 1\\
dizziness   & 1 & 1 & 1 & & 1 & 2 &  \\
bearability & 3 &   &   & & 1 & 2 &  \\
comfort     & 2 & 1 &   & & 1 & 2 &  \\
\hline
 \          & 9 & 2 & 1 & & 4 & 7 & 1 \\
\end{tabular}
\end{table}


The PBWS scores for metrics sickness, dizziness, bearability and comfort are summarized in \tabref{tab:pbws_scores}.
The PBWS scores show a similar trend as the AWS ratings. In the first comparison the RM cueing (C1) is preferred in 9 out of 12 possible preferences, but in the second comparison COHAM cueing (C2) is preferred in 7 out of 12. This is some evidence that whereas the default Desdemona RM tuning worked quite well when experienced for a shorter time, the COHAM cueing was more bearable and could be dealt with by our subjects for a longer time.

In their verbal and written comments, subjects reported that their comments and ratings on sickness and dizziness were mainly based on their experiences during high G-onsets, that is in all maneuvers that required G(un)-loading. The Dive-In (canyon entry, A) and Climb-Out (canyon exit, B) maneuvers were of less importance for their ratings. No differences were reported in this respect between the two motion filters.

\subsubsection{G-Alignment Scores}

Assessment of the G-alignment, using ALAC and AMS, is shown in \tabref{tab:alac_scores} and \tabref{tab:ams_scores}, respectively. 

\begin{table}[ht]
\centering
\caption{Accumulated G-ALignment Coherency Scores (ALAC).}
\label{tab:alac_scores}
\small
\begin{tabular}{lccccccc}
\hline
\ & \ & \multicolumn{2}{c}{R1} & \multicolumn{2}{c}{R2} & \multicolumn{2}{c}{\em ``best''}\\
\ & subject & C1 & C2 & C1 & C2 & R1 & R2 \\
\hline
ALAC & S1 & 0 & 0 & 0 & 0 & -  & -  \\
\    & S2 & 1 & 1 & 1 & 3 & -  & C1 \\
\    & S3 & 3 & 4 & 3 & 4 & C1 & C1 \\
\hline
\    & \  & 4 & 5 & 4 & 7 &    & \\
\end{tabular}
\end{table}

\begin{table}[ht]
\centering
\caption{G-Alignment Mismatch Scores (AMS).}
\small
\label{tab:ams_scores}
\begin{tabular}{llccccc}
\hline
\       & \          & Body Tilt \\
subject & evaluation & experienced? & Forward & Backward & Left & Right\\
\hline
S1 & R1 & -        & -       & -        & - & - \\
\  & R2 & C1       & -       & -        & - & $0-5^o$ \\
S2 & R1 & C1 \& C2 & $0-5^o$ & $0-5^o$  & - & - \\
   & R2 & C2       & $0-5^o$ & $0-5^o$  & - & - \\
S3 & R1 & C2       & -       & $5-10^o$ & - & - \\
   & R2 & C1 \& C2 & $0-5^o$ & $0-5^o$  & - & - \\
\hline
\end{tabular}
\end{table}

The ALAC scores table has the same structure as the AWS scores. It can be seen that, apart from Subject 1 who judged the G-alignment in both motion conditions to be `spot-on', Subjects 2 and 3 experienced some issues with C1 (total score 8) but especially C2 (total score 12). Adding ALAC scores is a questionable step and merely yields a qualitative, first indication. Nonetheless, COHAM was not always perceived to be coherent in terms of the G-vector alignment. Subject 3 rated the RM cueing to be `too far tilted' (ALAC level 3) while the G-alignment was in fact correct all the time. He rated the COHAM cueing to be `not true most of the time' (ALAC level 4). This indicates the difficulty to assess ALAC well.

The lower coherence with COHAM is corroborated by the AMS scores, summarized in \tabref{tab:ams_scores}. Subject 1 was the only one to report some slight (0-5 degrees) mismatch to his right, with C1. Subjects 2 and 3 reported slight forward and backward misalignments in C1, but especially in C2, with a 5-10$^o$ backward misalignment for C2 as maximum score.


\subsubsection{Overall preference}

All subjects reported the necessity of the second session (R2) for their definitive verdict. All subjects had an initial slight preference for the Rocket Man cueing, in R1. Subjects 1 and 3 switched their preference to the COHAM cueing in R2, mainly because they found this cueing to cause less sickness and dizziness and also being smoother at G-onsets during sharp turns. Recall that our subjects did not know what cueing was given in what round. The trend in the overall preference verdict corroborates the findings discussed in the AWS and PBWS scores. Apparently, the more often perceived misalignment of the G-vector with COHAM (in ALAC and AMS) did {\em not} lead subjects to prefer the RM cueing. Tentatively, they accepted the misalignments because the overall G environment was more bearable.

\subsubsection{Verbal comments}

Subject S2 reported a higher G-load experienced during a straight and level flight (1.4 G baseline) for the sessions with COHAM. Apparently, he was able to sense the offset in the cabin angle \gls{thetacoham} introduced by COHAM, discussed in \secref{sec:TTPF}, and he interpreted the higher cabin angle rotation ($\approx$ 49 versus 44 degrees) as a higher G-load, whereas the G-load was in fact correct. Clearly, for this subject, the CAZ threshold of 5 degrees was too large, and the perceived `additional tilt' relative to the RM cueing was interpreted as `higher G', which makes sense as when simulating higher G-levels the centrifuge cabin needs to be rotated more inwards to properly align the gravity vector.

\subsection{Objective Measurements}

\subsubsection{F-16 Model and Simulator G-levels}

\begin{figure}[!ht]
\centering
\begin{subfigure}[b]{.45\linewidth}
\vskip 0pt
\centering
\includegraphics[width=\linewidth]{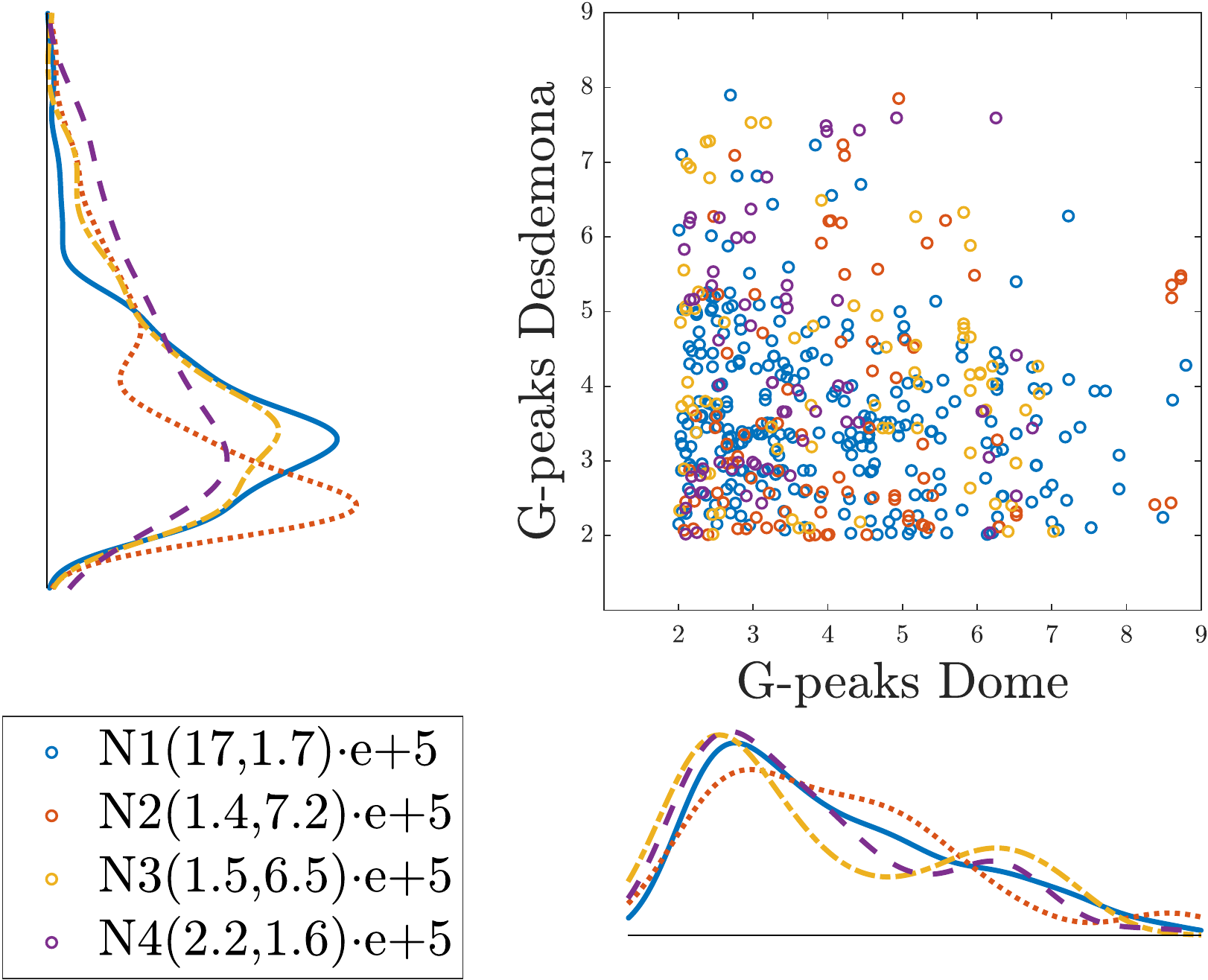}
\caption{Subject S1, days N1-N4}
\label{fig:sc_dodesd_s1}
\end{subfigure}%
\quad
\begin{subfigure}[b]{.45\linewidth}
\vskip 0pt
\centering
\includegraphics[width=\linewidth]{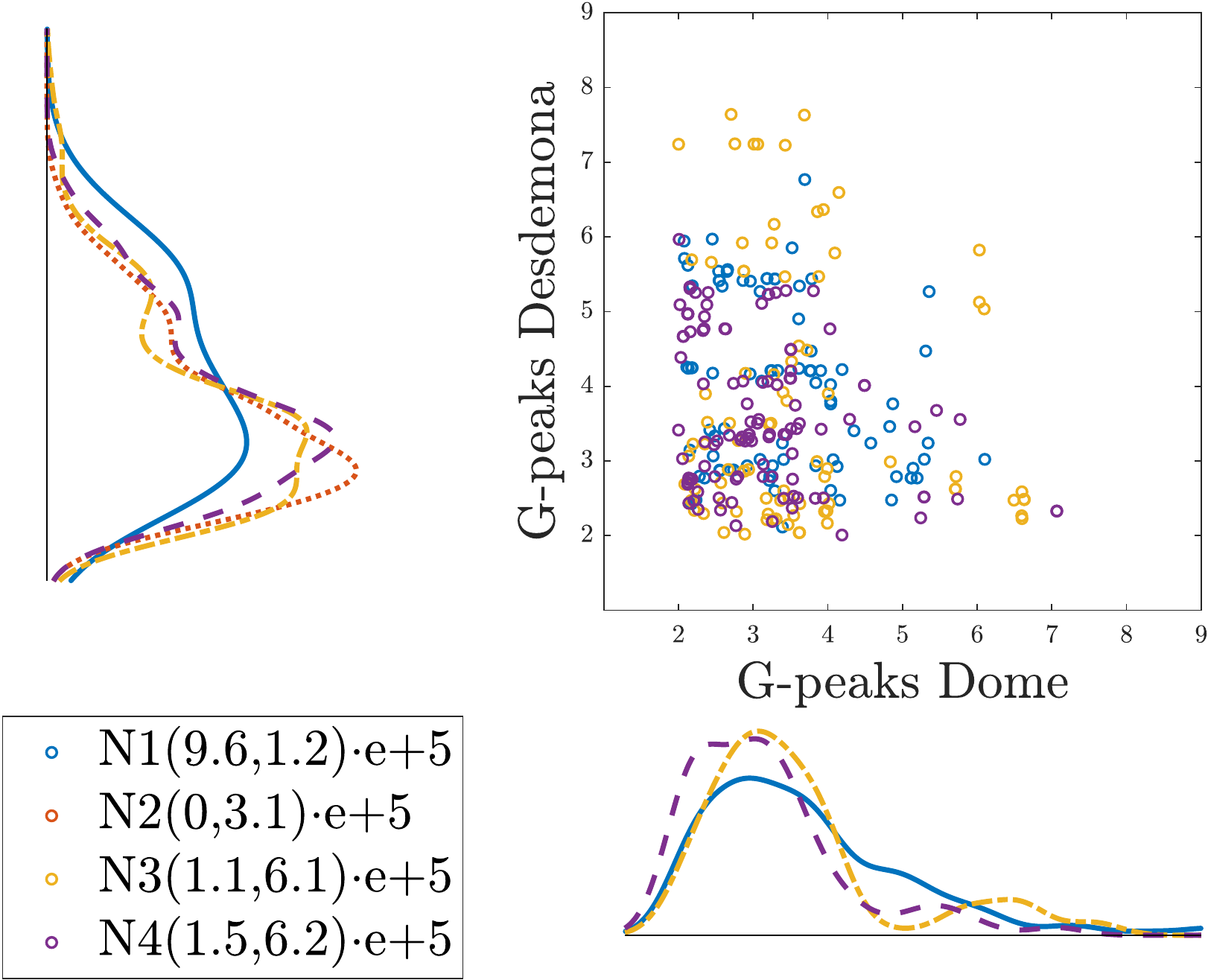}
\caption{Subject S2, days N1-N4}
\label{fig:sc_dodesd_s2}
\end{subfigure}
\newline
\begin{subfigure}[b]{.45\linewidth}
\vskip 0pt
\centering
\includegraphics[width=\linewidth]{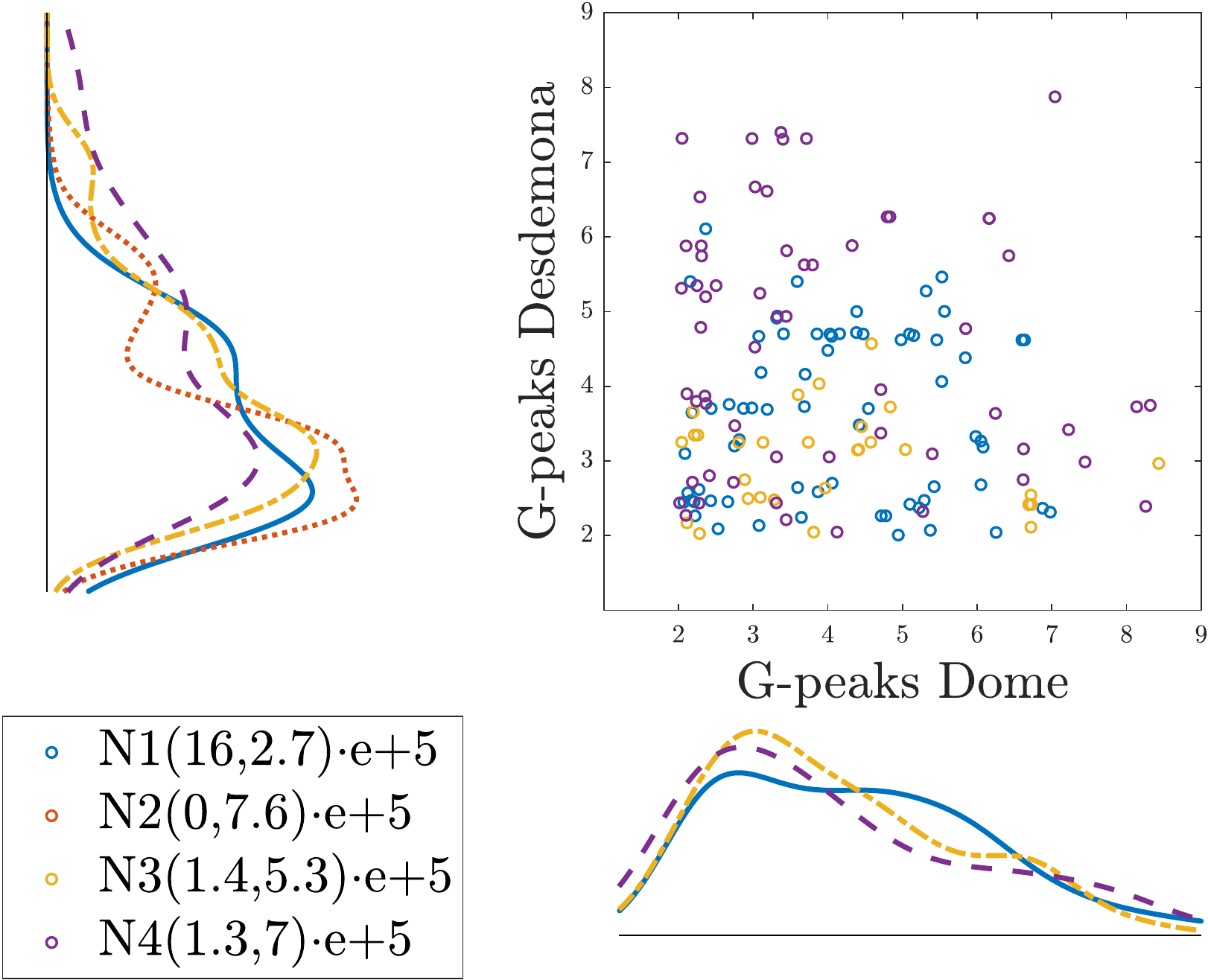}
\caption{Subject S3, days N1-N4}
\label{fig:sc_dodesd_s3}
\end{subfigure}%
\quad
\begin{subfigure}[b]{.45\linewidth}
\vskip 0pt
\centering
\includegraphics[width=\linewidth]{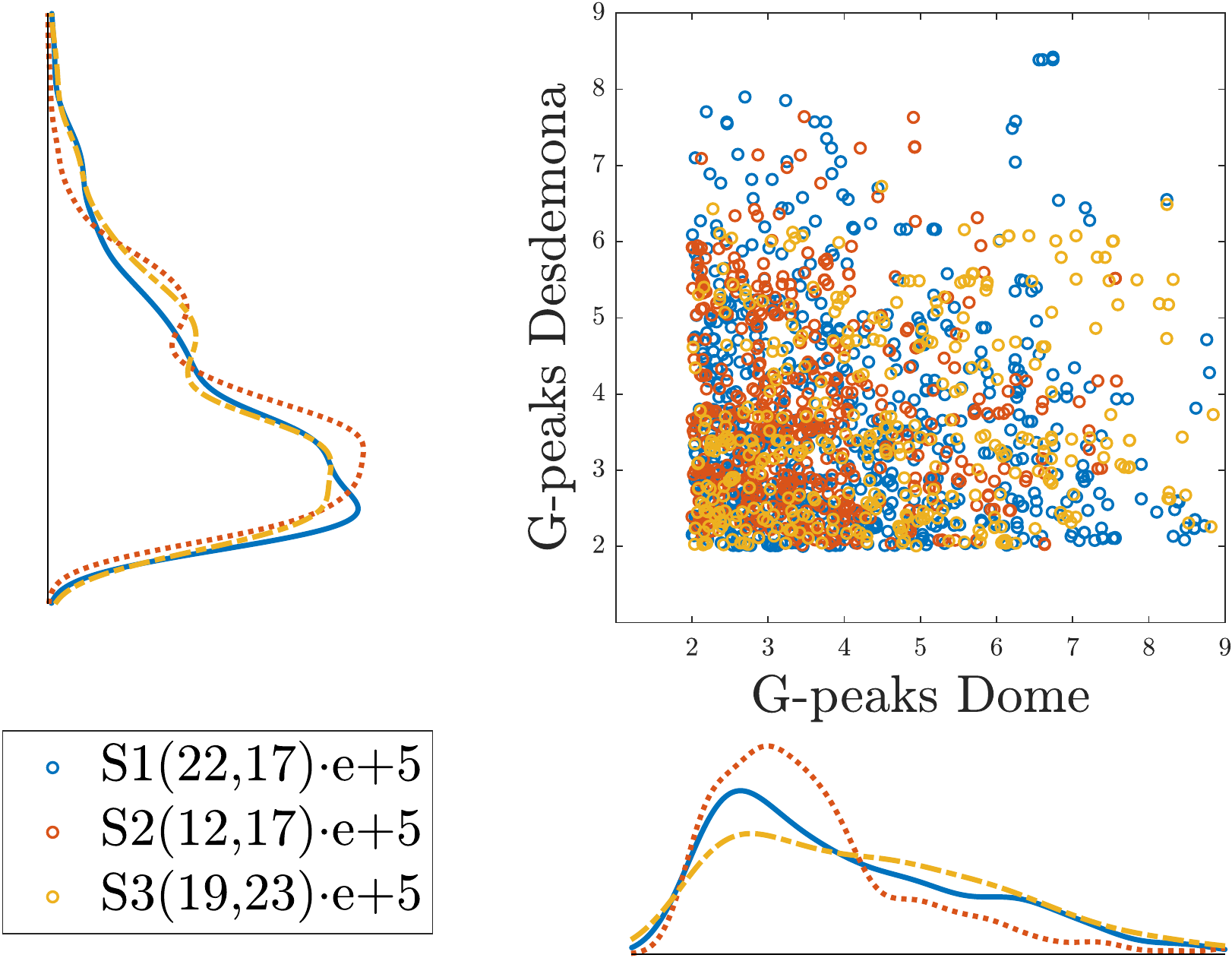}
\caption{All subjects, all days}
\label{fig:sc_dodesd_all}
\end{subfigure}%
\caption{Comparison Dome versus Desdemona with Scatter-histograms of F-16 model G-peaks.}
\label{fig:sc_dodesd}
\end{figure}

The G-levels of the F-16 model were collected in both the stationary Dome and the centrifuge Desdemona simulators across all training (N1--N3) and evaluation (N4) days. To investigate potential differences between the simulators, these G-levels were processed first. Because the subjects spent most of their time at the baseline G-level, we focused our analysis on their `willingness' to venture into higher G-regions. That is, all data below 2 G were discarded, and we analyzed the resulting `G-peak' data using scatter-histograms, illustrated in \figref{fig:sc_dodesd}.

These G-peaks reflect the high G-levels pulled during sharp turns, the Dive-In and Climb-Out instances. \figref{fig:sc_dodesd} shows a comparison of the G-peaks measured in the Dome (horizontal-axis) versus the Desdemona (vertical-axis) per subject. Whereas Figures~ \ref{fig:sc_dodesd_s1}-\ref{fig:sc_dodesd_s3} show the distributions for the three subjects S1-S3 averaged per experimental day (N1-N4), 
\figref {fig:sc_dodesd_all} shows the data averaged over all days for the three subjects.\footnote{The numbers next to the legends indicate the number of data points (Dome, Desdemona) for that day, e.g. N1-(17,1.7)$\cdot$e+5 in the top left.} 

\begin{figure}[ht]
 \centering
 \includegraphics[width=0.45\textwidth]{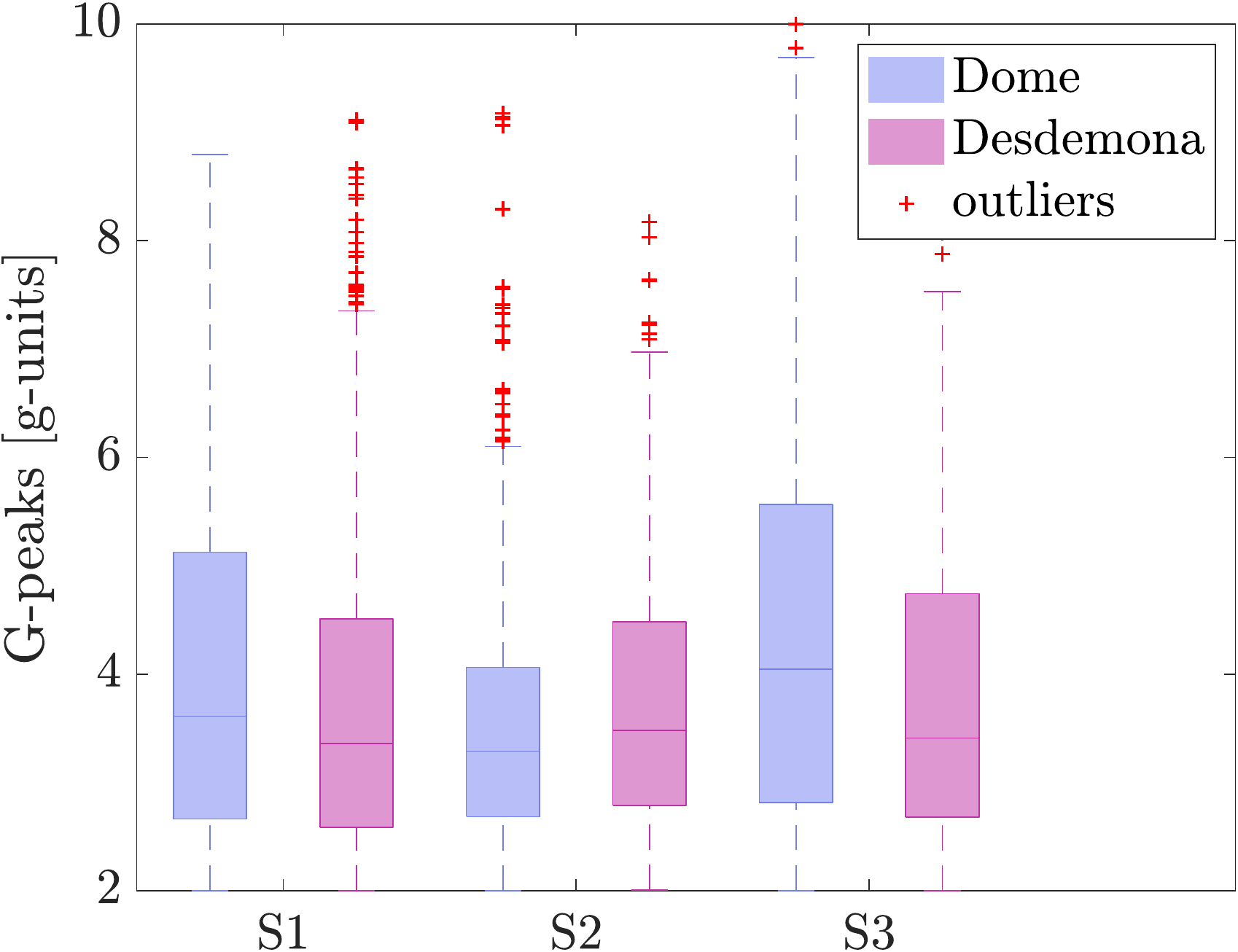}
 \caption{Boxplots of F-16 model G-peaks (Dome versus Desdemona, all subjects, day N4).}
 \label{fig:dome_desdG}
\end{figure}

Examining the distributions in \figref {fig:sc_dodesd_all}, and the boxplots of the data obtained on the evaluation day N4, \figref{fig:dome_desdG}, the differences between the Dome and Desdemona simulators are small, suggesting that the G-profiles flown by our subjects were indeed comparable. The high actual G-levels in the centrifuge did not lead the three pilots to reduce the intensity of their maneuvering. For both simulators the distributions are skewed, their modes occur around 3 - 3.5 G; recall that these are the G-levels of the simulated F-16 {\em model}, not the actual G-levels in Desdemona.

\begin{figure}[ht]
 \centering
 \includegraphics[width=0.45\textwidth]{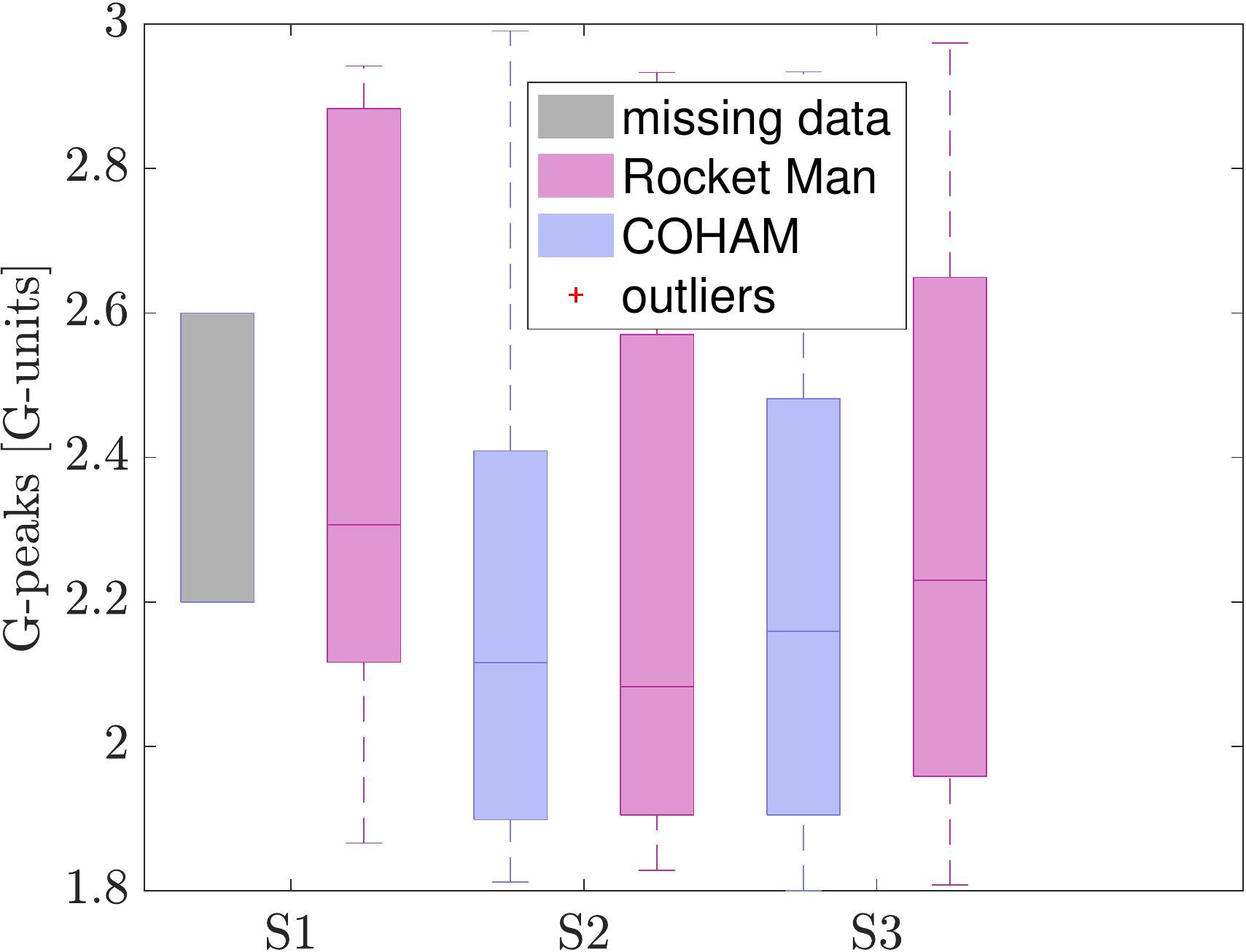}
 \caption{Boxplots of simulator G-peaks in Desdemona (RM versus COHAM, all subjects, day N4).}
 \label{fig:g_comparison}
\end{figure}

The actual G-levels in Desdemona were also recorded. \figref{fig:g_comparison} illustrates these G-levels for the RM and COHAM filters for all three subjects. Unfortunately, the data for Subject S1 in the first round (RM) were lost. These boxplots show that the G-levels were comparable in both cueing settings.

\subsubsection{Cabin Rotations}

\begin{figure}[ht]
 \centering
 \includegraphics[width=0.5\linewidth]{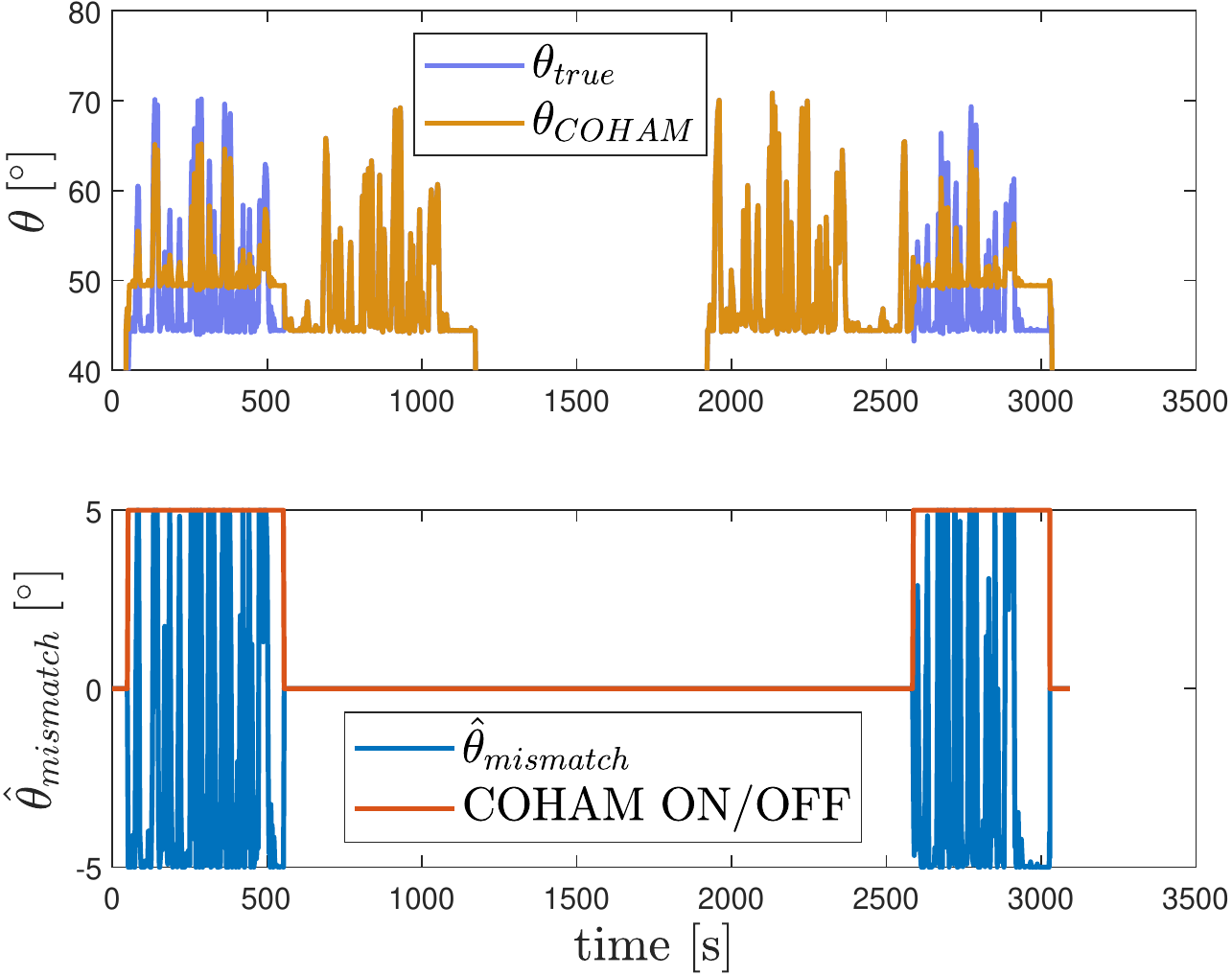}
 \caption{Cabin coordination angles \gls{thetacoham} versus $\theta_{true}$ (Subject S2, day N4) (top), and the mismatch (bottom), when the COHAM filter was active, or not.}
 \label{fig:theta_traces}
\end{figure}

\begin{figure}[ht]
 \centering
 \begin{subfigure}[b]{.4\linewidth}
 \includegraphics[width=\linewidth]{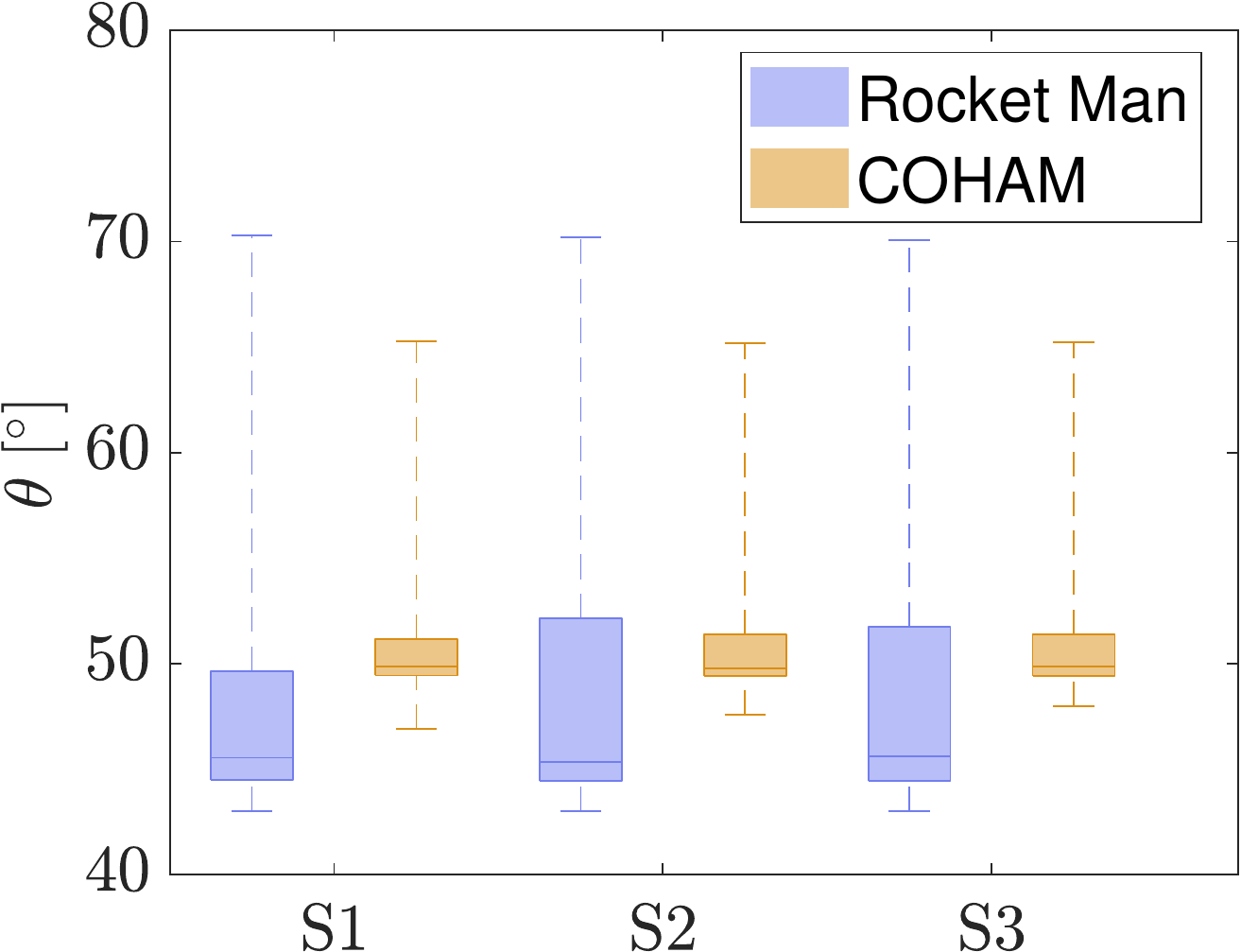}
 \caption{Rotation}
 \label{fig:bp_theta}
 \end{subfigure}%
 \begin{subfigure}[b]{.4\linewidth}
 \includegraphics[width=\linewidth]{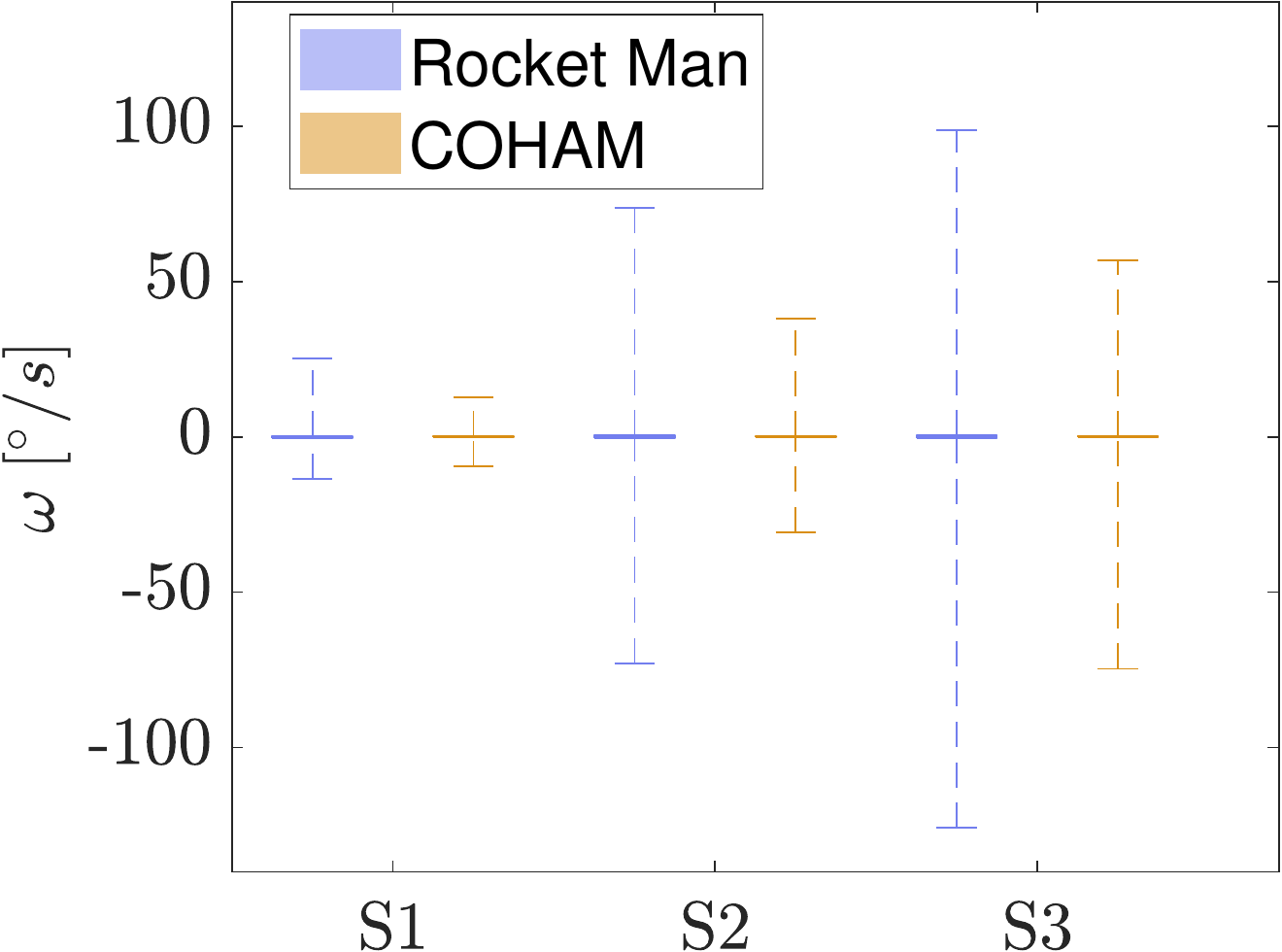}
 \caption{Rotation rate}
 \label{fig:bp_dtheta}
 \end{subfigure}%
 \caption{Boxplots of cabin rotation and rotation rate.}
 \label{fig:boxplots_th_thdot}
\end{figure}

\figref{fig:theta_traces} shows time histories of the true cabin rotation angle $\theta_{true}$ and $\theta_{COHAM}$ for subject S2, in the order C2 C1 C1 and C2, see \tabref{tab:schedule}. Similar to the cabin rotation, also the rotation rates $\omega$ (not shown, but see \figref{fig:pitch_rate}) are smaller with COHAM.
\figref{fig:boxplots_th_thdot} shows boxplots of the cabin rotation (left) and rotation rates (right) for all three subjects, measured on the evaluation day (N4) for RM and COHAM. These clearly show that COHAM considerably reduces the cabin rotations, which in turn (see \eqref{eq:holly}) reduces the Coriolis effects (not shown). This reduction indeed yielded lower sickness and dizziness in our subjects, especially when time progressed, however, the extent of mismatch coordination needs to be further investigated as two subjects reported issues related to the G-misalignment.

\section{Discussion}
\label{sec:disc}

A G-cueing filter was developed, the Coherent Alignment method (COHAM), to reduce Coriolis effects in human centrifuges. A four-day evaluation campaign with three fighter pilots was conducted to test its effectiveness. Pilots performed a high G-maneuver F-16 control task in the centrifuge-based motion simulator Desdemona, preceded by a training phase in the fixed-based Dome simulator. In the following, we will reflect on the experimental results using the four hypotheses stated in \secref{sec:hypos} and then discuss some recommendations for future work.

\subsection{Reflection on Experiment and Hypotheses}

The experiment, albeit performed with only three subjects, provided some fair evidence for supporting the first hypothesis. The lower rotations and rotation rates with COHAM led our subjects to ultimately prefer the COHAM cueing, while achieving comparable G-levels as with the conventional cueing. Although our subjects quickly became accustomed to the high G-levels, as witnessed by the steep decline of the MISC ratings over the three days, these ratings were smaller for COHAM until the last runs, comparison R2. The AWS, PBWS and Overall Preference scores were all in favor of COHAM after R2. The AWS scores illustrate that sickness and especially dizziness increase less with COHAM as compared to the conventional RM cueing. Overall, COHAM was considered more `bearable'.

Our second hypothesis, that COHAM is coherent, is rejected. The G-alignment scores (ALAC, AMS) all show that COHAM performed worse than RM, and two of our subjects reported the cabin misalignments with COHAM. This lack of coherence was interpreted differently, however, as the slightly higher cabin rotation led one of our subjects to believe that the simulator G loads were higher. The other subject also reported misalignments for RM, whereas there were none. Apparently, it is difficult to perceive the G-vector alignment in dynamic G maneuvering tasks. Despite the perceived misalignments with COHAM, it was still the preferred cueing setting.

The distribution of G-levels was used as an objective measure to assess whether the centrifuge would affect the subjects' performance. The distributions of G-peaks do not show differences between the Dome and centrifuge G-levels, despite the fact that the pilots were operating under considerable G-strain in Desdemona and high yaw rates up to (150 deg/s). This leads us to accept the third hypothesis, which has important implications for training, as apparently the fixed base Dome simulator did prepare the subjects well enough to produce a set of prescribed G-maneuvers.

For the given setup and subject profile, a systematic centrifuge training of approximately one hour a day over the course of three days has shown to produce nearly `exemplary' subjects that are able to cope with high frequency and magnitude of G-onsets, and thus large Coriolis effects, for a considerably long piloting task. The substantial decrease of the MISC ratings for all subjects over the three days shows that they became better capable of enduring almost a full hour of centrifuge training. Pilots indeed learn to cope with the G-environment relatively quickly, supporting hypothesis four. In this respect, the trends in almost all ratings are similar, and show that the COHAM cueing indeed is more bearable than the conventional cueing. The effectiveness of COHAM becomes more salient for longer centrifuge sessions.

The recovery of the MISC ratings, clearly seen over the progression of the runs from day N2 to N4, suggests that pilots have a buffer to sustain  certain levels of instantaneous G-onsets. The increase of MISC at the end of the sessions suggests that this buffer will eventually deplete, and subjects become less able to recover from a rush of nausea. 
While these observations are supportive of the positive adaptation effect in training \cite{Young2003,Newman2013}, the coping mechanism can also be attributed in part to adjustments in head movements. As minimal head movements reduce the active Coriolis effect, this negatively impacts the pilot's natural head-scanning behavior.

The setup of the experiment allowed us to investigate the effects of centrifuge training on our subjects' ability to adapt to high G-levels. The novel metrics (PBWS, AWS, ALAC, AMS) proved to be useful to characterize our subjects' well-being, endurance and witness their growing capacity to deal with the high G-onsets. For instance, the AWS allowed to test the motion cueing on different levels of well-being metrics, such as accumulated dizziness, sickness, comfort and bearability, that would otherwise remain implicit in the MISC.

\subsection{Recommendations}

An obvious recommendation is to continue investigating the tuning of COHAM, but with more pilots. Although some clear trends were present in the data, inter-subject differences exist, and the number of participants needs to be considerably higher to draw any firm conclusions. 

Regarding the Coherence Alignment Zone (CAZ), a number of recommendations can be made. First, in this work the CAZ thresholds were obtained using naive subjects, not using experienced fighter pilots. The reported mismatches by two of our three subjects suggests that their CAZ thresholds may be considerably lower. Second, the COHAM mismatch of 5 degrees represents a {\em static} CAZ threshold, which might depend on the experienced G-level, and we only measured it for relatively low G levels. Perhaps a more {\em dynamic}, i.e., G-level-dependent threshold could be determined and used in COHAM. Our evaluation suggests that the effects of a {\em prolonged exposure} to a static cabin mismatch, and the resulting perceived G-level, needs to be further investigated. That is, while maneuvering, do pilots obtain information to `calibrate' the alignment, and with that start to notice any misalignments? 

Regarding COHAM, the fixed tilt offset which was noticed by one subject is not a real requirement of the filter, it was only implemented to further limit the cabin rotations. COHAM can be easily adapted to maintain true coordination in the baseline, which would, however, render it less effective. An alternative, although more complex, could be to develop a mechanism that detects `prolonged phases of steady flight', and during these periods bring the cabin alignment back to the baseline. For simulation training sessions where the trajectory to be flown, or the tasks to be conducted, are to some extent predictable, this could be a viable option.

In this respect, for high-performance maneuvering with repetitive trajectory pursuit tasks, (i.e., where subjects are asked to track or produce a specific G-profile) our study suggests that systematic training can help pilots to achieve consistent performance in the centrifuge. However, for tasks with an unpredictable outcome (e.g., chase, dogfight) the differences may be larger. In addition, when pilots are instructed to carry out concentration demanding tasks, or simultaneously make difficult decisions, their performance might be affected under high G-load \cite{gtrack}. It is recommended to further investigate this and study how the increased endurance with COHAM can be better exploited.

Finally, the correspondence of our subjects' verbal reports with their AWS scores suggests that for future studies, these scales could indeed provide a convenient and simple way of assessing the subject's accumulated well-being in centrifuge training. Various coping mechanisms can be at work here, and it is recommended to investigate ways to clarify what mechanisms work best, whether they can be related to subjective ratings, and how they can be trained.

\section{Conclusions}
\label{sec:conc}

A novel motion cueing filter for centrifuge-based simulators is presented, COHAM, which aims to mitigate adverse Coriolis effects through reducing cabin rotations. Results show that for prolonged centrifuge sessions, in particular featuring high G-maneuvers, the filter was found to reduce sickness and dizziness, and increased the subjects' bearability and comfort. Objective performance metrics, such as average G-levels, did not change relative to a fixed base simulation. The novel filter may allow for longer exposure of pilots to elevated G-levels in centrifuge-based simulators. Future research focuses on whether dynamic thresholds, changing with actual simulated G-levels, can lead to improvements.

\bibliography{bibliography/bibliography.bib}

\end{document}